%% file: Reporting_Bias_Caliendo_et_al.tex
\documentclass[11pt]{article}
\usepackage[a4paper,left=2.5cm,right=2.5cm,top=2.5cm,bottom=2.5cm]{geometry}
\usepackage[T1]{fontenc}
\usepackage[utf8]{inputenc}
\usepackage[english]{babel}
\usepackage{lmodern,textcomp}
\usepackage[x11names]{xcolor}
\usepackage{copyrightbox}
\usepackage{booktabs}
\usepackage{graphicx}
\usepackage[normalem]{ulem} 
\usepackage{comment}

\usepackage[compress]{natbib} \setcitestyle{aysep={,},yysep={,}}
\usepackage{setspace} 

\usepackage[colorlinks=true,linkcolor=blue,citecolor=black]{hyperref}

\usepackage{csquotes}
\usepackage{graphicx}
\usepackage{amsmath}
\usepackage{amssymb}
\usepackage{multirow}

\usepackage{caption}
\usepackage{subcaption}
\usepackage{dsfont}
\usepackage{authblk}

\usepackage{float}

\usepackage[title]{appendix}

\usepackage{threeparttablex}

\newcommand{\figtext}[1]{
    \captionsetup{justification=justified,font=footnotesize}
    \caption*{#1}
}

\newcommand{\fignote}[1]{\figtext{\emph{Note:~}~#1}}


\newcommand\Katrin[1]{\textbf{\color{magenta}Katrin: \textit{#1}}}
\newcommand\Marco[1]{\textbf{\color{red}Marco: \textit{#1}}}

\newcommand\Jakob[1]{\textbf{\color{olive}Jakob: \textit{#1}}}
\newcommand\new[1]{\textbf{\color{orange}To Do: \textit{#1}}}

\begin{document}
\pagestyle{plain}

\renewcommand{\baselinestretch}{1.12}

\title{\vspace{-8mm}\textbf{On the Extent, Correlates, and Consequences of Reporting Bias in Survey Wages}\thanks{We thank Jonas Jessen, Nikolas Mittag, Markus Nagler, Michael Oberfichtner, Nico Thurow, Erwin Winkler, participants at the annual meetings of EALE (2023), and Verein für Socialpolitik (2023), and seminars at the University of Potsdam, and the University of Trier for their valuable comments and suggestions. We are grateful to Maximilian Biller and Jessica Fuchs for excellent research assistance. Support by the Deutsche Forschungsgemeinschaft (DFG, German Research Foundation) through CRC TR 224 (Project A02) is gratefully acknowledged.}}

\author{\textbf{Marco Caliendo}\thanks{%
		e-mail: \texttt{caliendo@uni-potsdam.de}}\\ \vspace{-4mm}
	\textit{University of Potsdam, CEPA, BSoE, IZA, DIW, IAB} \\
	\textbf{Katrin Huber}\thanks{Corresponding author. Address: University of Potsdam, August-Bebel-Str. 89, 14482 Potsdam, Germany, e-mail: \texttt{katrin.stephanie.huber@uni-potsdam.de}} \\
	\textit{University of Potsdam, CEPA, IZA, BSoE} \\
	\textbf{Ingo E. Isphording}\thanks{e-mail: \texttt{isphording@iza.org}} \\
	\textit{IZA, CESifo} \\
	\textbf{Jakob Wegmann}\thanks{e-mail: \texttt{jakob.wegmann@uni-bonn.de}} \\
	\textit{University of Bonn} \\\vspace{0.15cm}
}

\date{Working Paper\\ \today \vspace{-8mm}}
\maketitle
\thispagestyle{empty} 
\renewcommand{\abstractname}{Abstract}
\begin{abstract} 
\begin{singlespace}

\noindent Surveys are an indispensable source of data for applied economic research; however, their reliance on self-reported information can introduce bias, especially if core variables such as personal income are misreported. To assess the extent and impact of this misreporting bias, we compare self-reported wages from the German Socio-Economic Panel (SOEP) with administrative wages from social security records (IEB) for the same individuals. Using a novel and unique data linkage (SOEP-ADIAB), we identify a modest but economically significant reporting bias, with SOEP respondents underreporting their administrative wages by about 7.3\%. This misreporting varies systematically with individual, household, and especially job and firm characteristics. In replicating common empirical analyses in which wages serve as either dependent or independent variables, we find that misreporting is consequential for some, but not all estimated relationships. It turns out to be
inconsequential for examining the returns to education, but relevant for analyzing the gender wage gap. In addition we find that misreporting bias can significantly affect the results when wage is used as the independent variable. Specifically, estimates of the wage-satisfaction relationship are substantially overestimated when based on survey data, although this bias is mitigated when focusing on interpersonal changes. Our findings underscore that survey-based measures of individual wages can significantly bias commonly estimated empirical relationships. They also demonstrate the enormous research potential of linked administrative-survey data.\\[0.25cm] 
\thispagestyle{empty} 
\end{singlespace}

\end{abstract}

\newpage

\section{Introduction}

While economists often disagree on many issues \citep{kearl1979confusion,fuchs1997economists}, there is a consensus on the importance of wages and personal income. Individual wages are a core component of virtually every model in labor economics and related fields. Economic theory predicts close relationships between wages and individual decisions about labor supply \citep{Pencavel1986, Mulligan2008}, household investment \citep{Acemoglu1997}, consumer demand, or savings \citep{Sandmo1970}. Wage levels are considered an important measure of economic well-being \citep{deaton2008income}, and are closely linked to other economic outcomes such as education \citep{Becker1966}, health \citep{deaton2003health}, and social mobility, and determine tax liabilities and the eligibility for transfer payments. 

Applied researchers working on wage-related issues can choose between two main types of data. Register-based administrative data are considered highly reliable, often offering superior measurement quality, larger population coverage, and arguably higher consistency, precision, and more regular updating \citep{kunn2015challenges}. Survey-based self-reported data, on the other hand, provide access to interesting variables that are essential for many modern labor economics applications -- personality traits, risk preferences, and cognitive skills are examples -- but are typically not available in administrative data \citep{Borghans2008, Arni2014, kunn2015challenges}. Moreover, survey data can be flexibly adapted to the specific needs of researchers \citep{Stantcheva2022}. As a result, survey data complement or even replace administrative records, depending on the research question. However, the reliability and quality of self-reported wage information has often been questioned \citep{Bound2001}.

Against this background, this paper addresses three main research questions. First, we examine the extent to which survey respondents misreport their wages. Second, we analyze which observable characteristics can predict the extent of misreporting. Third, we examine the consequences that applied researchers face when working with these misreported wages. We use the novel and unique record linkage of the SOEP-ADIAB. This dataset combines the German Socioeconomic Panel (SOEP), one of the longest running and most widely used household surveys in the world \citep{Goebel2019}, with high quality administrative data from the Integrated Employment Biographies (IEB) of the Federal Employment Agency, covering all dependent employment in Germany. This unique data linkage provides us with 59,118 linked survey/administrative wage observations on about 5,500 individuals for up to 36 years. 

With this novel data setup, we contribute new answers to an established literature that dates back to early studies by \citet{Bound1991} and \citet{Pischke1995} and that has received continued attention from researchers \citep{Antoni2019,Angel2019,roth2021gender}. The unprecedented scope of data in the SOEP-ADIAB, both in terms of breadth of data and individual time series of linked subjective and administrative wage information, allows us to provide a more detailed characterization of the reporting bias in survey wages than previous studies. The linkage also provides us with a broad portfolio of available control variables at the individual, household, job and firm levels to examine a comprehensive set of predictive factors that goes beyond the existing literature. Finally, we re-examine common empirical relationships using wages as both dependent and independent variables, describing meaningful consequences of choosing self-reported survey wages over administrative records for typical empirical exercises in applied economics.

Our analysis consists of three steps. First, we quantify the extent of reporting bias. Mean wages differ moderately between SOEP and IEB, with SOEP respondents on average underreporting their individual wages by \texteuro186, with a mean relative reporting bias of 7.3\% . Consistent with previous research \citep[e.g.,][]{Bound1991, Bound2001, Kim2014, Bollinger2018}, we find evidence of a regression-to-the-mean component in the reporting bias. Low-income individuals in the bottom four ventiles overestimate their true wages on average. The reporting bias becomes negative from the fifth ventile and increases steadily with income, both in absolute and relative terms. Rounding contributes only modestly to the reporting bias, and changes in wages appear to be more accurately captured by survey responses than levels. The results are robust to several alternative wage concepts and definitions.

Second, we analyze which factors at the individual, household, job and firm levels can predict reporting bias. At the individual level, only a few characteristics such as gender and extroverted personality play a role in explaining the variation in reporting bias. At the household level, partners' income in particular in interaction with one's gender, predicts the direction and magnitude of the reporting bias, suggesting that social dynamics within households influence the reporting bias, as recently shown by \citet{roth2021gender}. The strongest predictors are job and firm level characteristics, with union membership, firm size, average wage level, and workforce composition strongly correlated with reporting bias.

Third, we examine the consequences of using survey rather than administrative wages in commonly analyzed economic relationships. We consider two settings in which wages are the dependent variable: returns to education and the gender wage gap. In the case of returns to education, the choice of data appears to be largely inconsequential; estimates based on SOEP and IEB wages are similar in magnitude, and a two-sample GMM t-test rejects significant differences in all but one specification. In the case of the gender wage gap, the significant relationship between gender and reporting bias leads to a small but significant overestimation of the gender wage gap that links to systematic variation of misreporting by gender. Estimates differ moderately, with the conditional gap shrinking from 10.7\% to 7.7\% based on IEB data, with a two-sample GMM t-test confirming significance of this difference.

To examine the sensitivity of the relationships with wages as the independent variable, we estimate wage effects on self-reported satisfaction. We show that using self-reported wages leads to a substantial overestimation of the effect of wages on subjective indicators such as satisfaction with personal income. For example, while satisfaction with personal income increases with an additional \texteuro 1,000 by about 25.4\% of a standard deviation when assessed on the basis of the SOEP, this relationship shrinks by a quarter to only 20.2\% of a standard deviation when assessed on the basis of IEB data. This bias is exacerbated with extended control variables, but mitigated by relying on inter-personal changes. The difference runs counter to the intuition of attenuation bias by classical measurement error, but rather hints at the role of systematic correlates of the measurement error \citep{bertrand2001people}. 

Taken together, our results should be read as a cautionary tale for researchers relying on self-reported measures of income and wages. While the reporting bias in wages turns out to be moderate, it is systematically related to several characteristics, and thus can -- depending on the application -- significantly bias commonly estimated empirical relationships. This is especially true when wages are used as explanatory variables to explain subjective measures such as well-being, concerns, or attitudes. In these cases, subjective measures of the same individual enter the estimation simultaneously as both dependent and explanatory variables. This may exaggerate problems of omitted variables driving the relationship, e.g. through individual-specific reporting behavior or reported wages reflecting desired or aspired instead of actual wages \citep{bertrand2001people,prati2017hedonic}. 

Despite these implications, representative survey datasets such as the SOEP remain indispensable for applied economics, underpinning many influential studies and reports. Survey data allows to augment potentially more reliable information from administrative sources with information on attitudes, preferences, non-cognitive skills, thus allowing for more far reaching explorations of the determinants of wages. However, given the sensitivity of results to misreporting bias, findings should be corroborated with more reliable income measures whenever possible. Where such measures are not available, the results should be interpreted with appropriate caution. We see the need for further research to extend our exemplary analyses of common empirical relationshops -- focusing on the returns to education, the gender gap, and the wage-satisfaction relationship -- to additional research questions. This will help to get a clearer picture of the importance of the problem of misreporting and will further highlight the potential benefits of linked administrative-survey data.

Our analysis contributes to a long-running literature using linked data sources to assess the reliability of survey wages \citep[see][for a comprehensive review]{Mooreetal2000, Bound2001}. In a seminal study, \citet{Duncan1985} describe substantial bias in the comparison of administrative records with employee survey responses from a manufacturing firm. Subsequent studies \citep[e.g.,][]{Bound1991, Bound1994, Pischke1995} extend this cross-sectional single-firm analysis to multiple firms over two waves. Overall, the results confirm the existence of reporting errors and show that they are correlated over time \citep{Bound1994}. Other studies have examined the reliability of survey responses beyond self-reported wages, e.g., hours worked \citep{Duncan1985}, occupation \citep{Isaoglu2010}, pensions and unemployment benefits \citep{Angel2019, bollinger2022income}, or loans and debts \citep{Madeira2022}. Several studies have extended the originally mostly US-based literature to test the reliability of German survey data \citep{Antoni2019,schmillen2024measurement,valet2018comparing}. We contribute to these studies by relying on an unusually large sample, both in number of observations and length of individual time series, from a fundamental dataset in economic research. The scope of our data allows for more in-depth analysis, a broader set of potential correlates of reporting bias, and the analysis of changes in reporting bias over time. Compared to previous results, we find established patterns of mean reversion \citep[e.g.,][]{Bound1994,Pischke1995}, while the extent of misreporting is rather moderate, nonetheless comes with substantial consequences for applied work.

Second, we complement several previous studies by examining which individual-, household-, job- and firm-level factors are correlated with reporting bias. \citet{Duncan1985} show a correlation between misreporting and job tenure, \citet{Kim2014} find that education, race, and wage level are associated with reporting bias. \citet{Antoni2019} explore whether reporting bias could be caused by socially desirable reporting and whether the error is affected by interview characteristics. \citet{Angel2019} also find evidence for social desirability as a possible explanation for reporting bias. Finally, \citet{roth2021gender} investigate whether partner income influences wage reporting. Most recently, \citet{Meyer2024race} provide a literature review summarizing patterns of greater measurement error among minorities. Because we are able to draw on the breadth of characteristics covered by the SOEP, in addition to the firm-level variables drawn from the IEB, we contribute to this area of research by jointly estimating the predictive power individual-, household-, job- and firm-level characteristics and examining the relative contribution of explained variance between these groups. 

Finally we contribute to a literature that assesses the consequences of misreported wages for the estimation of economic parameters. Already \cite{solon1992intergenerational} described the role of misreported income for the estimation of intergenerational mobility. \citet{Kim2005} found that reporting errors can lead to misleading conclusions about the procyclicality of real wages.  \citet{Gottschalk2005} and \citet{Gottschalk2010} analyze the effect of misreported survey wages on the probability of estimating nominal wage cuts. \citet{Hurstetal2014} show how misreporting may hinder the accurate estimation of self-employed income. \citet{meyer2019using} demonstrate substantial effects of misreported wages on the estimation of program participation. \citet{Bossler2020} show that misclassifications of minimum wage eligible workers based on survey wages can bias treatment effects by up to 30\%. \citet{roth2021gender} identify the consequences of misreporting due to social norms for estimating the gender wage gap. \cite{Stueber2023} find differences in aggregate measures of wage inequality between unlinked survey data and administrative data in Germany. We complement this literature by demonstrating the importance of misreporting for several commonly analyzed economic relationships relying on wage as either dependent or independent variable.

The rest of the paper is as follows. In chapter \ref{sec:data} we provide a detailed description of our unique dataset. We continue by showing the extent of the reporting bias in chapter \ref{sec:extent} and provide the empirical analysis of correlating factors in section \ref{sec:factors}. Chapter \ref{sec:impl} discusses the implications for economic analyses and chapter \ref{sec:concl} concludes. 

\section{Data} \label{sec:data}
\subsection{Data Sources}

Our analysis is based on a novel linkage of individual- and firm-level data from German social security records with survey data on individuals and households in the German Socio-Economic Panel (SOEP-ADIAB).\footnote{Our analysis relies on a pilot version of the SOEP-ADIAB linked data, which preceeded the now publicly available SOEP-CMI-ADIAB (\citet{Antoni2023}), and is largely identical. We use the SOEP Core v37 (\citet{SOEP2022}), and the SOEP IS 2019 (\citet{SOEPIS2020}).} Starting in 2019, SOEP respondents were asked for their consent to retrospectively link their survey responses with their employment histories (\textit{Integrierte Erwerbsbiographien}, IEB).

The IEB provide detailed and accurate data on the employment and unemployment histories of individuals in Germany, capturing labor market outcomes such as exact gross wages for each job held. The dataset covers employees subject to social security contributions, workers in marginal employment\footnote{Employees in Germany, with the exception of the self-employed and civil servants, and their respective employers are obliged to pay social security contributions. Employees earning up to a certain amount per month (in our observation period this threshold varied between approximately \texteuro200 in 1984 and \texteuro450 in 2020) are exempt from paying social security contributions. This type of employment is called marginal employment (or ``minijobs'').}, recipients of social security benefits, the registered unemployed, and participants in active labor market programs. It covers both individual- and firm-level data, such as the size of the establishment, the demographics of the employees, and industry sector. However, it does not cover civil servants or the self-employed. In addition, wage data in the IEB are top-coded at the assessment ceiling for social security contributions in Germany. Notably, top-coding affects only 5.2\% of all employment spells between 1975 and 2019, although this rate varies across subgroups and has increased over time.

The German Socio-Economic Panel (SOEP) has been conducted annually since its inception in 1984 and currently includes about 30,000 respondents from 15,000 households. The SOEP contains detailed information on self-reported gross wages. In addition to wages, it collects extensive self-reported data on individual and household characteristics such as personality traits, risk preferences, worries, beliefs, and household composition. 

By linking the IEB and SOEP, we can directly compare individuals' administrative and self-reported wages to assess the extent of reporting bias. In addition, we can examine factors that are correlated with reporting discrepancies at the individual, household, job and firm levels. In the following, we discuss details of the sampling process and data linkage underlying our empirical analysis.


\subsection{Data Cleaning, Linkage and Sample Restrictions}

\paragraph{Data Linkage} We prepare spells of the administrative information in the SOEP-ADIAB linkage data using standard procedures following \citet{dautheppelsheimer20}. The preparation consists of splitting overlapping spells, creating biographical variables, merging information from the Establishment History Panel (EHP), and cleaning occupational and educational information. To this data, all respondents to the SOEP are prepared for linkage who consented to this procedure. We include both individual respondents from the SOEP core sample and the SOEP IS sample, an independently drawn sample used to pilot new questions and running since 2011 \citep{RichterSchupp2015}. 
 
Unlike the annual SOEP data, the IEB data are stored in spell form. Each work spell is stored with a start date and an end date. We overcome this mismatch in the data structure by linking consenting SOEP observations to all administrative spells that either overlap with the SOEP interview or ended in the month prior to the interview. If SOEP interviews coincide with more than one parallel IEB spell, indicating multiple jobs at the same time, we sum wages from all concurrent spells. This linkage results in an unbalanced panel structure with one observation per individual per year. We link additional establishment-level information from the EHP to each respective administrative spell.\footnote{In the case of multiple spells, we set establishment-level information to missing, leading to slightly reduced numbers of observation in some analyses.}


\paragraph{Sample Restrictions} Table \ref{table1} provides details on the number of successful linkages and the subsequent sample restrictions required. Of the 23,525 observations being asked about linkage in 2019, 15,012 individuals consented.\footnote{Table \ref{tab:descr_cons} in the Appendix describes differences in covariates by consent. Individuals who consented to be linked and those who refused to be linked do not differ systematically on most basic characteristics, and when they do, the differences are only very small. Similar findings are reported by \citet{SakshaugKreuter2012} for the case of linkage of IEB to the ``German Labour Market and Social Security'' (PASS) study. Selective consent appears to be small, especially for substantive variables like employment, income and benefit recipiency.} Of the consenting respondents, 14,983 can be successfully linked to their social security records. These linked individuals have at least one spell in the IEB data set, i.e., they were in dependent employment, active labor market policies or officially registered as unemployed at least once during their working life. Rows 4 to 9 document how many observations we can observe, conditional on the necessary sample restrictions. Among the 14,983 successfully linked individuals, 7,614 respondents have a concurrent work spell in the IEB data in 2019 and thus serve as the basis for our sample. In the SOEP questionnaire, individuals are asked to report their last month's wage income. Accordingly, we restrict the sample to those individuals who, at the time of the interview, have a work spell that started at least 30 days before the interview to ensure that this survey and the administrative wage measure the same job and period.

We impose several additional sample restrictions. First, we drop individuals who report only self-employment in the SOEP but still have income from dependent employment in the IEB. 
For these individuals, we do not know whether their SOEP wage includes only income from self-employment or whether they have also income from dependent employment. Since income from self-employment cannot be measured in the IEB, this would mechanically inflate the SOEP wage. Second, we drop individuals below the age of 20 or above the age of 65 in order to restrict our sample to the working age population. We condition the sample on individuals with SOEP and IEB wages greater than zero, thus entirely focusing on misreporting bias on the intensive margin. We acknowledge, though, that income nonresponse is a further source of misreporting \citep{Bollingeretal2019}.\footnote{The IEB wage is reported as 0 in cases of ``employment interruptions''. During these periods, the employment relationship continues legally, but no remuneration is paid, e.g., illness after the end of continued pay, maternity leave and sabbaticals \citep{Frodermann2021}.}$^,$\footnote{We also drop individuals in informal employment, i.e., who are listed with a zero wage in the administrative data, but have a positive wage from non-self-employment in the SOEP. This amounts to only about 0.35\% of the spells in the merged sample before any other restrictions are imposed.}  Finally, we restrict our analysis to individuals with neither censored administrative wages nor survey wages that exceed the assessment limit for the social security contributions.\footnote{Following \cite{dautheppelsheimer20}, we drop incomes with an error margin of \texteuro120 surrounding the assessment limit.} 
These restrictions leave us with 6,136 individuals for whom we observe survey and administrative wages, with about 10\% based on the shorter-running SOEP IS sample. On average, we observe respondents for about ten waves, with the longest running individual linked histories going back to the very first SOEP wave in 1984 (Figure \ref{fig:periods} in the Appendix). This leaves us with 59,118 individual-year observations. The number of observations in various steps of the subsequent analysis may very further due to item non-response in the covariates.


\begin{center}
	[Insert Table \ref{table1} about here]
\end{center}

\paragraph{Definition of the Reporting Bias} We define reporting bias as the difference at the individual level between equivalent measures of gross monthly wage in both the SOEP and the IEB. In the SOEP, respondents are asked to report the gross monthly wage they have earned in the month prior to the interview. During the interview, respondents are also allowed to look at their pay slip if they do not know their wage. They are asked not to include special payments such as vacation pay or back pay, but to add overtime compensation.\footnote{The exact wording in the 2019 English version of the SOEP is as follows: ``\textit{What did you earn from your work last month? Please state both: gross income, which means income before deduction of taxes and social security and net income, which means income after deduction of taxes, social security, and unemployment and health insurance. If you received extra income such as vacation pay or back pay, please do not include this. Please do include overtime pay. If you are self-employed: Please estimate your monthly income before and after taxes.}''} Based on this information, we construct the gross monthly survey wage. In robustness checks, we construct alternative survey wage measures that include information on annual wages instead of monthly wages, as well as information on special payments (for details, see section \ref{sec:extent} and Appendix \ref{sec:app_A}).

The IEB data include administrative information on gross \textit{daily} wages. These are calculated by dividing the employer's information on the total wage of the employment period by the number of calendar days in the same period. 
In the case of parallel spells linked to a single SOEP observation, we compute the sum of gross daily wages, as the SOEP survey question aims at income from any work and is thus not restricted to wage income from the main job. While bonus payments are included in the IEB wages, the reporting of special payments to employees is not mandatory. 
In order to check the extent to which such extra payments might affect our results, we perform robustness checks as described in Appendix \ref{sec:app_A}. Two limitations of the IEB wage information have to be taken into account: First, until 1998, wages that were below the marginal employment threshold were not reported and appear as a zero wages in the administrative data. Consequently, the corresponding SOEP observation is also dropped. Second, wages are winsorized at the assessment limit for social security contributions. While there are ways to impute wages above the assessment ceiling, we refrain from doing so because this imputation procedure would artificially affect the measurement error as our key variable of interest. Instead, winsorized observations are excluded from the analysis. 

To construct a \textit{monthly} measure of wages from the reported daily wages, we follow the common practice in IEB-based studies of multiplying the daily wage by $\frac{365.25}{12}=30.4375$, which is the average number of days in each month of a year, taking into account leap years. We report the \textit{absolute reporting bias} as the simple difference between the wage measured in the SOEP and the wage measured in the IEB data. We also construct a \textit{relative reporting bias} by dividing the absolute reporting bias by the administrative gross wage at the individual level. In addition, we classify individuals according to whether they  ``overreport'' or ``underreport'' their survey wage relative to the administrative wage. In this classification, we omit individuals who are close to reporting correctly, defined as deviating by less than 2.5\% based on the annuals distributions of reporting bias. In robustness checks, we vary this definition from 1.25\% to 5\% around zero bias, and the results remain qualitatively and quantitatively unchanged.


\section{Empirical Analysis} \label{sec:empanalysis}

\subsection{The Extent of the Reporting Bias} \label{sec:extent}

We first document the extent and heterogeneity of the reporting bias between the SOEP and the IEB. Figure~\ref{fig:descr_wages}a shows that the administrative and self-reported wages follow similar but not perfectly matched distributions. The means of both distributions differ only slightly between the IEB (\texteuro2,292) and the SOEP (\texteuro2,106). Both distributions show a clustering of wages at time-varying earnings thresholds for jobs exempt from social security contributions (marginal employment). Self-reported SOEP wages appear to be more centered around the mean than administrative wages, with fewer people reporting very high or very low wages. Figure~\ref{fig:descr_wages}b illustrates the distribution of reporting bias which is closely centered around zero. Around 71\% of SOEP respondents underreport their wages, with an average underreporting of about \texteuro 186 (dashed line).\footnote{Only for the graphical representation of the reporting bias in Figure \ref{fig:descr_wages}b biases outside the 1st and 99th percentiles are binned to the 1st and 99th percentiles for clarity.} Despite extreme cases of underreporting and overreporting by more than \texteuro5,000, such outliers are rare and do not significantly affect the average reporting bias. Restricting the analysis to biases between the 1st and 99th percentiles of each year's distribution does not significantly change the mean reporting bias, which is then \texteuro190. 

Figure \ref{fig:descr_wages}c shows the reporting bias across the IEB wage distribution in both absolute and relative terms. In particular, there is significant overreporting at very low wages (below the 4th ventile). The mean bias then becomes negative and steadily larger along the income distribution. This is consistent with previous findings in the literature that find evidence for regression-to-the-mean in different settings \citep[e.g.,]{Bound1991,Kapteyn2007,Bollinger2018}, which is to some degree mechanical, with more scope for over-reporting among low income individuals, and more scope for under-reporting among high-income individuals.\footnote{Figure \ref{fig:wage_diff_age} in the Appendix shows no systematic age-related trends in wage reporting bias; it remains stable across different ages.} 

\begin{center}
	[Insert Figure \ref{fig:descr_wages} about here]
\end{center}

Table \ref{table2} provides further insight into the reporting bias by IEB wage ventile to quantify the graphical representation of Figure \ref{fig:descr_wages}. Column (1) shows the mean IEB wage of each ventile, expressed in Euros. For each wage ventile, the table shows the minimum, maximum, mean and different quantiles (5th, 50th, 95th) in Euros. It also shows the proportion of over- and underreporting individuals. These results indicate a mean reversion in misreporting, with low-wage individuals overreporting and high-wage individuals underreporting their income. The mean absolute reporting bias is positive in the first four wage ventiles and then becomes negative from the fifth ventile onward. In relative terms (column 8), the reporting bias is most pronounced in the very low income group, with individuals on average overreporting their wages by more than 200\% in the first ventile and by more than 20\% in the second ventile. However, this pronounced mean reporting bias among very low wage individuals is largely due to outliers who overreported by more than \texteuro5,000. 
Across the wage distribution, the mean relative reporting bias increases only moderately, up to -13.5\% for individuals in the highest wage ventile. Accordingly, the proportion of individuals who overreport their wages decreases from 57.4\% in the first wage ventile to 6.7\% in the last ventile. Overall, Table \ref{table2} corroborates the graphical representation of Figure \ref{fig:descr_wages} and highlights an overall modest reporting bias that nevertheless varies along the income distribution. The first central result of our analysis is thus a modest but economically significant difference in mean reported wages between the SOEP and the IEB, which averages 7.3\% (see column (8) in Table \ref{table2}, last line).

\begin{center}
	[Insert Table \ref{table2} about here]
\end{center}

Figure \ref{fig:descr_wages_changes} shows the reporting bias in wage changes using first differences, thereby examining the accuracy of year-to-year changes in wage income. Panel (a) compares the distributions of these first differences between SOEP and IEB and shows a high degree of similarity. The similarity leads to a narrow and symmetric distribution of deviations, as shown in Panel (b). In Panel (c), we observe that this minimal reporting bias in changes remains consistently close to zero across the entire wage distribution. Consequently, we find that the reporting bias between subjective SOEP wages and administrative IEB wages is significantly smaller when assessing changes rather than levels, suggesting time-invariant reasons for the reporting bias. This contrasts with the results from the seminal study by \citet{Bound1991} and \citet{schmillen2024measurement}, who found that the reliability of matched CPS and administrative data declines when first differences are analyzed. As we will show later in Section \ref{sec:impl}, controlling for individual fixed effects thus turns out to be a remedy to mitigate the consequences of using misreported income. 

\begin{center}
	[Insert Figure \ref{fig:descr_wages_changes} about here]
\end{center}

\paragraph{Robustness to Other Wage Definitions} In Figure \ref{fig:wage_diff_concepts} in the Appendix we examine the robustness of our results to various changes in the SOEP wage definition. We find that adding one-time payments (e.g., Christmas bonuses or a 13th paycheck) or income from additional jobs, or subtracting other sources of income (e.g., self-employment or work in family businesses) does not change our conclusions quantitatively or qualitatively. We also show that using self-reported wage income in the previous calendar year, where participants are asked to report the months they worked as an employee and their average gross monthly wage, leads to comparable results.  

\paragraph{Rounding} One likely explanation for differences between self-reported and administrative wage data is rounding. Survey respondents may rely on rounded wages for convenience, which reduces cognitive load but leads to noisy information. Instead, administrative data are reported by the employer based on the official payroll. Wages based on administrative data should therefore be accurate and unbiased by rounding.

\begin{center}
	[Insert Figure \ref{fig:rounding_numbers} about here]
\end{center}

Figure \ref{fig:rounding_numbers} plots the fraction of SOEP and IEB wages, that surround full increments of \texteuro50 in panel (a), of \texteuro100 in panel (b) of \texteuro500 in panel (c) and of \texteuro1,000 in panel (d). The subfigures show that SOEP wages indeed show a strong excess probability for reporting ``round'' numbers, while IEB wages show evenly distributed probabilities. Panels (a) and (b) show that about 60\% of individuals round to the nearest \texteuro50 or \texteuro100, panel (c) shows that almost 25\% round to the nearest \texteuro500. Rounding to the nearest \texteuro1,000 is not more pronounced in either dataset.

\begin{center}
	[Insert Figure \ref{fig:rounding_II} about here]
\end{center}

To examine the extent to which rounding explains income misreporting, we plot the reporting bias after excluding ``rounded'' SOEP wages (Figure \ref{fig:rounding_II}). Rounding does indeed partially explain misreporting, with the resulting distribution being more closely centered around 0. However, the proportion of underreporting (overreporting) individuals decreases only slightly from 24.5\% to 23.1\% (70.5\% to 70.3\%). Thus, rounding contributes only slightly to the observed total misreporting. 

\subsection{Correlates of Reporting Bias} \label{sec:factors}

\subsubsection{Empirical Strategy}

In the previous section, we have described moderate but economically significant deviations between SOEP and IEB wages. We now proceed to analyze factors that may be predictive of these deviations. The richness of our linked dataset allows us to simultaneously examine several groups of such potential correlates of the reporting bias: individual-level, household-level, and firm-level correlates.\footnote{We also examine the role of survey methodology factors, as previously examined by \cite{Angel2018, Angel2019, Antoni2019}. In particular, we find little influence of panel tenure or interview mode. The results are summarized in Table \ref{tab:yearFE_survey} in the Appendix.} To investigate the correlation of these groups of variables, we estimate a simple linear regression:

\begin{equation} \label{equ:main_analysis}
Y_{it} = \beta_0 + \beta_1 X'_{i(t)} + \xi_{t} + u_{it}
\end{equation}

As the dependent variable $Y_{it}$, we use different operationalizations of the reporting bias. First, we define a binary indicator that takes the value of $1$ if the reporting bias is positive, i.e., if the individual overreports her wage in the SOEP compared to her administrative wage in the IEB. The indicator takes the value of $0$ in the case of correct reporting or underreporting. 
Correct reporting is defined as no or small reporting bias within a +/- 2.5 percentile range of the annual reporting bias.\footnote{Applying alternative bounds at the 1.25th or 5th percentile does not change our conclusions (see Tables \ref{tab:main_table_125} and \ref{tab:main_table_5}).} Second, we use the log reporting bias, the difference in log wages between SOEP and IEB. For this second outcome variable, and to facilitate interpretation of the estimated coefficients, we also run split-sample regressions separately for individuals who overreport and who underreport their wages in the SOEP.\footnote{The interpretation of the log difference per se is not straightforward, as a negative sign of the estimated coefficient can imply both a decrease in a positive bias and an increase in a negative bias. For this reason, we estimate this difference separately for individuals with a negative and a positive bias.} Our main parameters of interest are estimated as the coefficient vector $\beta_1$ for the range of potential correlates $X'_{i(t)}$ of the reporting bias, including individual, household, and firm characteristics. We also examine time-fixed effects $\xi_t$ for trends in the reporting behavior.

As individual-level correlates, we include socioeconomic characteristics: gender, age, migration background (none, direct, indirect), years of education, and region of residence. We also assess personality using standardized measures of the Big 5 (openness, conscientiousness, extroversion, agreeableness and neuroticism). All individual-level variables are drawn from the SOEP. 

As correlates at the household level, we include indicator variables for being married, having a partner (also interacted with gender), household size, and the number of children.

Finally, we include firm-level correlates. This set of variables includes indicators for the size of the establishment (less than 10, between 10 and 50, between 50 and 250 and more than 250) and the establishment's median wage (lower, middle, or higher tercile), the industry, and indicators for the composition of the workforce (average age of employees, share of highly qualified, female, full-time, and German employees in the establishment). Establishment-level variables are based on IEB administrative data aggregated to the establishment level. In addition, we further assess individual job characteristics using indicators for working hours (working less than 35 hours, between 35 and 44 hours, and more than 44 hours per week), an indicator for being a blue-collar or white-collar worker, and an indicator for trade union membership. These indicators are derived from the SOEP data.

\paragraph{Shapley Decomposition for Relative Variable Importance} To determine the relative importance of different groups of predictors, we apply a Shorrocks-Shapley decomposition to a regression based on Equation (\ref{equ:main_analysis}) with the log reporting bias as the dependent variable. This technique decomposes the coefficient of determination $R^2$ into contributions from individual or groups of explanatory variables.\footnote{The decomposition is implemented by using the \textit{Shapley2} post-estimation command to compute the decomposition of the R-squared of the model based on \cite{Juarez2012}.} Intuitively, the Shorrocks-Shapley decomposition considers the marginal effect on the outcome of eliminating each of the contributing factors in turn, and then assigns to each factor the average of it's marginal contributions in all possible elimination sequences \citep{Shorrocks2013}. Unlike a simple incremental $R^2$, which depends on the order in which the variables are added to the model, the Shorrocks-Shapley decomposition is robust to the order of the variables.

\subsubsection{Results}

In this subsection we discuss the sources of reporting bias estimated by Equation \ref{equ:main_analysis}. The results are presented in Table~\ref{tab:main_table}, Appendix Table~\ref{tab:yearFE_survey} additionally documents the coefficients for survey characteristics as well as year and sector dummies. In column (1) of Table~\ref{tab:main_table}, the dependent variable is an indicator of whether an individual has a positive reporting bias. The indicator is equal to one if the difference between the survey and administrative wage is larger than zero, i.e., if the individual overreports the own wage in the survey as compared to the administrative data. 

In column (2), the dependent variable is defined as log reporting bias. To facilitate the interpretation of reporting bias coefficients, in columns (3) and (4) we also report the estimates separately for overreporters and underreporters, excluding correctly reporting individuals. In addition, the bottom panel of Table~\ref{tab:main_table} summarizes the results of the Shapley decomposition, which are reported in more detail in Table \ref{tab:main_table_shapeley} in the Appendix.

\paragraph{Individual Characteristics} The Shorrocks-Shapley decomposition (Table \ref{tab:main_table_shapeley}) shows that individual-level variables have a relatively
small contribution to the explained variation in the log reporting bias: only
about 3.9\% of the R-squared can be attributed to these characteristics. The results in Table \ref{tab:main_table} can be read as follows: women are 2.1 percentage points less likely than men to overreport their wages (column 1), and have a smaller difference in log wages by about 1.6 percentage points (column 2). The coefficient on the female dummy is negative but insignificant in column (3), suggesting that there is no difference in the reporting bias for overreporting women and men. In column (4), however, we see that women who underreport their true wage, underreport by 0.72 percentage points more on average than men who underreport. Previous findings on gender differences in wage reporting have been inconclusive. Our results are consistent with the findings of \citet{,Bound1991,Bollinger1998,Angel2019,Antoni2019, roth2021gender}, but differ from \citet{Bricker2008,Kim2014,Angel2018}, who find no gender differences in reporting bias. 

For more educated individuals, we observe a lower but insignificant probability of overreporting (column 1), and a lower negative reporting bias (column 4). Again, previous findings on the relationship between education and reporting bias are mixed: \citet{Bricker2008,Angel2018,Angel2019} also find a positive correlation between education and the reporting bias, while \citet{Bound1991,GottschalkHuynh2005} find no correlation and \citet{Kim2014,Antoni2019} find a negative correlation. Our results also suggest little regional variation in the reporting bias. Beyond these basic socioeconomic variables, personality traits of the worker also have little predictive power for reporting bias. One exception is individuals with higher extroversion, who are more likely to overstate their wages, in line with the findings of \citet{Antoni2019}. 

\paragraph{Household characteristics} Similar to the individual predictors, household-level variables are, on average, weak predictors of reporting bias. Only 15.6\% of the explained variation in the log reporting bias can be associated with these characteristics (see Table~\ref{tab:main_table_shapeley}). The most important household characteristic is partner income. Conditional on having a partner, a difference of \texteuro1,000 in partner income increases the probability of being an overreporter by about half a percentage point (column 1). This effect may be linked to the findings of \citet{roth2021gender} for Austria and Switzerland, who found that couples in which women out-earn their partners systematically underreport their income to conform to the male breadwinner model \citep{bertrand2015gender}. Instead, our results suggest a tendency to equalize reported income, instead of having a tendency to maintain income rankings in the household.

\paragraph{Job and Firm Characteristics} Job and firm characteristics have the strongest predictive power, accounting for by far the largest share of explained variance, about 69.2\% (see Table \ref{tab:main_table_shapeley}). Virtually every observed factor at the job or firm level shows strong correlations with reporting bias. At the job level, white-collar versus blue-collar employment is negatively correlated with the likelihood of overreporting. Union membership reduces the probability of overreporting by 3.3 percentage points (column 1) and reduces the log reporting bias (column 2). Looking at over- and underreporters separately in columns (3) and (4), we find that union membership has no effect on the magnitude of a positive reporting bias (overreporting), but increases negative reporting bias by 0.7 percentage points, i.e., union members tend to underreport their wages more than non-union members. Working full-time reduces the probability of overreporting wages by 6.4 percentage points compared to those working less than 35 hours per week. 

At the firm level, larger firm size is negatively correlated with the likelihood of wage overreporting. Workers in firms with more than 250 employees are 6.2 percentage points less likely to overreport than workers in small firms with up to nine employees, and show a 6.4\% lower reporting bias, driven by both lower overreporting and higher underreporting. A similar picture emerges for the average wage level of a firm. Workers employed by firms in the 3rd tercile of the firm premium distribution are 19.3 percentage points less likely to overreport their wages than workers in the 1st tercile, mostly driven by stronger underreporting. The composition of the workforce (in terms of qualification, gender, nationality and age, as well as the share of full-time employees) are strong predictors for the reporting bias. Higher proportions of women, German nationals and full-time employees all predict lower levels of wage misreporting in the firm. 

We can only speculate about the reasons for this strong relationship, which may be explained by heterogeneity in organizational culture and transparency in compensation practices. For example, larger and higher-paying firms may have more formalized pay structures and human resource policies, which may reduce the propensity to overreport wages by providing clearer benchmarks for employee compensation.\footnote{Table \ref{tab:yearFE_survey} in the Appendix additionally shows results for the predictive power of several survey characteristics for the reporting bias, which turn out to have little explanatory power for the reporting bias. An exception are in-person interviews, which appear to reduce the amount of misreporting.}

\begin{center}
	[Insert Table \ref{tab:main_table} about here]
\end{center}

\paragraph{Reporting Bias over Time} Finally, we look more closely at the year fixed effects included in Equation (\ref{equ:main_analysis}) to understand potential dynamics in the reporting bias. The year fixed effects account for about 10.6\% of the explained variance and thus add non-negligible to the explanation of the reporting bias. Figure \ref{fig:year_FE} shows that the average reporting bias has increased since the beginning of the SOEP in 1984, but has leveled off since about 2010. We acknowledge, however, that we are only observing a selected sample of panel survivors for the early years, as we can only link information for those respondents who have consented since 2019. Thus, the observed pattern may also represent an effect of panel tenure.   

\subsection{Consequences for Economic Analysis} \label{sec:impl}

So far, we have established a modest but economically significant extent of reporting bias between SOEP and IEB (Section \ref{sec:extent}). This reporting bias can be predicted to some extent by firm-level factors. Household and individual characteristics have little predictive power. In this section, we examine the implications of the reporting bias for common empirical applications in applied economic research. 

\subsubsection{Wage as Dependent Variable}

We first examine the consequences of using self-reported wages rather than administrative wages as the dependent variable in the empirical analysis. We estimate simple versions of two commonly studied economic relationships: the returns to education and the gender wage gap. These topics are of high societal importance, and have received extensive attention from applied economists, partly based on the SOEP data itself.  

\paragraph{Returns to Education} Our first case study examines how estimates of returns to education are affected by the use of self-reported versus administrative data. Labor economists have traditionally studied the relationship between education and subsequent labor market outcomes to show how investments in education or training can lead to higher future earnings despite initial opportunity costs such as forgone earnings. This concept, first proposed by \cite{Becker1962} and extensively studied using variations of the Mincer equation \citep{Mincer1974}, relates wages as the dependent variable to education and experience as explanatory factors. The coefficient on education, usually measured in years, is interpreted as an estimate of the annual return to education, i.e., the financial benefit of an additional year of education. Through this empirical analysis researchers can address questions about the private returns associated with additional years of education and how these returns vary over time, across different levels of education, and for individuals with different background characteristics.

We examine the sensitivity of using either self-reported SOEP or administrative IEB wages as the dependent variable by estimating the following equation: 

\begin{equation}
  \label{equ:rte}
  ln(Y) _{it} = \alpha + \beta_1 YED_{it} + \gamma X'_{i(t)} + \varepsilon_{it} 
  \end{equation}

  where $ln(Y) _{it}$ is the natural logarithm of the monthly wage taken from either the SOEP or the IEB, and $YED_{it}$ is the years of education attained measured by the SOEP. The control variables $X'_{i(t)}$ consist of a minimum (age, age squared, firm tenure in years, a dummy for being full-time employment ($>35$ hours), year dummies) or an extended set (basic controls plus a dummy for being female, a dummy for having migration background, full-time and unemployment experience in years, industry dummies, county of residence dummies, firm size indicators, a dummy for having children under 16 in the household, marital status indicators) of individual-specific characteristics. Sample restrictions remain similar to those described in Table \ref{table1}, i.e., we consider individuals aged 20-65 in any dependent full- or part-time employment for whom all relevant controls are non-missing. We cluster standard errors at the individual level.\footnote{The results remain qualitatively and quantitatively similar when we cluster at the household level instead in order to correct for a potential correlation of reporting bias within households (see Figure \ref{fig:wage_gaps_and_premia_hh} in Appendix \ref{sec:app_B}).}

 Figure \ref{fig:wage_gaps_and_premia}, Panel (a), shows the estimated coefficients of $\beta$ as a measure of the returns to an additional year of education, based on either SOEP or IEB. We estimate three different specifications: a ``plain vanilla'' OLS specification with no additional controls, a ``basic controls'' specification, and an ``extended'' specification, as described above. In all specifications, the estimated returns to education are of the same order of magnitude, and the confidence intervals of the IEB and SOEP estimates overlap. An additional year of education significantly increases wages by 7\% to 7.8\% when using SOEP wages and by 6.8\% to 7.3\% when using IEB wages. Different sets of controls have only little effect on these differences. When estimating the two regression equations jointly with GMM, a two-sample t-test does not detect significant differences in all but the extended specification.\footnote{The respective p-values of the test for significance are 0.30 for the regression without controls, 0.82 with basic controls and 0.03 with extended controls.} We therefore conclude that the choice of data in the case of the returns to education appears to be largely inconsequential and does not affect its interpretation in an economically relevant sense.

\paragraph{Gender Wage Gap} Perhaps no other measure of labor market inequality receives as much public attention as the gender wage gap. The gender wage gap describes the average difference in earnings between women and men in the labor force. It is typically expressed as a percentage of men's earnings to highlight the difference in pay for comparable work. It can be measured as a raw hourly or monthly wage gap, or as an adjusted gender wage gap, which takes into account factors such as occupation, industry, work experience, education, hours worked, and job tenure. Its extent and development is described over time and contexts \citep{Goldin14, blau2017gender}. The gender wage gap has received renewed attention in recent years through the emerging literature on the ``child penalty'' as a primary reason for these persistent inequalities \citep[e.g.,][]{Bertrandetal10, Goldin14,Addaetal17, Klevenetal19b, Klevenetal19a}. 

To measure the sensitivity of estimates of the gender wage gap to the reliance on self-reported versus administrative wage data, we estimate the following equation: 

\begin{equation}
 \label{equ:gwg}
 ln(Y) _{it} = \alpha + \beta_2 Female_{i} + \gamma X'_{i(t)} + \varepsilon_{it} 
 \end{equation}

which relates the natural logarithm of the monthly wage $ln(Y) _{it}$, taken from either SOEP or IEB, to an indicator $Female_{i}$ that takes the value of 1 for women. The coefficient $\beta$ approximates of the gender wage gap. Control variables in $X'_{i(t)}$ are equivalent to those used in equation \ref{equ:rte}, additionally controlling for an individual's education level. Again, the sample restrictions correspond to those described in Table \ref{table1}. In addition, we impose a restriction on full-time workers in order to identify wage differences rather than differences in working hours.

Figure \ref{fig:wage_gaps_and_premia}, Panel (b) shows that the estimated gender wage gap is substantially larger in all specifications when based on self-reported SOEP data. Without controls, it is about 15.1\% based on SOEP, and only about 13.6\% based on IEB data.\footnote{Both figures are substantially smaller than those reported  by the German Statistical Office for the same period (around 19-24\% (6-9\%) for West (East) Germany in the years 2006-2022, including part-time workers) \citep{destatis}. The relatively small gender wage gap in our sample is driven by the fact that we only consider full-time employees and that later years, in which the gender wage gap has already narrowed, are strongly overrepresented.} With extended controls, it shrinks to 10.7\% based on SOEP data and to 7.7\% based on IEB data. However, the relative difference between the estimates based on the different data sources increases from 11\% to 38\% with extended controls. This difference is directly related to the results observed in Section \ref{sec:factors}: The negative coefficient of female gender on the reporting bias thus translates into the observed difference in the gender wage gap. We conclude that the choice of data source is relevant for the gender wage gap. The confidence intervals of the coefficients remain overlapping, but a when estimating the two equations jointly with GMM, the t-test confirms significant differences in all specifications.\footnote{The respective p-values of the test for significance are 0.04 for the regression without controls, 0.01 with basic controls and 0.01 with extended controls.} Note that this difference runs counter the intuition of a mere attenuation bias by classical measurement error in survey wages, which would lead to a lower gender gap based on survey data.

\begin{center}
	[Insert Figure \ref{fig:wage_gaps_and_premia} about here]
\end{center}

\subsubsection{Wage as Independent Variable}

In a second step, we examine the consequences of using self-reported wages instead of administrative wages as the right-hand side explanatory variable. To this end, we replicate empirical exercises explaining individual-level subjective outcomes of satisfaction in different domains that have been shown to be affected by individual income \citep[e.g.,][]{Frijtersetal2004, Boyceetal2010, CheungLucas2015}. 

\paragraph{Estimation Strategy} To compare the estimates when using either the SOEP or the IEB wage as explanatory variables, we estimate the following equation: 

\begin{equation}
  \label{equ:rhs}
  SI_{it} = \alpha + \beta Wage_{it} + \gamma X'_{i(t)} + \delta_i + \varepsilon_{it}    
  \end{equation}

where $Wage _{it}$ is the monthly wage (in \texteuro1,000), based on either IEB or SOEP data. $SI_{it}$ is one of several indicators of subjective well-being: life satisfaction and satisfaction with job and household or personal income. Each of these items is measured by standard questionnaire items\footnote{The exact wording in the 2019 English version of the SOEP is the following: ``a) for satisfaction with job and income: \textit{How satisfied are you today with the following areas of your life? ... (if employed) with your job?... with your household income?	... with your personal income?}; Possible answers: 0 `completely dissatisfied' to 10 `completely satisfied'; b) for life satisfaction: \textit{In conclusion, we would like to ask you about your satisfaction with your life in general. How satisfied are you with your life, all things considered?}; possible answers: 0 `completely dissatisfied' to 10 `completely satisfied'.''} and are standardized to have a mean of zero and a standard deviation of one. Our main parameter of interest is $\beta$, which shows the association between wages and the subjective indicators. The control variables $X'_{i(t)}$ are similar to those described above. Standard errors are clustered at the individual level, sample restrictions are equivalent to those described in Table \ref{table1}.\footnote{The results remain qualitatively and quantitatively similar when we cluster at the household level instead in order to correct for a potential correlation of reporting bias within households (see Figure \ref{fig:satisfaction_wage_hh} in Appendix \ref{sec:app_B}).}

\paragraph{Results} Figure \ref{fig:satisfaction_wage} shoes the point estimates $\beta_{IEB}$ in black and $\beta_{SOEP}$ in gray with the corresponding 95\% confidence intervals. Again, we compare specifications that differ in the number of control variables. In addition, we examine a specification that focuses on individual changes by including individual fixed effects. 

\begin{center}
	[Insert Figure \ref{fig:satisfaction_wage} about here]
\end{center}

The results show that the influence of misreporting depends on the type of satisfaction assessed. The type of data source appears to be unrelated to general life or job satisfaction (Panels (a) and (d)), but significantly affects the estimated income effects for satisfaction with household income or personal income, i.e., in those dimensions where it is the variable of focal interest (Panels (b) and (c)). In the latter cases, point estimates based on self-reported data consistently exceed those based on administrative data. The confidence intervals do not overlap. For example, satisfaction with personal income increases with an additional \texteuro1,000 by about 25.4\% of a standard deviation when assessed on the basis of self-reported income, but by only 20.2\% of a standard deviation when assessed on the basis of IEB data -- a difference of 26\%. We conclude that the reporting bias can lead to quantitatively different conclusions depending on the data source when examining income-related subjective outcomes. However, introducing individual fixed effects to rely on interpersonal changes mitigates the bias, reflecting the fact that misreporting appears to be less of an issue for changes than for levels. Further note that this difference cannot be explained by a mere attenuation bias in survey wages, which would hint at \textit{smaller} coefficients when working with survey data. Rather, our results may be explained by unobservable factors that simultaneously and similarly affect both subjective satisfaction and self-reported income \citep{bertrand2001people}. \cite{prati2017hedonic} argues that self-reported wages may be prone to a \textit{hedonic recall bias}, meaning that individuals tend to overreport their wages when they are satisfied and underreport when they are dissatisfied with their wages, leading to an upward bias in the relationship between income and satisfaction.

\section{Conclusion} \label{sec:concl}
Based on a novel record linkage of the SOEP-ADIAB, which links the German Socioeconomic Panel (SOEP) to administrative wage information from the Integrated Employment Biographies (IEB), this paper examines the extent to which survey respondents misreport their wages, which observable characteristics may predict the extent of misreporting, and the consequences for applied researchers working with these misreported wages. 

Our analysis yields three sets of results. First, with respect to the magnitude of the reporting bias, the IEB and SOEP wage distributions are closely aligned, with the average SOEP respondent underreporting individual wages by a modest but economically significant 7.3\% of the mean. We find evidence of a mean-reverting component to the reporting bias, with respondents in the very low tails of the wage distribution overreporting their wages and a steadily increasing underreporting from there. The magnitude of misreporting appears to be smaller in terms of changes than in terms of levels. While patterns of mean reversion are comparable to previous studies, the extent of misreporting, as well as the rather good representation of wage changes seem to be a particularity of the SOEP data.

Second, with respect to the correlates of reporting bias, we find that job and firm characteristics are particularly predictive of the extent to which individuals misreport their wages. In contrast, household characteristics and especially individual characteristics contribute only weakly to explaining the variance in reporting bias. 

Third, turning to the consequences of choosing survey over administrative wages, we examine the sensitivity of prominent economic relationships commonly estimated in applied economic studies. Using income as the dependent variable, we find no difference in the estimated returns to education and an economically meaningful and significant difference in the gender wage gap. Using income as an explanatory variable to estimate the relationship between satisfaction and income, the use of misreported survey data leads to a significant overestimation, especially in the dimensions of household income and personal income satisfaction.  

Taken together, our findings suggest that researchers may need to be more cautious in interpreting survey-based measures of individual income in empirical research and the empirical relationships estimated on them. In our exemplary analysis, misreporting was largely inconsequential for examining the returns to education, but appeared to be more relevant for analyzing the gender wage gap -- reflecting systematic variation in misreporting along the gender dimension. In addition, we found that misreporting bias can significantly affect the results when wage is used as the independent variable, as we did in estimating the wage-satisfaction relationship. These consequences of misreported income extend beyond simple attenuation bias due to classical measurement error in survey wages. Our findings underscore that, at least in some cases, survey-based measures of individual wages can significantly bias commonly estimated empirical relationships. They also demonstrate the enormous research potential of linked administrative-survey data \cite[see, e.g., also the project by][]{mittag2023} which would moreover enable researchers to create a measure of ``true'' wage income that could be a hybrid version of the survey and the administrative information \citep{Meijeretal2012, AbowdStinson2013}. Extending our analysis to different research questions will help to get a clearer picture of the importance of the misreporting problem. It may be the case that even modest misreporting biases in wages, influenced by various individual, household, job, and firm characteristics, as shown in our paper, can significantly bias (or not bias) common empirical exercises. This is an interesting avenue for future research.

\clearpage\newpage 

\begin{singlespace}
\setlength{\baselineskip}{9pt}
\bibliographystyle{ecca}
\bibliography{literature_wagebeliefs.bib}

\end{singlespace}

\newpage\clearpage
\begin{appendix}
	
\section*{Tables and Figures}	
\input{Tables}

\input{Figures}
\renewcommand{\thesection}{A}
\renewcommand{\thetable}{A.\arabic{table}}
\setcounter{table}{0}
\renewcommand{\thefigure}{A.\arabic{figure}}
\setcounter{figure}{0}

\section{Additional Wage Concepts} \label{sec:app_A}

\begin{figure}[H] 
    \centering
  \begin{subfigure}{.49\textwidth}
    \centering
    \includegraphics[width=0.9\textwidth]{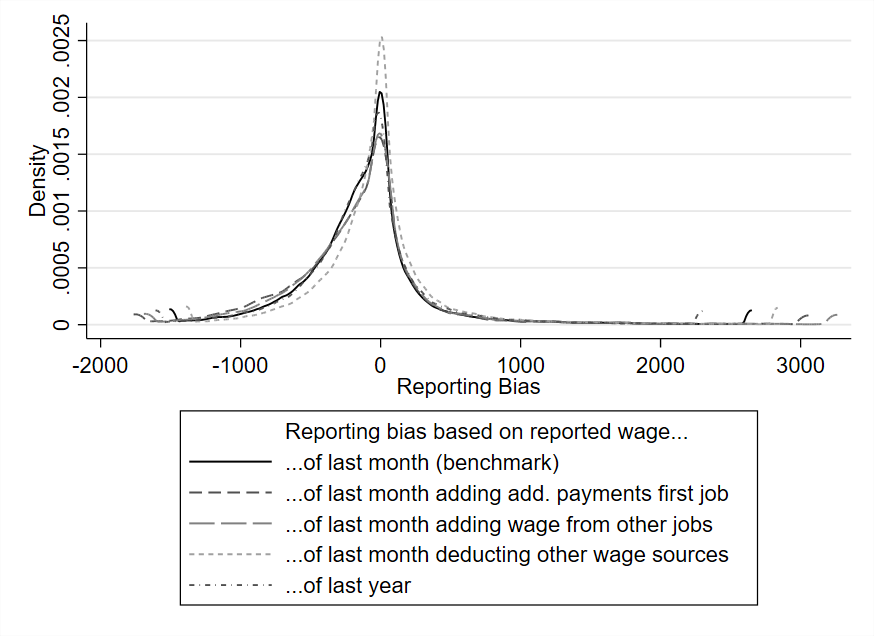}
    \caption{\footnotesize Distribution of the reporting bias for different wage concepts}
  \end{subfigure}
  \begin{subfigure}{.49\textwidth}
    \centering
    \includegraphics[width=0.9\textwidth]{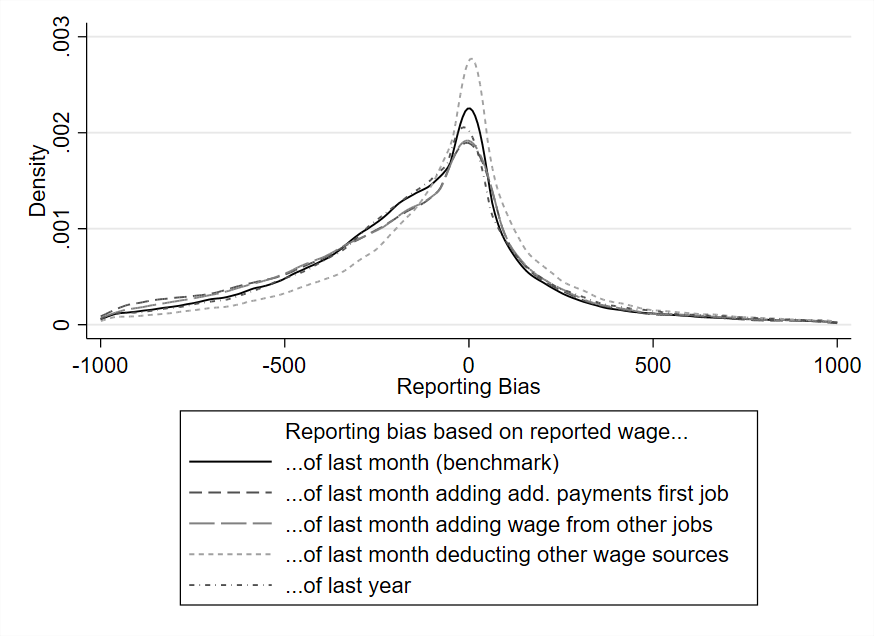}
    \caption{\footnotesize Zoom in: restricted to a reporting bias of   $+$ / $-$ \texteuro1,000 }
  \end{subfigure}
    \caption{Reporting bias for different wage measures.}
        \fignote{Graph shows the density distribution of the reporting bias for different wage concepts. Here, we vary the definition of SOEP wages. Income from additional sources includes reported vacation pay, Christmas pay, and the 13th and 14th monthly wage payment. For the exclusion of additional payment we deduct reported income from other employment. This variable is surveyed in SOEP only for the survey waves from 2017 to 2019, so the reporting bias is only plotted for these years. The yearly measure is based on a different set of survey items that asks about the income in the year before, specifically the monthly wage income and the number of months the individual received the wage income. Reporting biases that lie below the 1st percentile or above the 99th percentile of the distribution of differences are binned into the 1st or 99th percentile respectively.}
    \label{fig:wage_diff_concepts}
\end{figure}

Figure \ref{fig:wage_diff_concepts} investigates whether taking into account additional information from the SOEP changes the distribution of the reporting bias. Panel a) shows the whole distribution of the reporting bias. For presentational reasons, panel b) zooms in and restricts the distribution to a reporting bias of plus/ minus \texteuro1,000.

First, we will consider additional sources of wage from the primary job. The reporting bias could potentially mislead if the question is perceived as solely pertaining to wage income from the main job. Consequently, we will incorporate all declared additional income sources, including vacation pay, Christmas pay, 13th, and 14th monthly wage payments. The resulting effect is small, with a shift towards a more uniform distribution centered around 0.

Second, we supplement our wage variable from wages earned from other jobs in regular employment as reported in the SOEP. The phrasing of the primary question (``Arbeitsverdienst im letzten Monat'') implies that these wages are already accounted for. We verify this. The graphs indicate that participants either incorporate or exclude wages from other jobs in the primary SOEP inquiry regarding last month's wage. Unfortunately, this survey item was only introduced in 2017, so the analysis only analyses responses for the years 2017, 2018 and 2019.

Third, we subtract reported wages from other sources, specifically self-employment, work in family businesses, and other non-regular employment jobs, from the SOEP wage variable. Again, we do not detect a economically relevant difference.

Also, Figure \ref{fig:wage_diff_concepts} displays the distribution of the reporting bias based on a SOEP questionnaire. Each year, individuals are asked not only about their income from the previous month but also about their wages earned during the previous calendar year. Participants are specifically prompted to report the months they worked as an employee and their average gross monthly wage in each month. The Figure reveals that the reporting bias that is based on yearly income is very similar to the reporting bias that is based on the reported income the month before.

\renewcommand{\thesection}{B}
\renewcommand{\thetable}{B.\arabic{table}}
\setcounter{table}{0}
\renewcommand{\thefigure}{B.\arabic{figure}}
\setcounter{figure}{0}
\newpage
\section{Supplementary Figures and Tables}  \label{sec:app_B}
\begin{figure}[H]
\centering
\includegraphics[width=.7\linewidth]{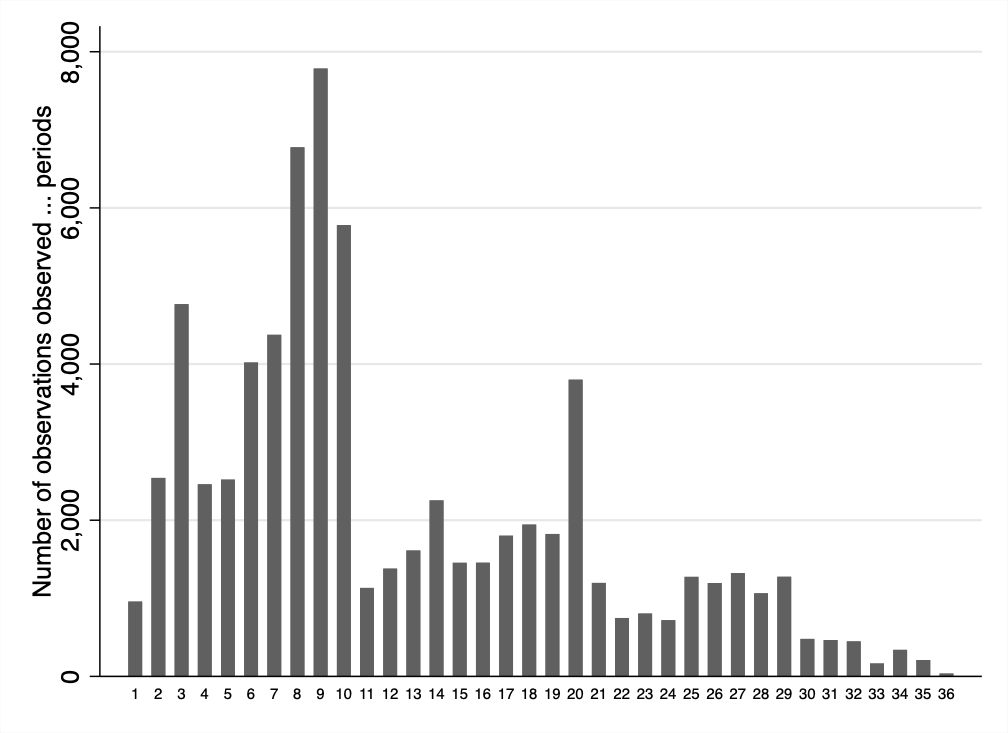}
\caption{Number of individuals observed over x periods of observation}
\fignote{This figure summarizes the panel tenure of individuals in our sample. Bars show the number of individuals that are observed for the number of years depicted on the x - axis. A large number of individuals is observed for 8-10 years.}
\label{fig:periods}
\end{figure}
\vspace{1cm}

\begin{figure}[H] 
		\centering
		\includegraphics[width=.7\linewidth]{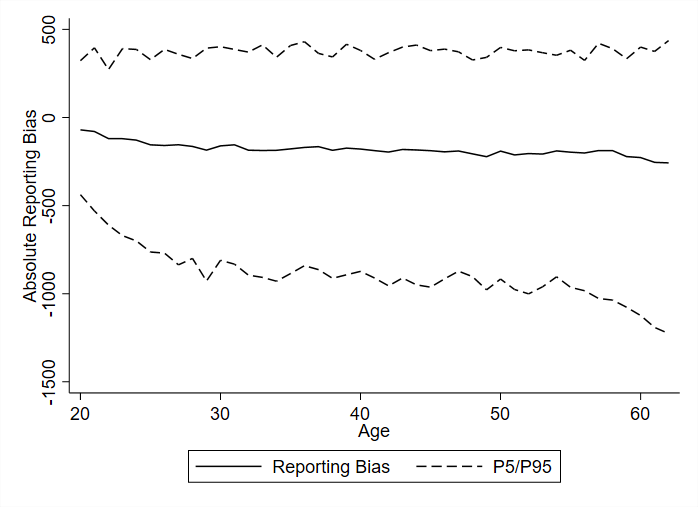}
		\caption{Distribution of difference SOEP - IEB over age}
    \fignote{This figure shows the mean reporting bias in absolute terms (black solid line) and the respective 5th and 95th percentile (dashed lines) over age.}
		\label{fig:wage_diff_age}
\end{figure}

\begin{figure}[H]
	\centering
	\begin{subfigure}{.45\textwidth}
		\centering
		\includegraphics[width=0.9\textwidth]{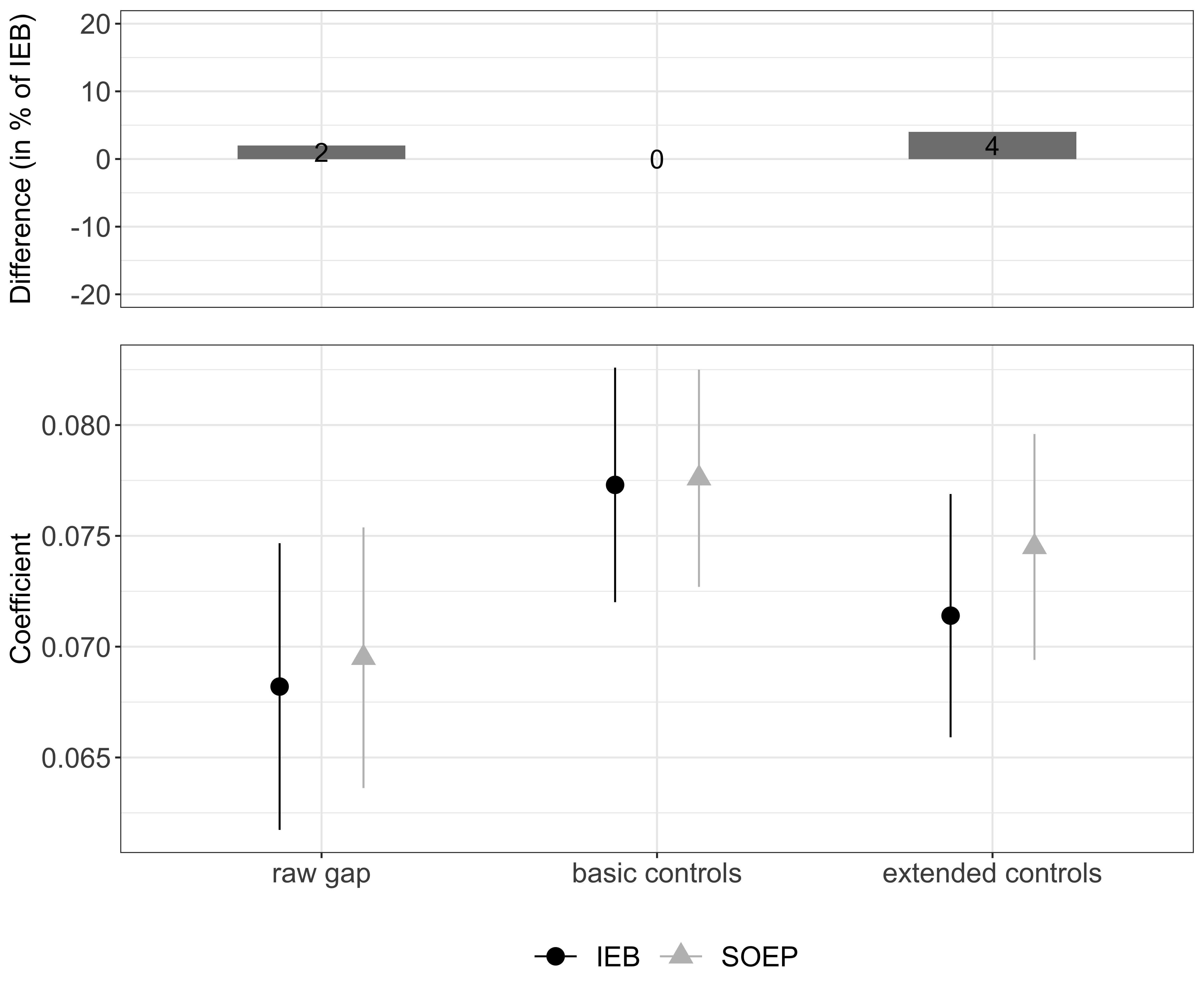}
		\caption{\footnotesize Returns to education}
	\end{subfigure}
	\begin{subfigure}{.45\textwidth}
		\centering
		\includegraphics[width=0.9\textwidth]{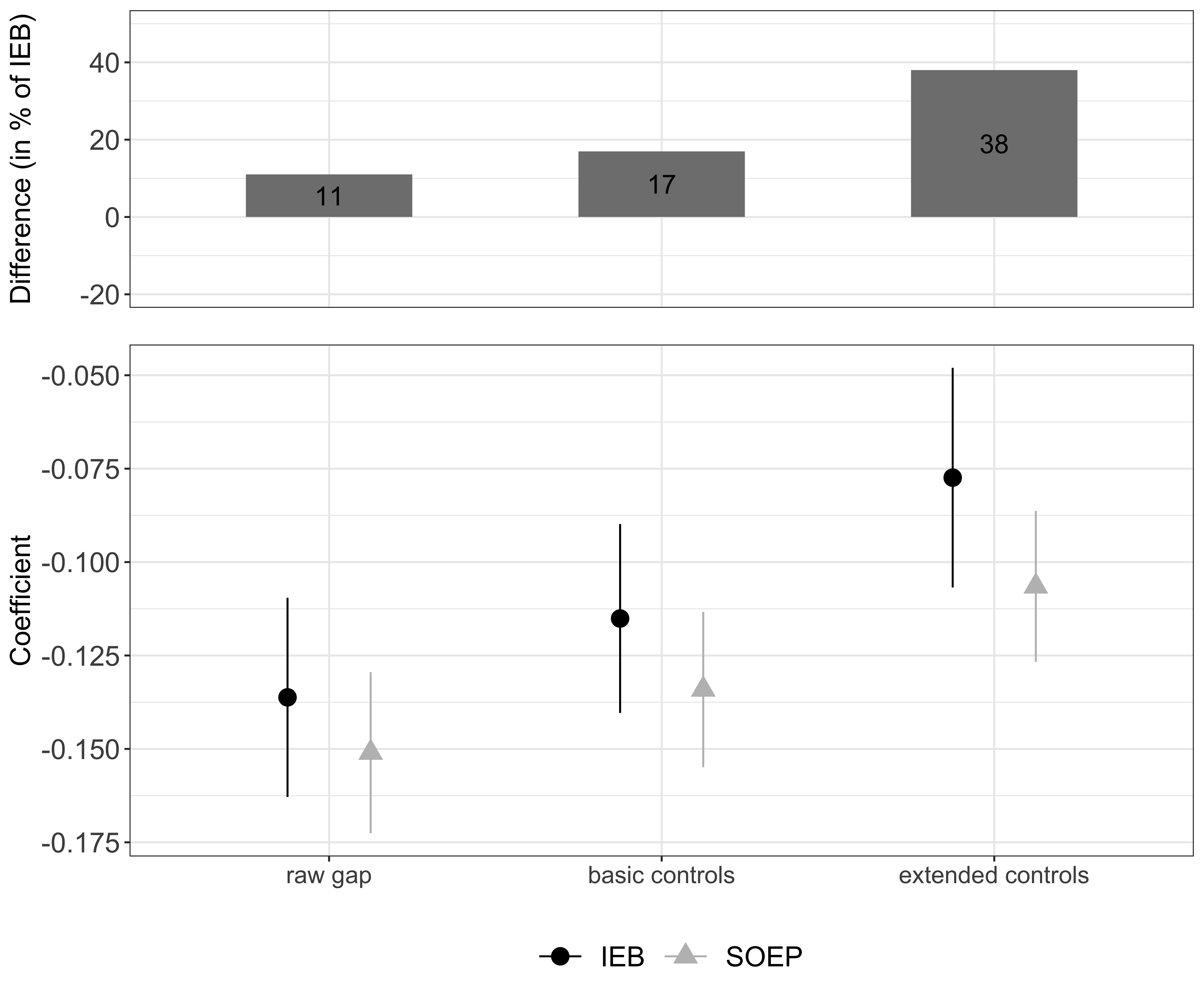}
		\caption{\footnotesize Gender wage gaps}
	\end{subfigure}
	\caption{Returns to education and gender wage gap based on SOEP and IEB wages}
	\fignote{This figure summarizes results of regressions of log wages, based on either SOEP or IEB data, on various sets of explanatory variables. Panel (a) summarizes estimates of the returns to education $\beta_1$ based on equation \ref{equ:rte}. Panel (b) summarizes estimates of the gender wage gap $\beta_2$ based on  equation \ref{equ:gwg}. Raw, basic and extended specifications differ in control variables as described in the text. Bar graphs illustrate the relative size of the difference by dividing the difference between the SOEP and IEB estimate by the IEB estimates.  Restricted to full-time for Panel (b).  The graph displays 95\% confidence intervals, based on standard errors clustered at the household level.
	}
	\label{fig:wage_gaps_and_premia_hh}
\end{figure}

\begin{figure}[H]
	\centering
	\begin{subfigure}{.45\textwidth}
		\centering
		\includegraphics[width=0.85\textwidth]{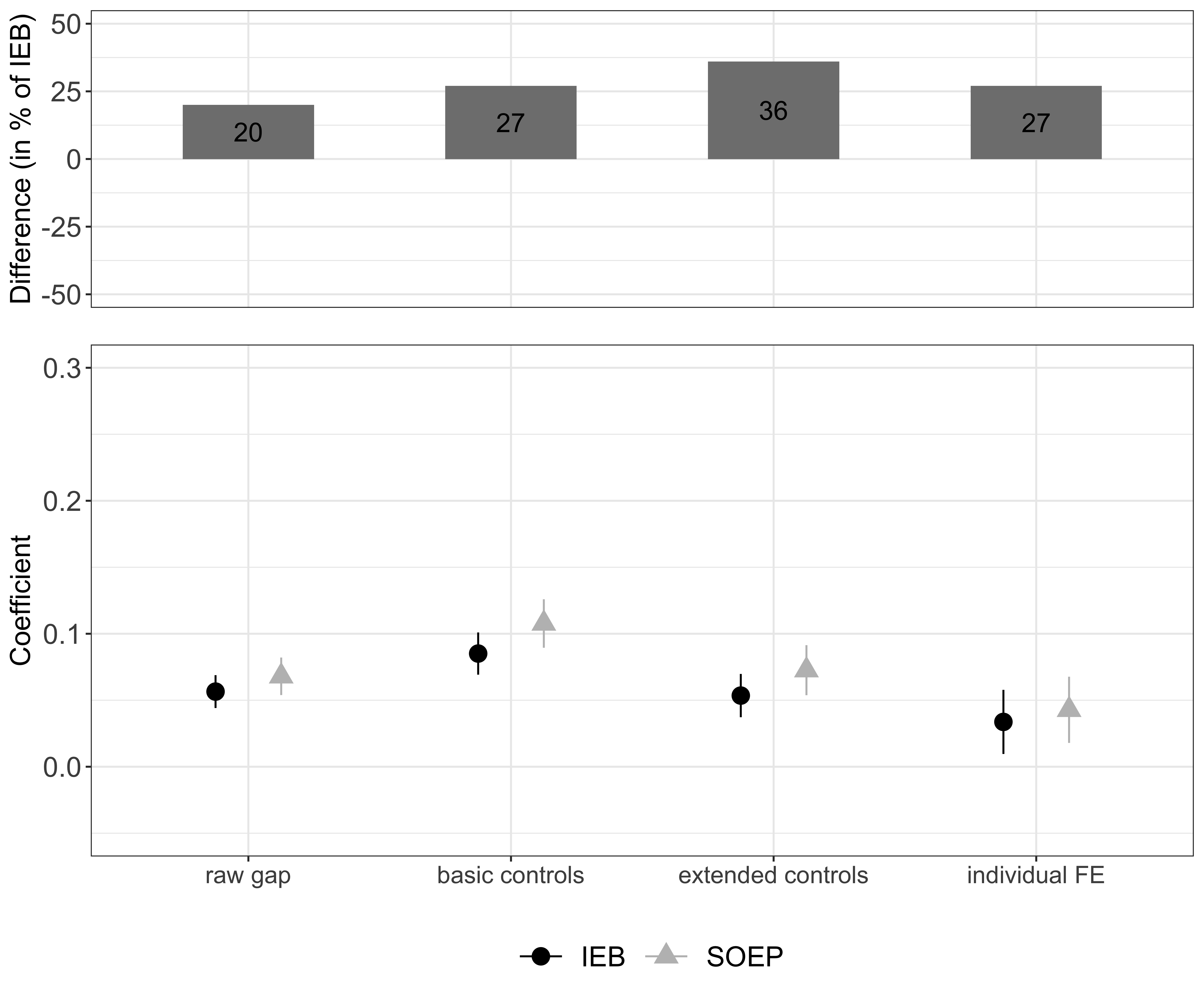}
		\caption{\footnotesize Life satisfaction \\\hspace{0.5\textwidth} }
	\end{subfigure}
	\begin{subfigure}{.45\textwidth}
		\centering
		\includegraphics[width=0.85\textwidth]{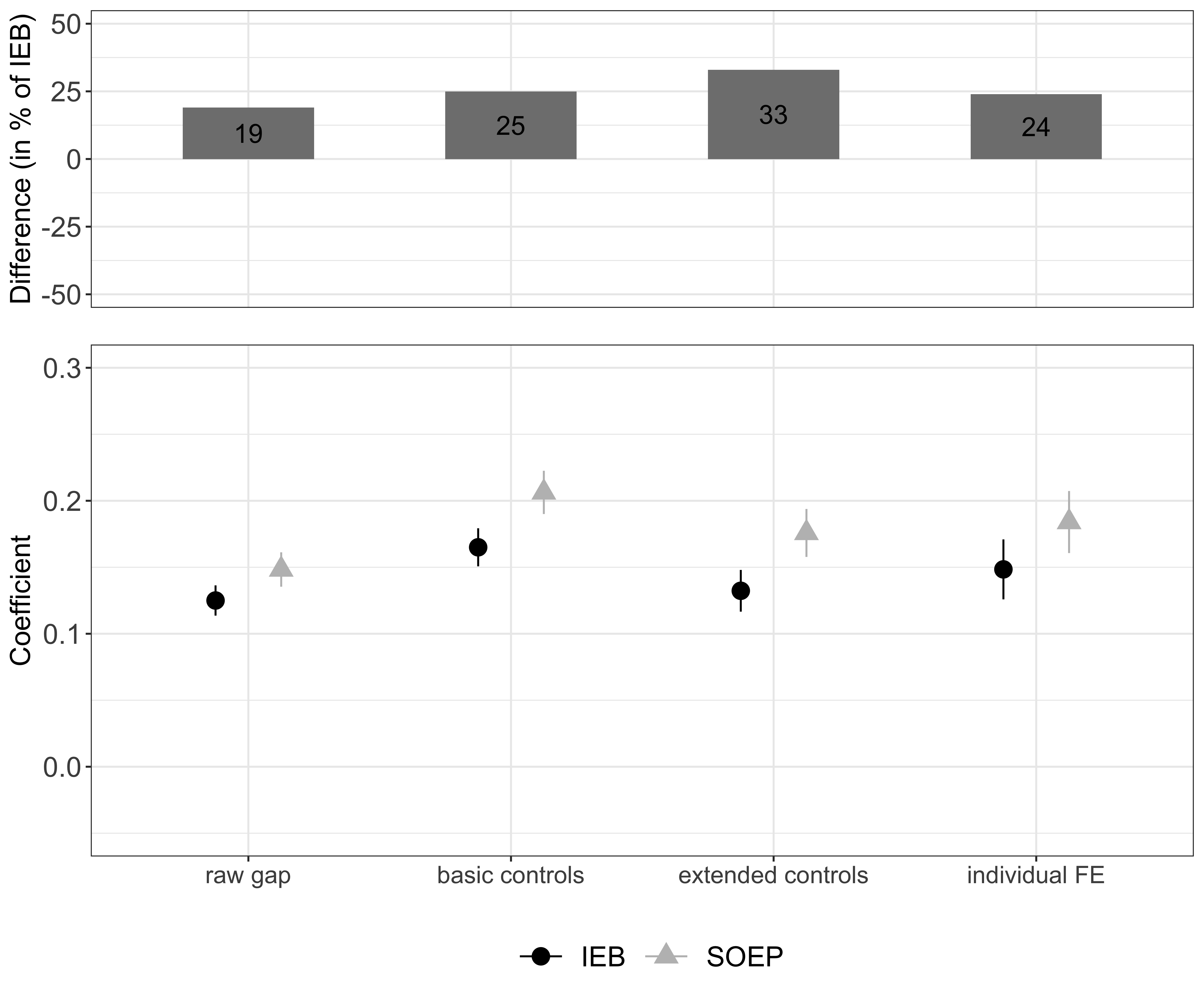}
		\caption{\footnotesize Satisfaction with household income \\\hspace{0.5\textwidth} }
	\end{subfigure}
	\begin{subfigure}{.45\textwidth}
		\centering
		\includegraphics[width=0.85\textwidth]{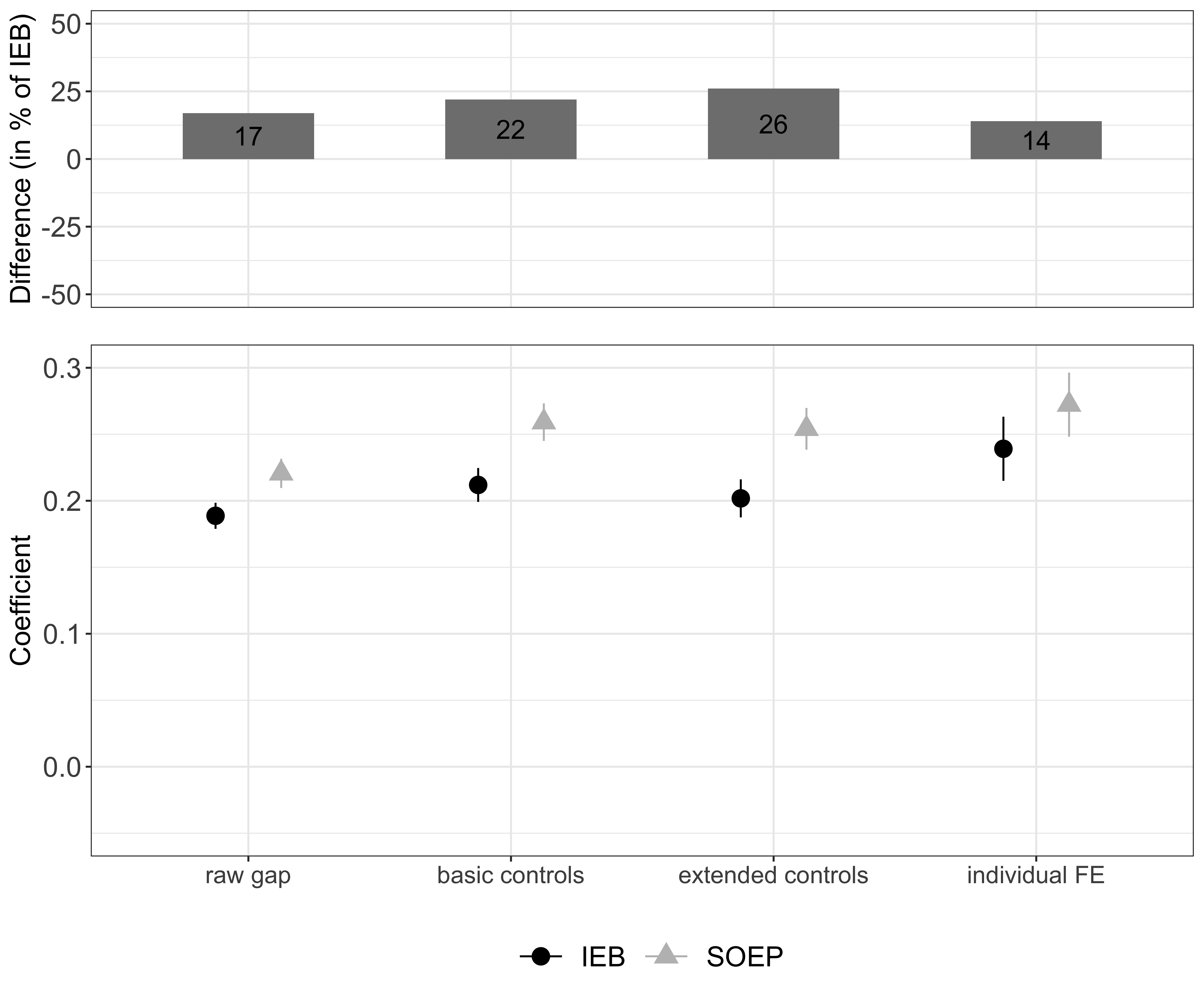}
		\caption{\footnotesize Satisfaction with personal income \\\hspace{0.5\textwidth} }
	\end{subfigure}
	\begin{subfigure}{.45\textwidth}
		\centering
		\includegraphics[width=0.85\textwidth]{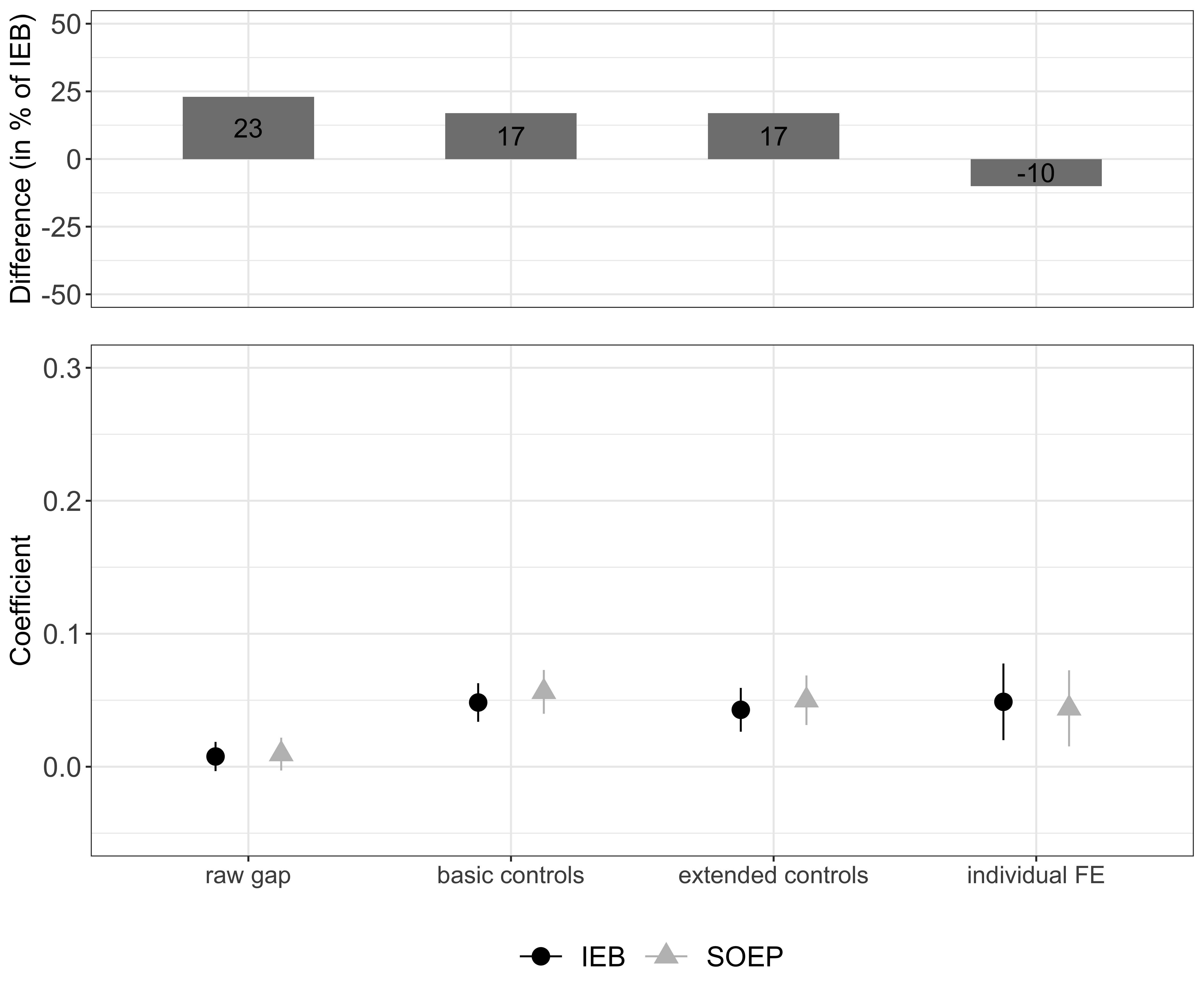}
		\caption{\footnotesize Satisfaction with job \\\hspace{0.5\textwidth} }
	\end{subfigure}
	\caption{Relation between (life) satisfaction and wages}
	\fignote{This figure summarizes results of regressions of subjective satisfaction indicators on log wages and control variables based on equation \ref{equ:rhs}, with varying sets of control variables. Outcome variables are standardized to a mean of zero and a standard deviation of one. Raw, basic and extended specifications differ in control variables as described in the text. Bar graphs illustrate the relative size of the difference by dividing the difference between the SOEP and IEB estimate by the IEB estimates. The graph displays 95\% confidence intervals, based on standard errors clustered at the household level.}
	\label{fig:satisfaction_wage_hh}
\end{figure}

\newpage
\setstretch{0.5}
\begin{table}[H]
	\footnotesize
	\centering
	\caption{Differences between Consenters and Non-Consenters, Cross-Section 2019} \label{tab:descr_cons}
		\input{tables/A1_Descriptives_Consent}

\end{table}
\setstretch{1.5}

\setstretch{0.5}
\begin{table}[H] \addtocounter{table}{-1}
	\footnotesize
	\centering
	\caption{Differences between Consenters and Non-Consenters, Cross-Section 2019 - continued} 
		\input{tables/A1_Descriptives_Consent_2}
		\fignote{
		Cross-section 2019. Column 3 shows the difference in means between column 1 and 2. Max. Experience Fulltime, Parttime and Unemployed is measured conditional on ever having a positive value for fulltime, parttime or unemployment. All individuals, both SOEP IS and SOEP Core, before any further sample restrictions are made. Compared to Table \ref{table1}, the number of consenters deviates by 2 observations, which are missing in the IAB remote application we are using for linked dataset. Standard errors in parenthesis. * \(p<0.1\), ** \(p<0.05\), *** \(p<0.01\)}
\end{table}
\setstretch{1.5}

\setstretch{0.5}
\begin{table}[H]\footnotesize
    \caption{Factors of reporting bias, year effects \& survey characteristics}
    \label{tab:yearFE_survey}
    \centering
    \input{tables/A4_yearFE_survey.tex}
\end{table}

\begin{table}[H]\footnotesize \addtocounter{table}{-1}
    \caption{Factors of reporting bias, year effects \& survey characteristics - continued}
    \label{tab:yearFE_survey_p2}
    \centering
    \input{tables/A4_yearFE_survey_2.tex}
\end{table}

\begin{table}[H]\footnotesize \addtocounter{table}{-1}
    \caption{Factors of reporting bias, year effects \& survey characteristics - continued}
    \label{tab:yearFE_survey_p3}
    \centering
    \input{tables/A4_yearFE_survey_3.tex}
    \fignote{This table summarizes coefficients of various operationalisations of the reporting error on potential correlates on the survey mode and year effects based on Equation \ref{equ:main_analysis}. These coefficients were omitted from Table \ref{tab:main_table} for clarity reasons. Dependent variables are a binary indicator for overreporting (column 1), the difference in log wages (column 2), and the difference in log wages split between overreporting (column 3) and underreporting individuals (column 4). 
Correct reporting is defined as no or small reporting bias within a +/- 2.5 percentile range of the annual reporting bias.
All regressions include year-fixed effects. Standard errors in parentheses. * \(p<0.1\), ** \(p<0.05\), *** \(p<0.01\)}
\end{table}
\setstretch{1.5}

\setstretch{0.5}
\begin{table}[H]\footnotesize
    \caption{Factors of reporting bias, zero bias defined as $+/- 1.25$\% around 0}
    \label{tab:main_table_125}
    \centering
    \input{tables/A2_zerobias_125_1_july2024.tex}
\end{table}

\begin{table}[H]\footnotesize \addtocounter{table}{-1}
    \caption{Factors of reporting bias, zero bias defined as $+/- 1.25$\% around 0 - continued}
    \label{tab:main_table_125_p2}
    \centering
    \input{tables/A2_zerobias_125_2_july2024.tex}
    \fignote{This table summarizes coefficients of various operationalisations of the reporting error on potential correlates on the individual, household- and firm level, based on Equation \ref{equ:main_analysis}. Dependent variables are a binary indicator for overreporting (column 1), the difference in log wages (column 2), and the difference in log wages split between overreporting (column 3) and underreporting individuals (column 4). 
Correct reporting is defined as no or small reporting bias within a +/- 1.25 percentile range of the annual reporting bias.
All regressions include year-fixed effects. Standard errors in parentheses. * \(p<0.1\), ** \(p<0.05\), *** \(p<0.01\)}
\end{table}
\setstretch{1.5}

\setstretch{0.5}
\begin{table}[H]\footnotesize
    \caption{Factors of reporting bias, zero bias defined as $+/- 5$\% around 0}
    \label{tab:main_table_5}
    \centering
    \input{tables/A3_zerobias_5_1_july2024.tex}
\end{table}

\begin{table}[H]\footnotesize \addtocounter{table}{-1}
    \caption{Factors of reporting bias, zero bias defined as $+/- 5$\% around 0 - continued}
    \label{tab:main_table_5_p2}
    \centering
    \input{tables/A3_zerobias_5_2_july2024.tex}
    \fignote{This table summarizes coefficients of various operationalisations of the reporting error on potential correlates on the individual, household- and firm level, based on Equation \ref{equ:main_analysis}. Dependent variables are a binary indicator for overreporting (column 1), the difference in log wages (column 2), and the difference in log wages split between overreporting (column 3) and underreporting individuals (column 4). 
Correct reporting is defined as no or small reporting bias within a +/- 5 percentile range of the annual reporting bias.
All regressions include year-fixed effects. Standard errors in parentheses. * \(p<0.1\), ** \(p<0.05\), *** \(p<0.01\)}
\end{table}
\setstretch{1.5}

\setstretch{0.5}
\begin{table}[H]\footnotesize
	\caption{Detailed Shapley decomposition}
	\label{tab:main_table_shapeley}
	\centering
	\input{tables/04_shapeley_values_v5.tex}
	\fignote{Table shows the results of the Shapley decomposition based on an OLS regression with the log of reporting bias (log) as dependent variable and all individual, household, and firm variables included in Table \ref{tab:main_table} as explanatory variables. Shapley values are computed for the groups of variables.}
\end{table}
\setstretch{1.5}

\end{appendix}

\end{document}

%% file: Tables.tex
\setstretch{0.5}
\begin{table}[H]
	\footnotesize
	\centering
	\caption{Cross-section of individuals interviewed in 2019} \label{table1}		
	\input{tables/01_Sampling_SOEPADIAB_v3}
	\fignote{This table summarizes the number of observations in the linked SOEP - ADIAB dataset at different sampling-stages. Observation numbers are displayed separately for the SOEP Core sample and the SOEP Innovation Sample (IS). Rows 4 to 9: cross-section 2019. }
\end{table}
\setstretch{1.5}

\setstretch{0.5}
\begin{table}[H]
	\scriptsize
	\centering
	\caption{Reporting Bias per Wage Ventile} \label{table2}		
	\input{tables/02_Descr_Vintiles}
	\fignote{This table reports summary statistics by IEB wage ventiles. Column (1) depicts the mean IEB wage of each ventile, measured in \texteuro. Columns (2) - (11) show various information on the reporting bias, separately for each wage ventile. The minimum reporting bias in column (2), the 5th percentile value of the reporting bias in (3), the mean and median reporting bias in (4) and (5), the 95th percentile value of the reporting bias in column (6) and the maximum reporting bias in (7). Column (8) shows the average relative reporting bias as compared to the mean wage of each ventile in percent. Column (9) to (11) depict the proportion of individuals overstating their wage, understating their wage and reporting correctly within a +/- 2.5 \% range around the zero bias. Column (12) gives the number of observations for each ventile.}
\end{table}
\setstretch{1.5}

\setstretch{0.5}
\begin{table}[H]\footnotesize
	\caption{Factors of reporting bias}
	\label{tab:main_table}
	\centering
	\input{tables/03a_main_table_july2024_1.tex}
\end{table}

\begin{table}[H]\footnotesize \addtocounter{table}{-1}
	\caption{Factors of reporting bias - continued}
	\label{tab:main_table_p2}
	\centering
	\input{tables/03a_main_table_july2024_2.tex}
	\fignote{This table summarizes coefficients of various operationalisations of the reporting error on potential correlates on the individual, household- and firm level, based on Equation \ref{equ:main_analysis}. Dependent variables are a binary indicator for over-reporting (column 1), the difference in log wages (column 2), and the difference in log wages split between over-reporting (column 3) and under-reporting individuals (column 4). Shapley decomposition values indicate shares of explained variation attributed to groups of variables and are reported for the estimates in column (2). All regressions include year-fixed effects. Standard errors in parentheses. * \(p<0.1\), ** \(p<0.05\), *** \(p<0.01\)}
\end{table}
\setstretch{1.5}

%% file: tables/01_Sampling_SOEPADIAB_v3.tex
{
\def\sym#1{\ifmmode^{#1}\else\(^{#1}\)\fi}
\begin{tabular}{clrrr}
\toprule
\multicolumn{5}{c}{Unique Individuals by Sampling Step}               \\
Sampling  stage& Restriction & Core & IS & Total \\
\midrule
1. & Contacted & 20,525 & 3,000  & 23,525  \\             
2. & Have consented & 13,166 & 1,846  & 15,012  \\            
3. & Linked & 13,154 & 1,829  & 14,983 \\            
4. & Working spell within 30 days &  6,903 & 711 & 7,614\\
5. & No pure self-employment spell (SOEP) & 6,835 & 711 & 7,546 \\
6. & Age 20-65  & 6,497 & 680 & 7,177   \\
7. & SOEP and IEB wages $>$ 0 & 5,997& 622 & 6,619  \\
8. & IEB wages $<$ assessment limit - \texteuro{}120  & 5,592 & 584 & 6,176\\
9. & SOEP wages $<$ assessment limit - \texteuro{}120 & 5,557 & 579  & 6,136 \\
\midrule
& \multicolumn{1}{c}{Unique individual x year observations} & \multicolumn{3}{c}{59,118}               \\
\bottomrule
\end{tabular}
}

%% file: tables/02_Descr_Vintiles.tex
{
\def\sym#1{\ifmmode^{#1}\else\(^{#1}\)\fi}
\begin{tabular}{ccccccccccccc}
\toprule
& &   \multicolumn{8}{c}{Reporting Bias} & & &   \\ \cmidrule(lr){3-12}
Ventile & Mean IEB & Min & P5& Mean& Median & P95 & Max & Avg. rel. &  Over &  Under & Zero Bias & N\\ 
& wage in \texteuro & in \texteuro & in \texteuro& in \texteuro & in \texteuro & in \texteuro & in \texteuro& in \% &  in \% &  in \% &  in \% & \\ \addlinespace
& (1) & (2) & (3) & (4) & (5) & (6) & (7) & (8) & (9) & (10) & (11) & (12)\\
\midrule
1    &    249   &      -513     &     -94      &    232     &      23     &    1,356    &     5,127    &     242.5     &    57.4      &     25.4     &       17.2   &   2,985\\
2      &    439    &     -590    &     -127     &      91      &      1    &      594     &    5,151       &   22.9     &    47.0       &     28.8      &      24.2    &   2,989\\
3      &    682    &     -761     &    -289     &      53      &     -6     &     637    &     4,807    &     10.1     &    37.7      &     50.5      &     11.7   &    2,912\\
4     &     973    &    -1,101      &   -357     &       1      &    -37     &     532      &   3,114    &     0.3      &   34.6        &    58.7       &    6.7   &   2,952\\
5      &   1,207    &    -1,203    &     -411     &     -36     &     -59     &     391    &     4,121    &    -2.8    &     30.1       &   64.9     &     5.0    &   2,957\\
6     &    1,413    &    -1,352    &     -412     &     -44     &     -66    &      377      &   3,453   &     -3.2    &     30.0       &    64.8     &   5.2  &    2,960\\
7      &   1,593    &    -1,452     &    -463     &     -64     &     -72    &      349     &    2,784    &    -4.0    &     29.5        &    65.0     &   5.6    &   2,953\\
8      &   1,772     &   -1,670     &    -536     &    -101    &     -101      &    329     &    2,987    &    -5.7    &     24.7       &    71.3     &     4.0   &   2,959\\
9     &    1,948    &    -2,186     &    -669     &    -138     &    -124    &      387     &    3,026    &    -7.1    &     24.9      &     71.7     &      3.4      & 2,960\\
10    &    2,130     &   -2,424    &     -654   &      -152     &    -145    &      361     &   3,230     &   -7.1     &    25.5     &      72.1     &      2.4    &   2,944\\
11     &    2,310    &    -2,188    &     -702     &    -173    &     -170      &    341      &   2,458    &    -7.6     &    21.8     &       75.6   &       2.6    &   2,962\\
12     &    2,495    &    -2,342     &    -707    &     -186    &     -188      &    347   &      2,763   &     -7.5    &     20.0      &     77.8   &       2.1     &  2,950\\
13      &   2,680    &    -2,770    &     -777     &    -226     &    -221      &    337    &     2,492    &    -8.6    &     17.7     &      80.2     &      2.1    &   2,963\\
14    &    2,876     &   -2,297     &    -843     &    -265    &     -267     &     317     &    1,862     &   -9.3     &   15.5        &    83.1       &     1.5   &    2,951\\
15     &    3,073    &    -3,136   &     -942     &    -295     &    -297      &    325    &     2,105    &    -9.7      &   16.6       &  81.9       &   1.5    &   2,959 \\
16    &     3,296   &    -2,674   &     -1,018     &    -336     &    -324     &     327     &    2,263    &    -10.2     &    15.1         &   83.8       &    1.2     &  2,954\\
17     &    3,557    &    -2,406    &    -1,091     &    -377      &   -369     &     286    &     1,852    &    -10.7    &    13.4       &    85.9      &     0.8    &   2,958\\
18     &    3,879    &    -3,571    &    -1,266      &   -461     &    -414     &     210     &    1,577     &   -11.9      &   11.6       &     87.7       &     0.7    &   2,954\\
19     &    4,314    &    -4,321    &    -1,462      &   -566     &    -511      &    173     &    1,318     &   -13.2     &    9.7       &     89.5      &      0.7    &  2,958\\
20     &    5,001    &    -5,121   &     -1,670     &    -679     &    -605      &     70     &     774     &   -13.5    &  6.7        &    92.9      &      0.4    &   2,938\\
\midrule
All & 2,293& -5,121 & -921 & -186 & -138 & 373 & 5,151 & 7.3 & 24.5 & 70.5 & 5.0 & 59,118 \\
\bottomrule
\end{tabular}
}

%% file: tables/03a_main_table_july2024_1.tex
{
	\def\sym#1{\ifmmode^{#1}\else\(^{#1}\)\fi}
	\begin{tabular}{l*{4}{c}}
		\toprule
		&\multicolumn{1}{c}{(1)} &\multicolumn{1}{c}{(2)} &\multicolumn{1}{c}{(3)} &\multicolumn{1}{c}{(4)}  \\  \addlinespace
		&\multicolumn{1}{c}{Over Y/N}&\multicolumn{1}{c}{Diff of Logs}&\multicolumn{1}{c}{Diff of Logs ($+$)}&\multicolumn{1}{c}{Diff of Logs ($-$)}\\
		\midrule
		\textbf{Individual Characteristics} & & &  \\ \addlinespace
Age            &        -0.0005         &         0.0000         &         0.0002         &         0.0002         \\
               &       (0.0003)         &       (0.0002)         &       (0.0007)         &       (0.0001)         \\
Woman          &        -0.0207\sym{*}  &        -0.0160\sym{*}  &        -0.0195         &        -0.0072\sym{*}  \\
               &       (0.0087)         &       (0.0067)         &       (0.0265)         &       (0.0029)         \\
Years of education&        -0.0015         &         0.0007         &         0.0001         &         0.0017\sym{***}\\
               &       (0.0010)         &       (0.0009)         &       (0.0031)         &       (0.0004)         \\
\textit{Region (ref. cat: North)} & & & \\
East           &         0.0022         &        -0.0080         &        -0.0383\sym{*}  &         0.0002         \\
               &       (0.0104)         &       (0.0056)         &       (0.0151)         &       (0.0035)         \\
South          &        -0.0032         &         0.0092         &         0.0217         &         0.0037         \\
               &       (0.0104)         &       (0.0070)         &       (0.0221)         &       (0.0035)         \\
West           &         0.0103         &         0.0168\sym{*}  &         0.0236         &         0.0083\sym{**} \\
               &       (0.0096)         &       (0.0065)         &       (0.0208)         &       (0.0032)         \\
\addlinespace
\textit{Migration background (ref. cat: none)} & & & \\
Direct migration background&        -0.0156         &        -0.0098         &        -0.0168         &        -0.0031         \\
               &       (0.0124)         &       (0.0066)         &       (0.0155)         &       (0.0048)         \\
Indirect migration background&        -0.0105         &        -0.0019         &        -0.0017         &        -0.0010         \\
               &       (0.0163)         &       (0.0091)         &       (0.0228)         &       (0.0071)         \\
		\textit{Personality Traits} & & &  \\
Openness (std.)&         0.0015         &        -0.0016         &        -0.0037         &        -0.0014         \\
               &       (0.0045)         &       (0.0027)         &       (0.0080)         &       (0.0015)         \\
Conscientiousness (std.)&         0.0033         &        -0.0008         &        -0.0089         &         0.0008         \\
               &       (0.0046)         &       (0.0030)         &       (0.0090)         &       (0.0018)         \\
Extraversion (std.)&         0.0136\sym{***}&         0.0067\sym{**} &         0.0071         &         0.0005         \\
               &       (0.0041)         &       (0.0026)         &       (0.0082)         &       (0.0015)         \\
Agreeableness (std.)&        -0.0084         &        -0.0049         &        -0.0007         &        -0.0018         \\
               &       (0.0045)         &       (0.0031)         &       (0.0100)         &       (0.0015)         \\
Neuroticism (std.)&         0.0008         &        -0.0003         &        -0.0019         &        -0.0003         \\
               &       (0.0042)         &       (0.0027)         &       (0.0084)         &       (0.0013)         \\
	\textbf{Household Characteristics} & & & \\ \addlinespace
Married        &         0.0004         &         0.0022         &         0.0226         &        -0.0060\sym{*}  \\
               &       (0.0091)         &       (0.0046)         &       (0.0123)         &       (0.0029)         \\
Household size &        -0.0075\sym{*}  &        -0.0019         &         0.0147         &        -0.0038\sym{**} \\
               &       (0.0036)         &       (0.0027)         &       (0.0094)         &       (0.0013)         \\
Number of kids in HH&         0.0085         &         0.0023         &        -0.0170         &         0.0053\sym{***}\\
               &       (0.0045)         &       (0.0035)         &       (0.0119)         &       (0.0016)         \\
Partner        &         0.0048         &         0.0011         &        -0.0271         &         0.0094\sym{**} \\
               &       (0.0103)         &       (0.0056)         &       (0.0143)         &       (0.0034)         \\
Partner $\times$ Partner inc in \texteuro 1000 &         0.0057\sym{***}&         0.0031\sym{**} &         0.0020         &         0.0013\sym{*}  \\
               &       (0.0016)         &       (0.0012)         &       (0.0026)         &       (0.0005)         \\
		\bottomrule
		\multicolumn{4}{l}{\footnotesize Continued on next page}\\
	\end{tabular}
}						

%% file: tables/03a_main_table_july2024_2.tex
{
	\def\sym#1{\ifmmode^{#1}\else\(^{#1}\)\fi}
	\begin{tabular}{l*{4}{c}}
		\toprule
		&\multicolumn{1}{c}{(1)} &\multicolumn{1}{c}{(2)} &\multicolumn{1}{c}{(3)} &\multicolumn{1}{c}{(4)} \\  \addlinespace
			&\multicolumn{1}{c}{Over Y/N}&\multicolumn{1}{c}{Diff of Logs}&\multicolumn{1}{c}{Diff of Logs ($+$)}&\multicolumn{1}{c}{Diff of Logs ($-$)}\\
		\midrule
\textbf{Job and Firm Characteristics} & & & \\ \addlinespace
White collar worker&        -0.0493\sym{***}&         0.0031         &         0.0315         &         0.0104\sym{***}\\
               &       (0.0087)         &       (0.0062)         &       (0.0186)         &       (0.0030)         \\
Union memberhsip&        -0.0325\sym{***}&        -0.0112\sym{*}  &         0.0219         &        -0.0073\sym{**} \\
               &       (0.0075)         &       (0.0056)         &       (0.0218)         &       (0.0028)         \\
\multicolumn{2}{l}{\textit{Working Hours (ref. cat.: <35 hours)}} & &  \\
Full-time (35-44)&        -0.0643\sym{***}&        -0.0013         &         0.0060         &         0.0297\sym{***}\\
               &       (0.0080)         &       (0.0055)         &       (0.0184)         &       (0.0031)         \\
Over-time (45+)&        -0.0183         &         0.0129         &         0.0158         &         0.0276\sym{***}\\
               &       (0.0101)         &       (0.0067)         &       (0.0206)         &       (0.0035)         \\
		\textit{Firm Size (ref. cat.: $<$10 employees)} & & & \\
10-49 employees&        -0.0328\sym{**} &        -0.0374\sym{***}&        -0.0509         &        -0.0127\sym{***}\\
               &       (0.0116)         &       (0.0101)         &       (0.0270)         &       (0.0036)         \\
50-249 employees&        -0.0469\sym{***}&        -0.0487\sym{***}&        -0.0615\sym{*}  &        -0.0158\sym{***}\\
               &       (0.0118)         &       (0.0104)         &       (0.0289)         &       (0.0038)         \\
250+ employees &        -0.0627\sym{***}&        -0.0641\sym{***}&        -0.0788\sym{**} &        -0.0263\sym{***}\\
               &       (0.0124)         &       (0.0101)         &       (0.0275)         &       (0.0040)         \\
\textit{Pay Tercile Firm (ref. cat.: 1st Tercile)} & & &  \\
2. Tercile paying firm&        -0.1126\sym{***}&        -0.0598\sym{***}&        -0.0271         &        -0.0248\sym{***}\\
               &       (0.0085)         &       (0.0056)         &       (0.0143)         &       (0.0027)         \\
3. Tercile paying firm&        -0.1932\sym{***}&        -0.1083\sym{***}&        -0.0257         &        -0.0543\sym{***}\\
               &       (0.0101)         &       (0.0074)         &       (0.0215)         &       (0.0035)         \\
\textit{Workforce composition} & & & \\
Share of highly qualified employees&         0.0037         &         0.0155         &        -0.0198         &         0.0175\sym{**} \\
               &       (0.0191)         &       (0.0144)         &       (0.0532)         &       (0.0067)         \\
Share of female employees&        -0.0600\sym{**} &        -0.0629\sym{***}&        -0.1537\sym{***}&         0.0036         \\
               &       (0.0195)         &       (0.0151)         &       (0.0411)         &       (0.0057)         \\
Share of German employees&        -0.1288\sym{***}&        -0.0076         &         0.0515         &         0.0213         \\
               &       (0.0385)         &       (0.0212)         &       (0.0478)         &       (0.0134)         \\
Share of fulltime employees&        -0.1226\sym{***}&        -0.1113\sym{***}&        -0.2357\sym{***}&        -0.0016         \\
               &       (0.0174)         &       (0.0165)         &       (0.0448)         &       (0.0054)         \\
Mean age of employees in firm&        -0.0015\sym{*}  &        -0.0003         &         0.0002         &        -0.0001         \\
               &       (0.0007)         &       (0.0005)         &       (0.0013)         &       (0.0002)         \\
Constant        &         0.6368\sym{***}&         0.0812\sym{*}  &         0.3645\sym{***}&        -0.1859\sym{***}\\
               &       (0.0642)         &       (0.0369)         &       (0.1047)         &       (0.0213)         \\

		\midrule
		Firm sector controls & \checkmark  & \checkmark  & \checkmark  & \checkmark   \\
		Survey characteristics controls & \checkmark & \checkmark & \checkmark & \checkmark  \\
		Year FE & \checkmark & \checkmark & \checkmark & \checkmark  \\
		\midrule
Shapley Decomposition: &   &   &   &    \\
\hspace{3mm} Individual characteristics &   &  \multicolumn{2}{c}{3.93\%}    &    \\
\hspace{3mm} Household characteristics &   &  \multicolumn{2}{c}{15.61\%}    &    \\
\hspace{3mm} Job and firm characteristics &   &  \multicolumn{2}{c}{69.24\%}    &    \\
\hspace{3mm} Survey characteristics &   &  \multicolumn{2}{c}{0.61\%}    &    \\
\hspace{3mm} Year FE &   &  \multicolumn{2}{c}{10.59\%}    &    \\
		\midrule
N              &          42,046         &          42,046         &           9,591         &          31,732         \\
R2             &         0.0909         &         0.0734         &         0.0557         &         0.0603         \\
Mean dep. var. &         0.2281         &        -0.0678         &         0.1687         &        -0.1408         \\
		\bottomrule
	\end{tabular}
}

%% file: Figures.tex
\begin{figure}[H]
	\centering
	\begin{subfigure}{.45\textwidth}
		\centering
		\includegraphics[width=0.9\textwidth]{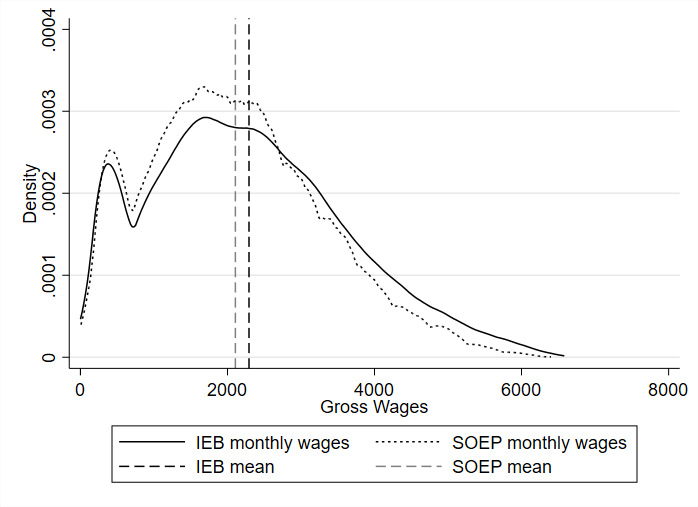}
		\caption{\footnotesize Distribution of gross monthly wages}
	\end{subfigure}
	\begin{subfigure}{.45\textwidth}
		\centering
		\includegraphics[width=0.9\textwidth]{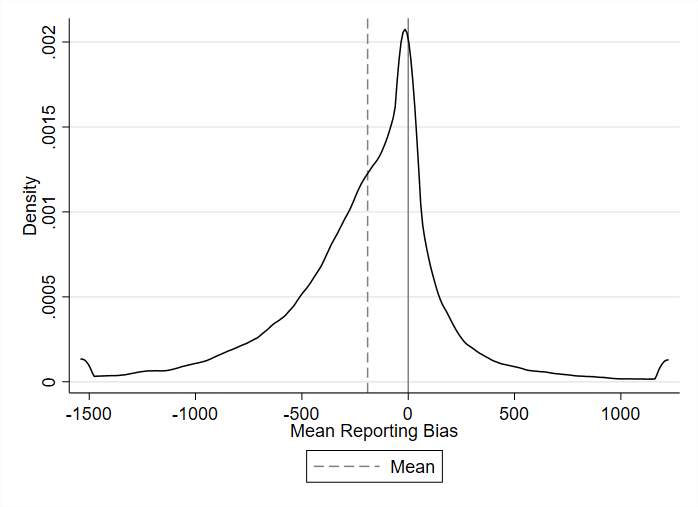}
		\caption{\footnotesize Distribution of the reporting bias}
	\end{subfigure}
	
	\begin{subfigure}{.45\textwidth}
		\centering
		\includegraphics[width=0.9\textwidth]{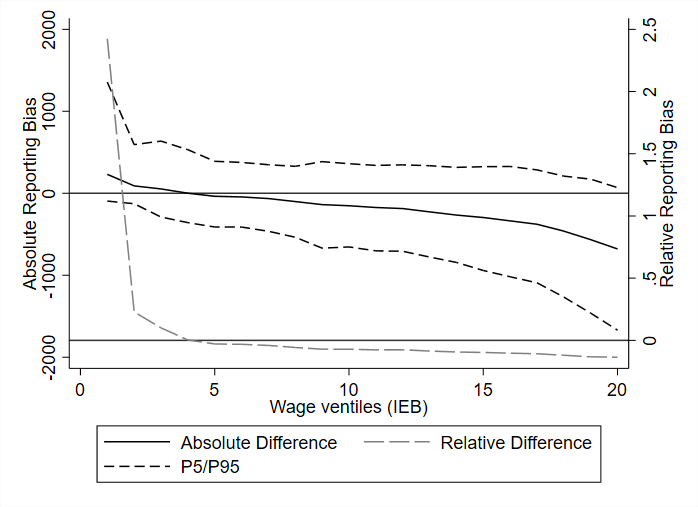}
		\caption{\footnotesize Reporting bias over the income distribution}
	\end{subfigure}
	\caption{Distribution of SOEP and IEB wages and the reporting bias}
	\fignote{Panel (a) plots the distribution of wages in SOEP and IEB (winsorized at \texteuro 7,000). The mean reporting bias in IEB data is \texteuro2,292, and \texteuro2,106 Panel (b) shows the density distribution of the reporting bias, defined by the difference of SOEP and IEB wages. Reporting biases that lie below the 1st percentile or above the 99th percentile of the distribution of differences are binned into the 1st or 99th percentile respectively. Panel (c) shows the mean reporting bias and the respective 5th and 95th percentile over each IEB wage ventile in absolute (black) and relative terms (grey).
	}
	\label{fig:descr_wages}
\end{figure}
\vspace{1cm}

\begin{figure}[H]
	\centering
	\begin{subfigure}{.45\textwidth}
		\centering
		\includegraphics[width=0.9\textwidth]{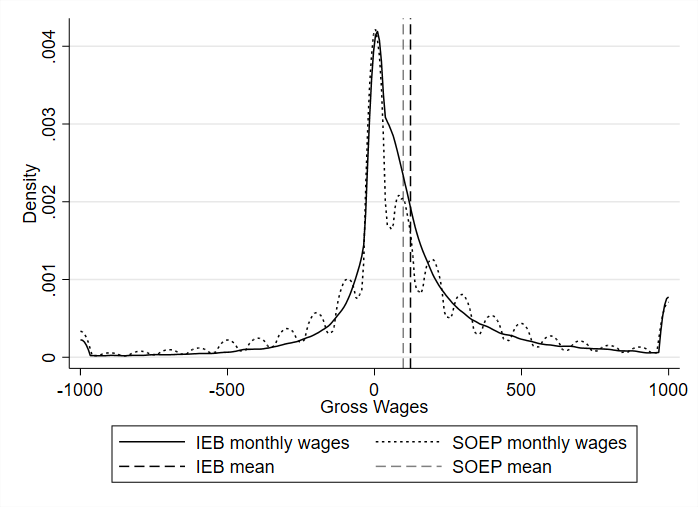}
		\caption{\footnotesize Distribution of gross monthly wage changes}
	\end{subfigure}
	\begin{subfigure}{.45\textwidth}
		\centering
		\includegraphics[width=0.9\textwidth]{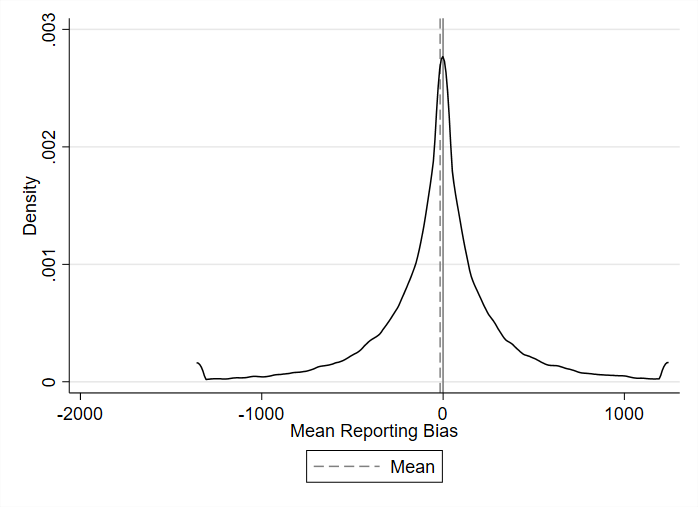}
		\caption{\footnotesize Distribution of the difference in wage changes}
	\end{subfigure}
	
	\begin{subfigure}{.45\textwidth}
		\centering
		\includegraphics[width=0.9\textwidth]{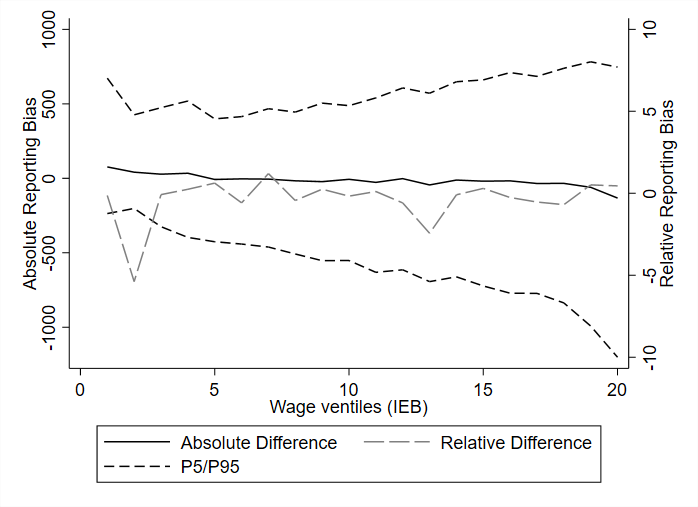}
		\caption{\footnotesize Reporting bias over the distribution in wage changes}
	\end{subfigure}
	\caption{Reporting bias in wage changes}
	\fignote{Panel (a) plots the distribution of changes in SOEP and IEB (winsorized at $\pm$ \texteuro1,000) from one year to an other. Panel (b) shows the density distribution of the difference in wage changes in SOEP and IEB. Reporting biases that lie below the 1st percentile or above the 99th percentile of the distribution of differences are binned into the 1st or 99th percentile respectively. The mean (median) difference in the change is \texteuro-16.1 (\texteuro-3.7). Panel (c) shows the difference in wage changes and the respective 5th and 95th percentile over each IEB wage ventile in absolute (black) and relative terms (grey). All three figures are based on 46,196 year-to-year differences.} 
	\label{fig:descr_wages_changes}
\end{figure}

\begin{figure}[H]
	\centering
	
	\begin{subfigure}{.35\textwidth}
		\centering
		\includegraphics[width=0.9\textwidth]{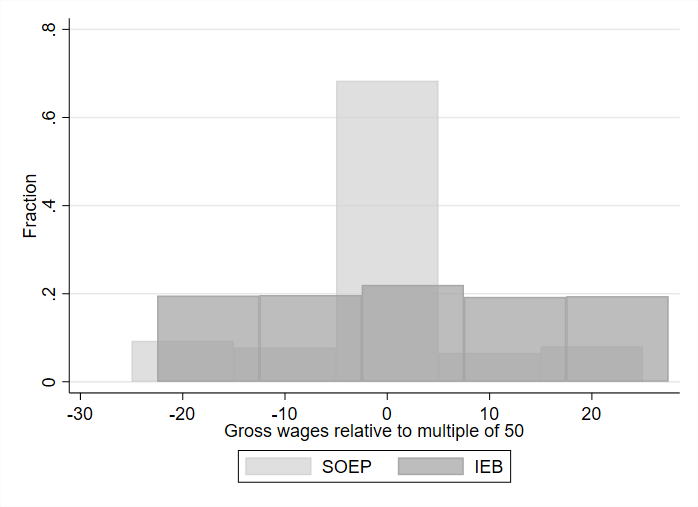}
		\caption{\footnotesize 50}
	\end{subfigure}
	\begin{subfigure}{.35\textwidth}
		\centering
		\includegraphics[width=0.9\textwidth]{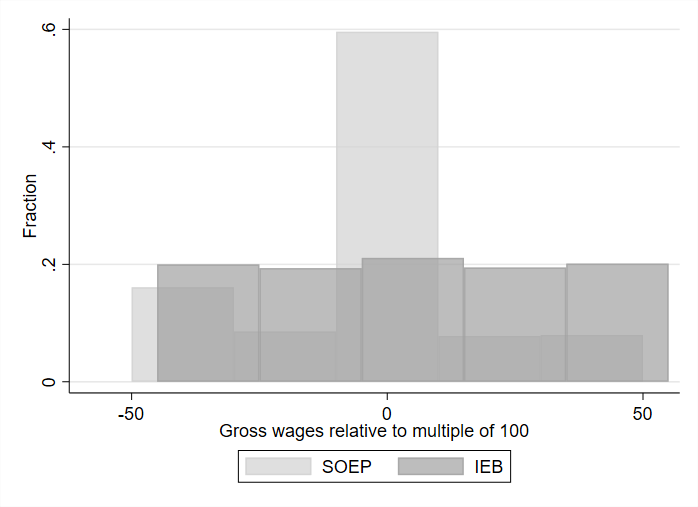}
		\caption{\footnotesize 100}
	\end{subfigure}
	\begin{subfigure}{.35\textwidth}
		\centering
		\includegraphics[width=0.9\textwidth]{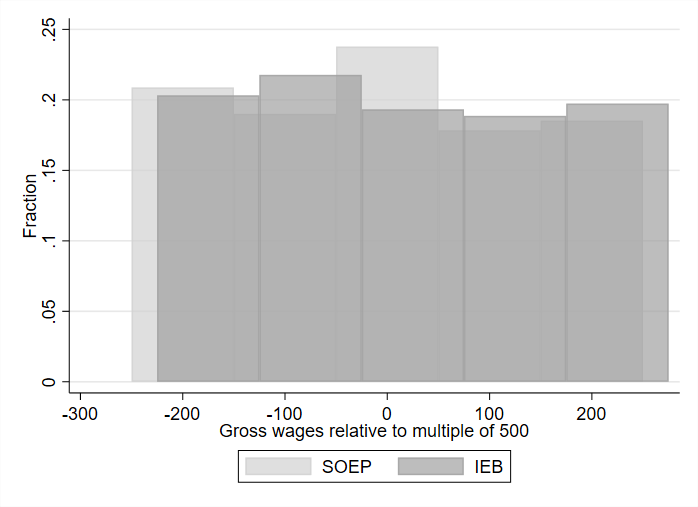}
		\caption{\footnotesize 500}
	\end{subfigure}
	\begin{subfigure}{.35\textwidth}
		\centering
		\includegraphics[width=0.9\textwidth]{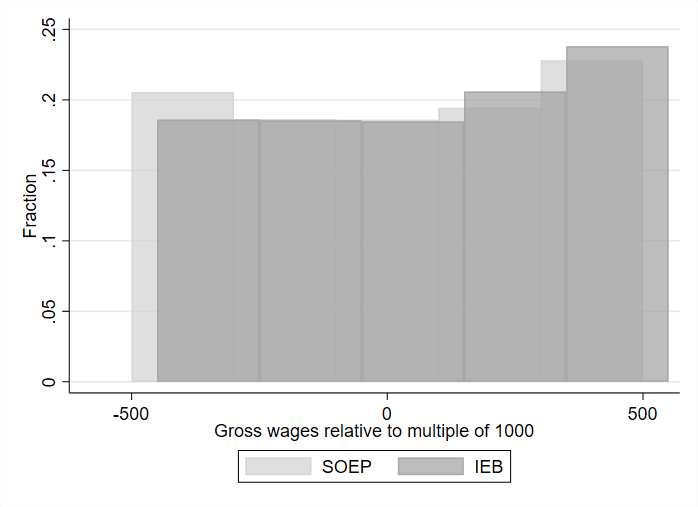}
		\caption{\footnotesize 1,000}
	\end{subfigure}
	
	\caption{Rounding of gross monthly income in both datasets}
	\fignote{This figure plots histograms of gross monthly income in SOEP (dark grey) and IEB (light grey) surrounding full increments of \texteuro50, 100, 500, 1000€. The underlying data is generated by taking the difference between the reported wage income rounded to the next \texteuro50/100/500/1,000 and the actually reported wage income.}
	\label{fig:rounding_numbers}
\end{figure}

\vspace{1cm}

\begin{figure}[H] 
	\centering
	\includegraphics[width=.49\linewidth]{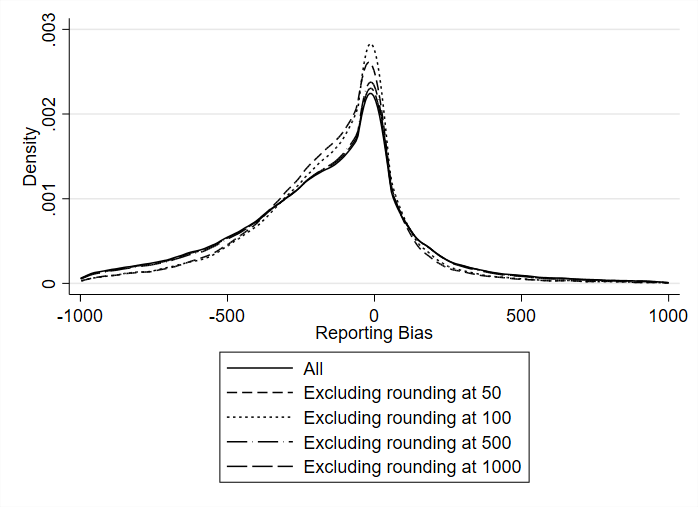}
	\caption{Distribution of reporting bias excluding rounding}
	\fignote{Graph shows the density distribution of the reporting bias for all differences and for differences excluding rounding at \texteuro 50, 100, 500 and 1,000. Observations with mean reporting bias larger than \texteuro1,000 are excluded for graphical reasons.}
	\label{fig:rounding_II}
\end{figure}

\begin{figure}[H] 
	\centering
	\includegraphics[width=.49\linewidth]{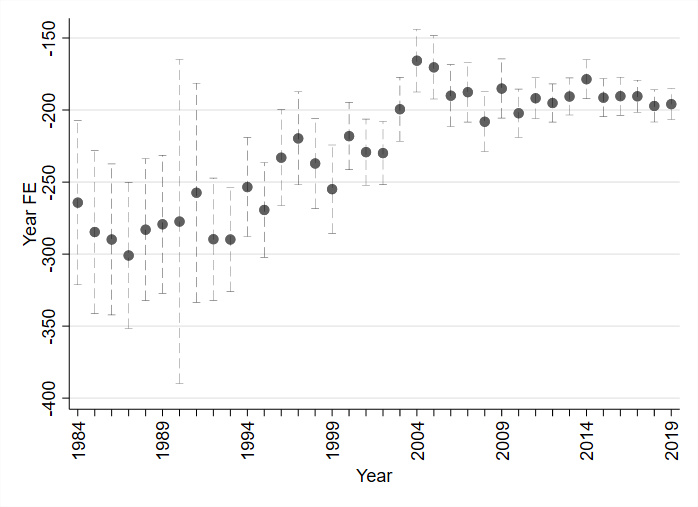}
	\caption{Year Fixed Effects}
	\fignote{Graph shows the year fixed effects estimated based on regression \ref{equ:main_analysis} with the difference between log SOEP and log IEB wages as dependent variable.}
	\label{fig:year_FE}
\end{figure}
\vspace{1cm}

\begin{figure}[H]
	\centering
	\begin{subfigure}{.45\textwidth}
		\centering
		\includegraphics[width=0.9\textwidth]{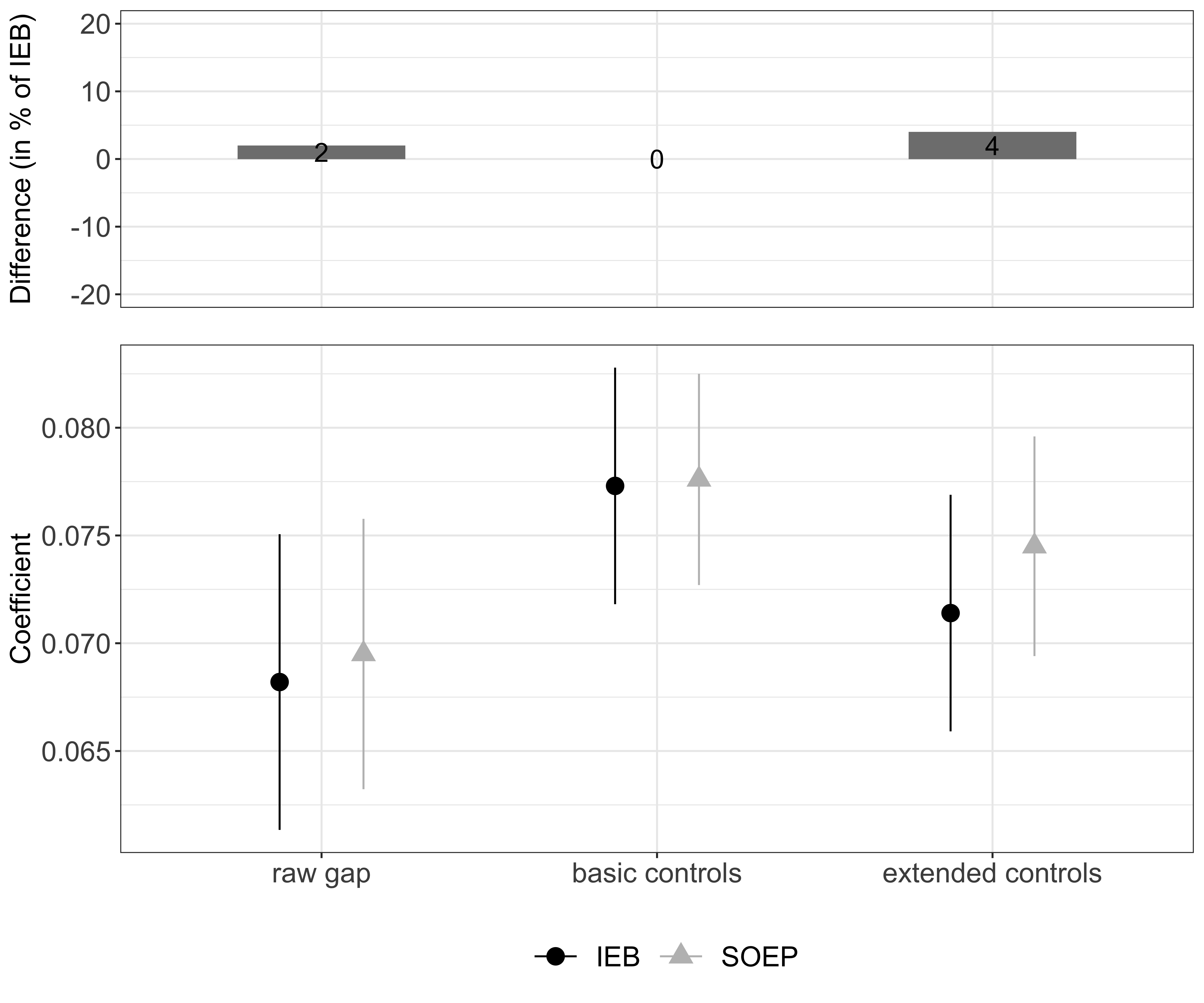}
		\caption{\footnotesize Returns to education}
	\end{subfigure}
	\begin{subfigure}{.45\textwidth}
		\centering
		\includegraphics[width=0.9\textwidth]{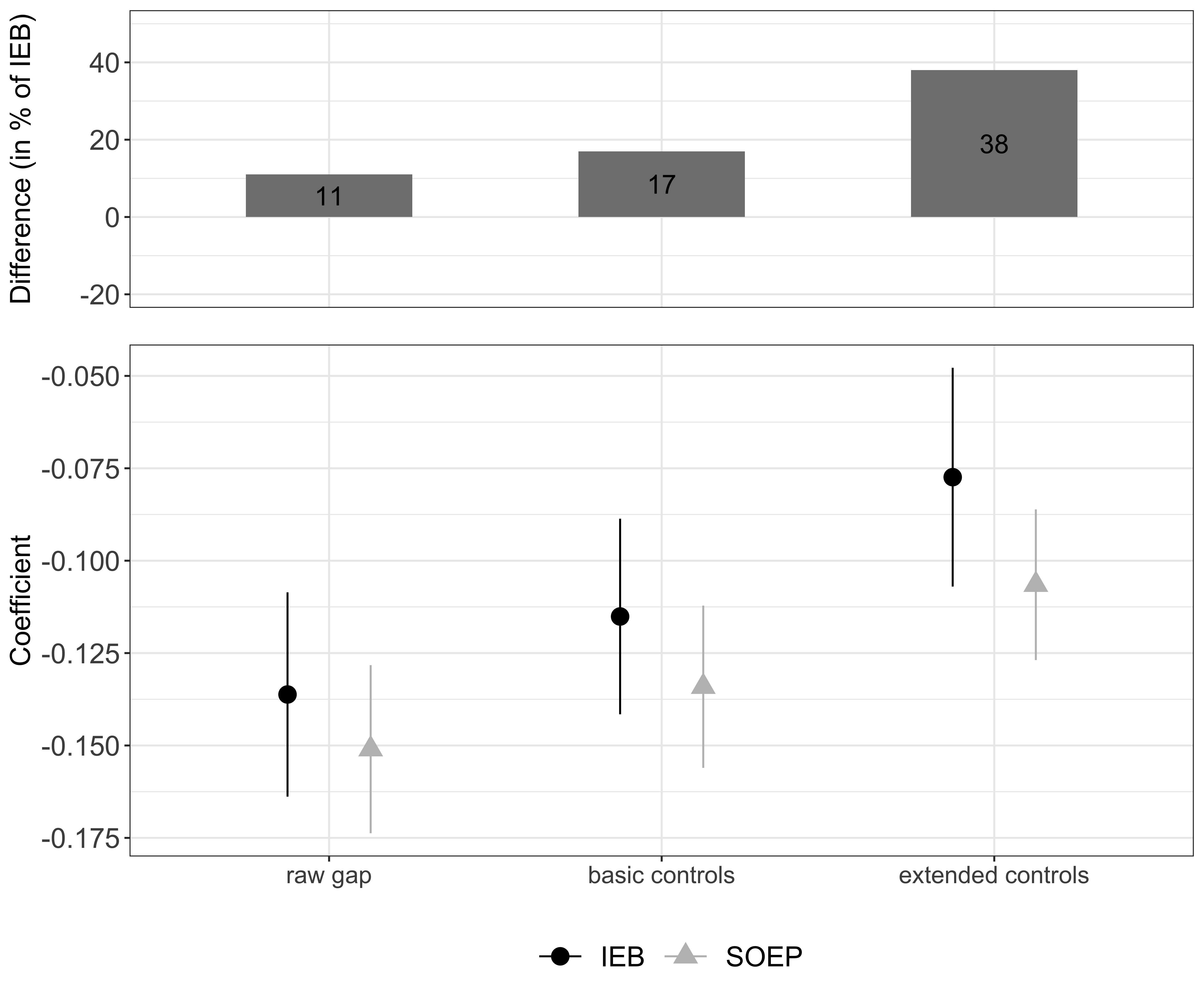}
		\caption{\footnotesize Gender wage gaps}
	\end{subfigure}
	\caption{Returns to education and gender wage gap based on SOEP and IEB wages}
	\fignote{This figure summarizes results of regressions of log wages, based on either SOEP or IEB data, on various sets of explanatory variables. Panel (a) summarizes estimates of the returns to education $\beta_1$ based on equation \ref{equ:rte}. Panel (b) summarizes estimates of the gender wage gap $\beta_2$ based on  equation \ref{equ:gwg}. Raw, basic and extended specifications differ in control variables as described in the text. Bar graphs illustrate the relative size of the difference by dividing the difference between the SOEP and IEB estimate by the IEB estimates.  Restricted to full-time for Panel (b).  The graph displays 95\% confidence intervals, based on standard errors clustered at the individual level.
	}
	\label{fig:wage_gaps_and_premia}
\end{figure}

\begin{figure}[H]
	\centering
	\begin{subfigure}{.45\textwidth}
		\centering
		\includegraphics[width=0.85\textwidth]{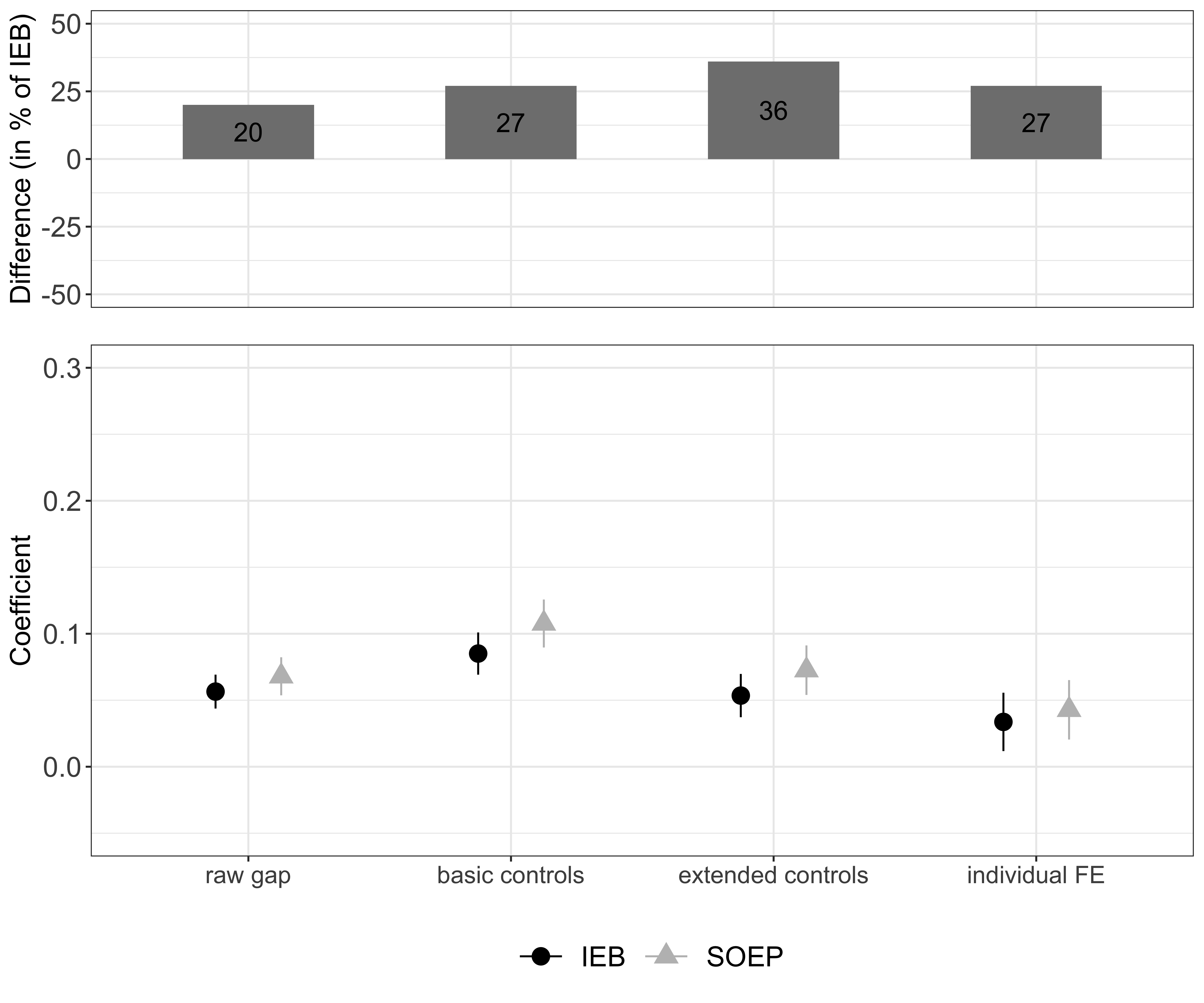}
		\caption{\footnotesize Life satisfaction \\\hspace{0.5\textwidth} }
	\end{subfigure}
	\begin{subfigure}{.45\textwidth}
		\centering
		\includegraphics[width=0.85\textwidth]{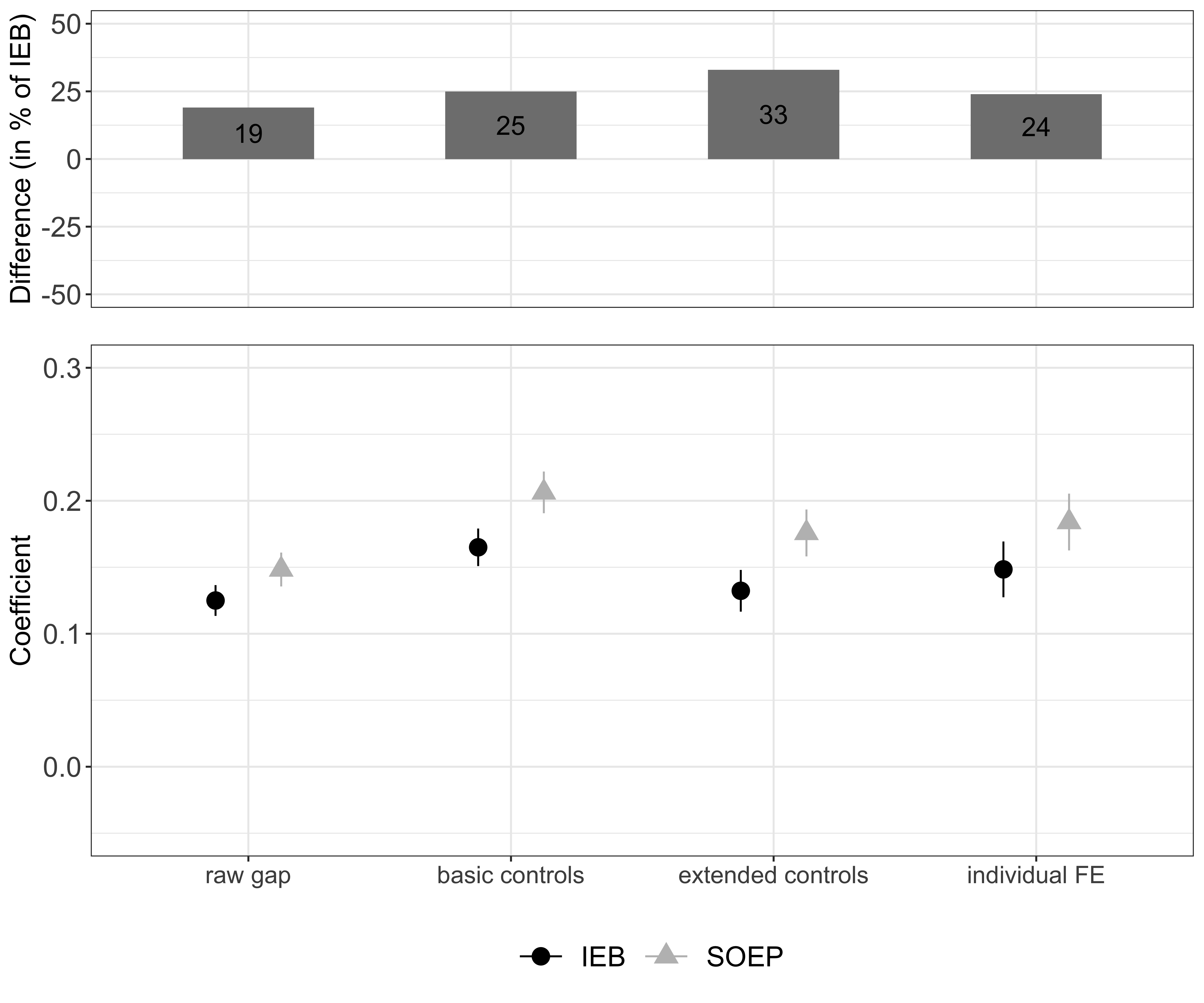}
		\caption{\footnotesize Satisfaction with household income \\\hspace{0.5\textwidth} }
	\end{subfigure}
	\begin{subfigure}{.45\textwidth}
		\centering
		\includegraphics[width=0.85\textwidth]{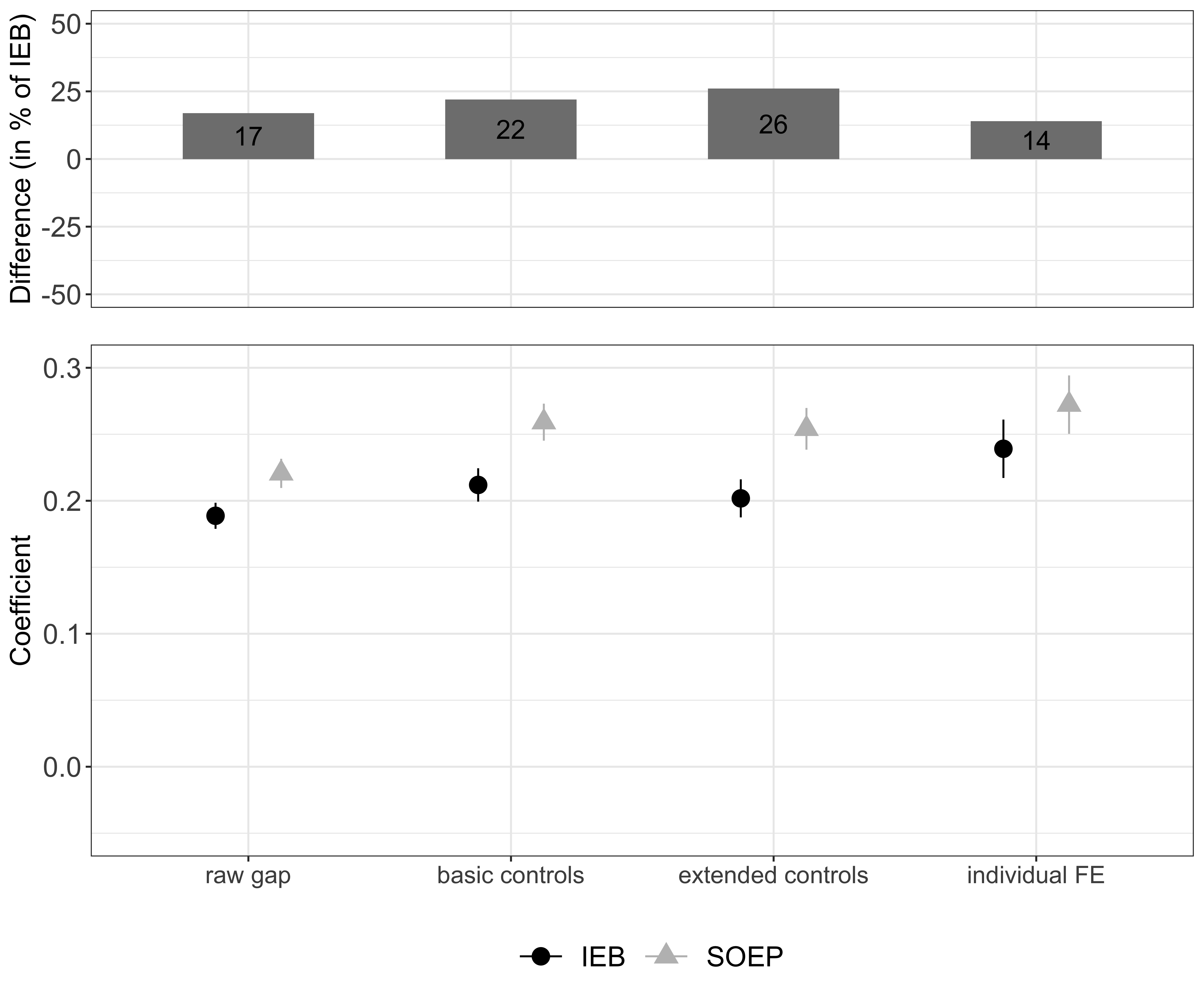}
		\caption{\footnotesize Satisfaction with personal income \\\hspace{0.5\textwidth} }
	\end{subfigure}
	\begin{subfigure}{.45\textwidth}
		\centering
		\includegraphics[width=0.85\textwidth]{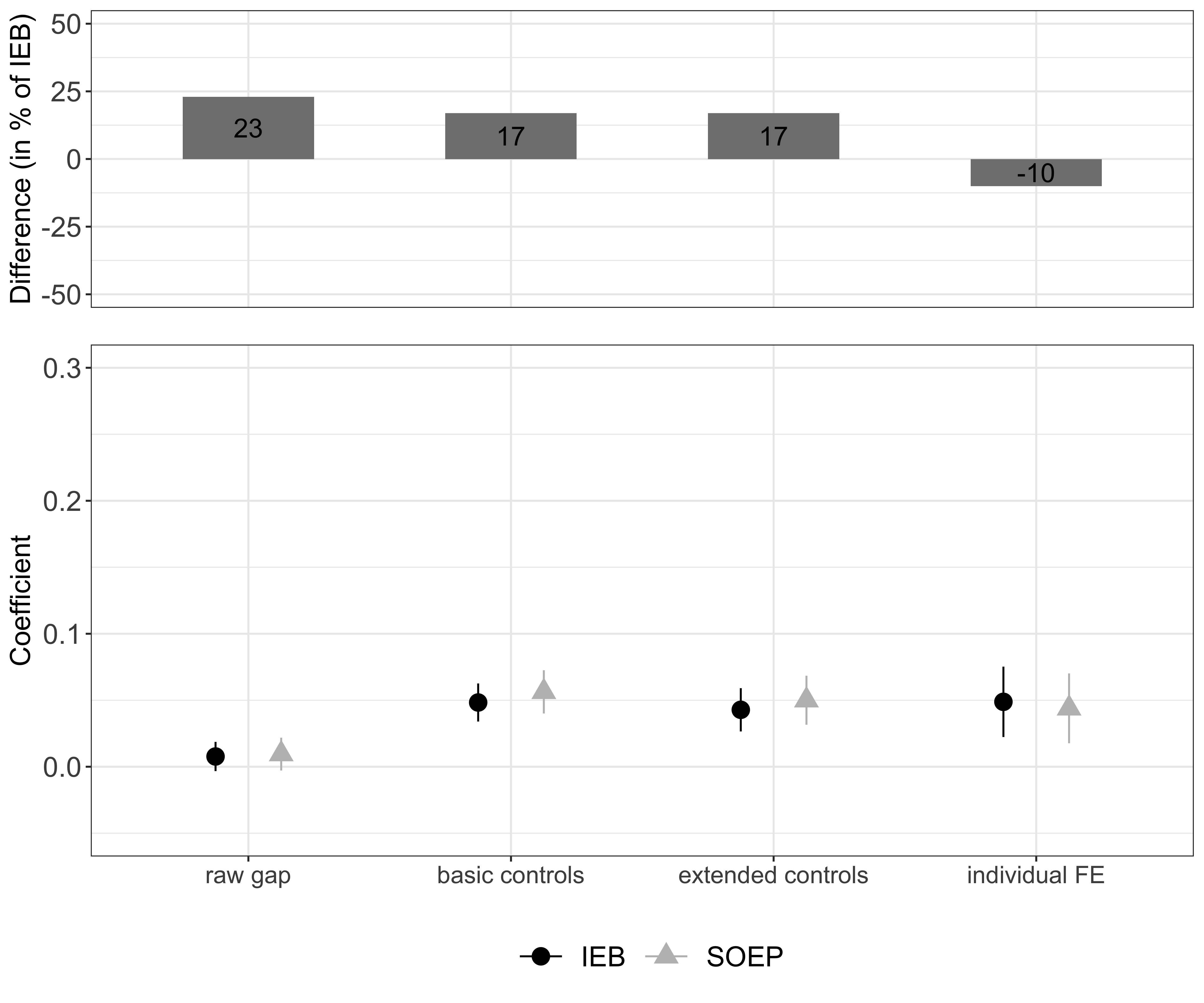}
		\caption{\footnotesize Satisfaction with job \\\hspace{0.5\textwidth} }
	\end{subfigure}
	\caption{Relation between (life) satisfaction and wages}
	\fignote{This figure summarizes results of regressions of subjective satisfaction indicators on log wages and control variables based on Equation \ref{equ:rhs}, with varying sets of control variables. Outcome variables are standardized to a mean of zero and a standard deviation of one. Raw, basic and extended specifications differ in control variables as described in the text. Bar graphs illustrate the relative size of the difference by dividing the difference between the SOEP and IEB estimate by the IEB estimates. The graph displays 95\% confidence intervals, based on standard errors clustered at the individual level.}
	\label{fig:satisfaction_wage}
\end{figure}

%% file: tables/A1_Descriptives_Consent.tex
\begin{tabular}{l*{3}{cc}}
\toprule
            &\multicolumn{1}{c}{Consent}&\multicolumn{1}{c}{No Consent}&\multicolumn{1}{c}{Diff C/NC}\\
               &     Mean/SD&     Mean/SD&   t-Test/SE         \\
\midrule
\multicolumn{4}{l}{\textbf{Individual Characteristics}} \\ \addlinespace
Age         &       50.70&       51.86&        1.16***\\
            &     (17.86)&     (18.44)&      (0.25)         \\
Female  (0/1)    &        0.54&        0.54&       -0.00         \\
            &      (0.50)&      (0.50)&      (0.01)         \\
German    (0/1)    &        0.96&        0.95&       -0.02***\\
 &      (0.19)&      (0.22)&      (0.00)         \\
Married   (0/1)    &        0.57&        0.58&        0.00         \\
            &      (0.49)&      (0.49)&      (0.01)         \\
Ever partner  (0/1) &        1.04&        1.11&        0.06***\\
            &      (0.82)&      (0.87)&      (0.01)         \\
Years of education&       11.83&       11.84&        0.01         \\
            &      (4.15)&      (4.50)&      (0.06)         \\
Father abitur  (0/1) &        0.16&        0.17&        0.01*  \\
            &      (0.37)&      (0.38)&      (0.01)         \\
Mother abitur  (0/1) &        0.11&        0.12&        0.01         \\
            &      (0.31)&      (0.32)&      (0.00)         \\ \addlinespace
\multicolumn{4}{l}{\textit{Personality traits}} \\
Optimistic (0/1) &        0.80&        0.77&       -0.03***\\
            &      (0.40)&      (0.42)&      (0.01)         \\
Risk lovingness (std)&        0.04&       -0.07&       -0.11***\\
            &      (0.90)&      (0.92)&      (0.01)         \\
IQ prediction (std)&        0.08&        0.13&        0.04*  \\
            &      (1.00)&      (0.99)&      (0.02)         \\
Openness (std)&        0.04&       -0.01&       -0.05***\\
            &      (0.85)&      (0.85)&      (0.01)         \\
Conscientiousness (std)&       -0.09&       -0.07&        0.01         \\
            &      (0.86)&      (0.87)&      (0.01)         \\
Extraversion (std)&        0.03&       -0.04&       -0.07***\\
            &      (0.90)&      (0.90)&      (0.01)         \\
Agreeableness (std)&       -0.02&       -0.05&       -0.04**\\
            &      (0.84)&      (0.83)&      (0.01)         \\
Neuroticism (std)&       -0.05&        0.00&        0.05***\\
            &      (0.88)&      (0.87)&      (0.01)         \\
Worriedness (std)&       -0.02&        0.03&        0.05**\\
            &      (0.99)&      (0.98)&      (0.01)         \\
		\bottomrule
		\multicolumn{4}{l}{\footnotesize Continued on next page}\\
	\end{tabular}

%% file: tables/A1_Descriptives_Consent_2.tex
\begin{tabular}{l*{3}{cc}}
\toprule
            &\multicolumn{1}{c}{Consent}&\multicolumn{1}{c}{No Consent}&\multicolumn{1}{c}{Diff C/NC}\\
               &     Mean/SD&     Mean/SD&   t-Test/SE         \\
\midrule
\multicolumn{4}{l}{\textbf{Job and Firm Characteristics}} \\ \addlinespace
Monthly wage SOEP&     2916.50&     3069.70&      153.20** \\
            &   (2514.09)&   (2594.24)&     (46.71)         \\
Unemployed >1 Year (0/1) &        0.21&        0.19&       -0.02***\\
            &      (0.41)&      (0.39)&      (0.01)         \\
Public sector empl  (0/1) &        0.28&        0.27&       -0.01         \\
            &      (0.45)&      (0.44)&      (0.01)         \\
Ever civil servant  (0/1) &        0.07&        0.08&        0.01** \\
            &      (0.25)&      (0.27)&      (0.00)         \\
Max. experience fulltime | >0&       19.41&       20.07&        0.66** \\
            &     (14.00)&     (13.91)&      (0.22)         \\
Max. experience parttime | >0&        7.85&        8.01&        0.16         \\
            &      (8.08)&      (8.35)&      (0.16)         \\
Max. experience unemployed | >0&        3.66&        3.12&       -0.54***\\
            &      (5.08)&      (4.45)&      (0.12)         \\
White collar worker&        0.78&        0.80&        0.02** \\
            &      (0.42)&      (0.40)&      (0.01)         \\
Union membership&        0.12&        0.11&       -0.01** \\
            &      (0.33)&      (0.31)&      (0.00)         \\
Firm size <=10&        0.15&        0.16&        0.01*  \\
            &      (0.35)&      (0.37)&      (0.01)         \\
Firm size < 100&        0.23&        0.24&        0.01         \\
            &      (0.42)&      (0.43)&      (0.01)         \\
Firm size < 2000&        0.30&        0.30&       -0.00         \\
            &      (0.46)&      (0.46)&      (0.01)         \\
Firm sie >=2000&        0.32&        0.30&       -0.02* \\
            &      (0.47)&      (0.46)&      (0.01)         \\
\midrule
\(N\)       &       15,014&        8,514&       23,528         \\
\bottomrule
\end{tabular}

%% file: tables/A4_yearFE_survey.tex
{
	\def\sym#1{\ifmmode^{#1}\else\(^{#1}\)\fi}
	\begin{tabular}{p{0.35\linewidth} *{4}{c}}
		\toprule
		&\multicolumn{1}{c}{(1)} &\multicolumn{1}{c}{(2)} &\multicolumn{1}{c}{(3)} &\multicolumn{1}{c}{(4)}  \\  \addlinespace
		&\multicolumn{1}{c}{Over Y/N}&\multicolumn{1}{c}{Diff of Logs}&\multicolumn{1}{c}{Diff of Logs ($+$)}&\multicolumn{1}{c}{Diff of Logs ($-$)}\\
		\midrule
\multicolumn{2}{l}{\textbf{Firm sector controls}}  & & \\ \addlinespace
\multicolumn{2}{l}{\textit{1-digit industries (ref. cat: Agriculture, hunting and forestry)}} & & \\
Fishing &        -0.2373         &        -0.0328         &        -0.1305\sym{**} &         0.0653\sym{***}\\
               &       (0.2261)         &       (0.0194)         &       (0.0440)         &       (0.0107)         \\
Mining and quarrying  &        -0.0501         &        -0.0096         &         0.1126         &        -0.0390         \\
           &       (0.0505)         &       (0.0291)         &       (0.1585)         &       (0.0534)         \\
Manufacturing &        -0.0480         &        -0.0240         &        -0.0199         &        -0.0153         \\
               &       (0.0344)         &       (0.0152)         &       (0.0264)         &       (0.0102)         \\
Electricity, gas and water supply &        -0.0970\sym{*}  &        -0.0525\sym{**} &        -0.0134         &        -0.0329\sym{*}  \\
               &       (0.0434)         &       (0.0199)         &       (0.0419)         &       (0.0156)         \\
Construction    &        -0.0360         &        -0.0345\sym{*}  &        -0.0707\sym{*}  &        -0.0102         \\
               &       (0.0361)         &       (0.0157)         &       (0.0280)         &       (0.0112)         \\
Wholesale and retail trade; repair of motor vehicles, motorcycles and personal and household goods &        -0.0911\sym{**} &        -0.0370\sym{*}  &        -0.0331         &        -0.0112         \\
               &       (0.0348)         &       (0.0158)         &       (0.0302)         &       (0.0103)         \\
Hotels and restaurants  &         0.0298         &         0.0368         &         0.0698         &        -0.0362\sym{*}  \\
               &       (0.0423)         &       (0.0292)         &       (0.0527)         &       (0.0153)         \\
Transport, storage and communication &         0.0159         &         0.0176         &         0.0179         &        -0.0014         \\
               &       (0.0378)         &       (0.0204)         &       (0.0408)         &       (0.0111)         \\
Financial intermediation &        -0.0621         &        -0.0473\sym{**} &        -0.0050         &        -0.0448\sym{***}\\
               &       (0.0381)         &       (0.0181)         &       (0.0374)         &       (0.0119)         \\
Real estate, renting and business activities &        -0.0323         &        -0.0092         &        -0.0025         &        -0.0070         \\
               &       (0.0347)         &       (0.0159)         &       (0.0297)         &       (0.0106)         \\

Public administration and defence; compulsory social security &        -0.1126\sym{**} &        -0.0384\sym{*}  &         0.0057         &        -0.0158         \\
               &       (0.0350)         &       (0.0164)         &       (0.0391)         &       (0.0107)         \\
Education &        -0.0818\sym{*}  &        -0.0212         &         0.0309         &        -0.0077         \\
               &       (0.0363)         &       (0.0182)         &       (0.0453)         &       (0.0114)         \\
Health and social work &        -0.0487         &        -0.0115         &         0.0133         &        -0.0075         \\
               &       (0.0355)         &       (0.0184)         &       (0.0437)         &       (0.0107)         \\
Other community, social and personal service activities &        -0.0598         &        -0.0006         &         0.0697         &        -0.0063         \\
               &       (0.0368)         &       (0.0207)         &       (0.0553)         &       (0.0114)         \\
Private households with employed persons &         0.0326         &         0.0343         &        -0.0649         &         0.0451\sym{***}\\
               &       (0.1050)         &       (0.0329)         &       (0.0707)         &       (0.0128)         \\
\textbf{Survey mode} \textit{(ref.cat.: in person)} & & &  \\ \addlinespace
Survey: Not In Person&        -0.0352\sym{***}&         0.0019         &         0.0082         &         0.0123\sym{***}\\
               &       (0.0079)         &       (0.0059)         &       (0.0230)         &       (0.0026)         \\
Survey: Unclear&        -0.0269         &        -0.0087         &        -0.0273         &         0.0065         \\
               &       (0.0145)         &       (0.0083)         &       (0.0295)         &       (0.0049)         \\
Number of SOEP interviews before&        -0.0006         &        -0.0004         &        -0.0014         &         0.0001         \\
               &       (0.0005)         &       (0.0003)         &       (0.0010)         &       (0.0002)         \\
		\bottomrule
		\multicolumn{4}{l}{\footnotesize Continued on next page}\\
	\end{tabular}
}

%% file: tables/A4_yearFE_survey_2.tex
{
	\def\sym#1{\ifmmode^{#1}\else\(^{#1}\)\fi}
	\begin{tabular}{l*{4}{c}}
		\toprule
		&\multicolumn{1}{c}{(1)} &\multicolumn{1}{c}{(2)} &\multicolumn{1}{c}{(3)} &\multicolumn{1}{c}{(4)} \\  \addlinespace
			&\multicolumn{1}{c}{Over Y/N}&\multicolumn{1}{c}{Diff of Logs}&\multicolumn{1}{c}{Diff of Logs ($+$)}&\multicolumn{1}{c}{Diff of Logs ($-$)}\\
		\midrule
\textbf{Year effects} \textit{(ref.cat.: 1984)} & & &  \\ \addlinespace
1985     &        -0.0013         &        -0.0220         &        -0.0361         &        -0.0205         \\
               &       (0.0322)         &       (0.0153)         &       (0.0653)         &       (0.0111)         \\
1986     &         0.0342         &        -0.0288\sym{*}  &        -0.0802         &        -0.0345\sym{***}\\
               &       (0.0273)         &       (0.0134)         &       (0.0508)         &       (0.0099)         \\
1987     &        -0.0169         &        -0.0266         &        -0.0783         &        -0.0174         \\
               &       (0.0284)         &       (0.0141)         &       (0.0589)         &       (0.0114)         \\
1988     &         0.0400         &        -0.0195         &        -0.0568         &        -0.0299\sym{*}  \\
               &       (0.0298)         &       (0.0154)         &       (0.0540)         &       (0.0131)         \\
1989     &         0.0494         &        -0.0082         &        -0.0372         &        -0.0227         \\
               &       (0.0313)         &       (0.0147)         &       (0.0513)         &       (0.0121)         \\
1990     &        -0.0149         &        -0.0377         &        -0.1256\sym{*}  &        -0.0362         \\
               &       (0.0545)         &       (0.0309)         &       (0.0554)         &       (0.0332)         \\
1991     &         0.0318         &         0.0043         &        -0.0706         &        -0.0121         \\
               &       (0.0420)         &       (0.0183)         &       (0.0658)         &       (0.0148)         \\
1992     &         0.0455         &        -0.0248         &        -0.0163         &        -0.0533\sym{***}\\
               &       (0.0295)         &       (0.0160)         &       (0.0619)         &       (0.0119)         \\
1993     &         0.0349         &        -0.0086         &        -0.0275         &        -0.0323\sym{**} \\
               &       (0.0286)         &       (0.0141)         &       (0.0532)         &       (0.0105)         \\
1994     &         0.0883\sym{**} &         0.0254         &        -0.0255         &        -0.0057         \\
               &       (0.0290)         &       (0.0140)         &       (0.0530)         &       (0.0104)         \\
1995     &         0.0576\sym{*}  &         0.0223         &        -0.0046         &        -0.0025         \\
               &       (0.0280)         &       (0.0137)         &       (0.0552)         &       (0.0102)         \\
1996     &         0.1284\sym{***}&         0.0444\sym{***}&        -0.0555         &         0.0136         \\
               &       (0.0297)         &       (0.0134)         &       (0.0494)         &       (0.0102)         \\
1997     &         0.1331\sym{***}&         0.0564\sym{***}&        -0.0229         &         0.0194         \\
               &       (0.0286)         &       (0.0134)         &       (0.0496)         &       (0.0102)         \\
1998     &         0.1276\sym{***}&         0.0463\sym{**} &        -0.0136         &         0.0082         \\
               &       (0.0283)         &       (0.0146)         &       (0.0553)         &       (0.0109)         \\
1999     &         0.1296\sym{***}&         0.0390\sym{**} &        -0.0317         &         0.0016         \\
               &       (0.0295)         &       (0.0142)         &       (0.0534)         &       (0.0105)         \\
2000     &         0.1754\sym{***}&         0.0688\sym{***}&         0.0166         &         0.0172         \\
               &       (0.0280)         &       (0.0145)         &       (0.0551)         &       (0.0099)         \\
2001     &         0.1616\sym{***}&         0.0699\sym{***}&         0.0538         &         0.0138         \\
               &       (0.0278)         &       (0.0155)         &       (0.0610)         &       (0.0099)         \\
2002     &         0.1541\sym{***}&         0.0573\sym{***}&         0.0331         &         0.0059         \\
               &       (0.0270)         &       (0.0146)         &       (0.0561)         &       (0.0101)         \\
2003     &         0.2144\sym{***}&         0.0775\sym{***}&        -0.0042         &         0.0185         \\
               &       (0.0281)         &       (0.0139)         &       (0.0507)         &       (0.0099)         \\
2004     &         0.2728\sym{***}&         0.0949\sym{***}&         0.0009         &         0.0171         \\
               &       (0.0287)         &       (0.0155)         &       (0.0546)         &       (0.0102)         \\
2005     &         0.2400\sym{***}&         0.0937\sym{***}&         0.0089         &         0.0258\sym{*}  \\
               &       (0.0286)         &       (0.0147)         &       (0.0528)         &       (0.0101)         \\
		\bottomrule
		\multicolumn{4}{l}{\footnotesize Continued on next page}\\
	\end{tabular}
}

%% file: tables/A4_yearFE_survey_3.tex
{
	\def\sym#1{\ifmmode^{#1}\else\(^{#1}\)\fi}
	\begin{tabular}{l*{4}{c}}
		\toprule
		&\multicolumn{1}{c}{(1)} &\multicolumn{1}{c}{(2)} &\multicolumn{1}{c}{(3)} &\multicolumn{1}{c}{(4)} \\  \addlinespace
			&\multicolumn{1}{c}{Over Y/N}&\multicolumn{1}{c}{Diff of Logs}&\multicolumn{1}{c}{Diff of Logs ($+$)}&\multicolumn{1}{c}{Diff of Logs ($-$)}\\
		\midrule
\textbf{Year effects} \textit{(ref.cat.: 1984)} & & &  \\ \addlinespace
2006     &         0.2430\sym{***}&         0.0811\sym{***}&        -0.0385         &         0.0224\sym{*}  \\
               &       (0.0285)         &       (0.0144)         &       (0.0522)         &       (0.0101)         \\
2007     &         0.2389\sym{***}&         0.0890\sym{***}&         0.0102         &         0.0190         \\
               &       (0.0285)         &       (0.0156)         &       (0.0567)         &       (0.0103)         \\
2008     &         0.2458\sym{***}&         0.0748\sym{***}&        -0.0227         &         0.0063         \\
               &       (0.0287)         &       (0.0158)         &       (0.0562)         &       (0.0108)         \\
2009     &         0.2388\sym{***}&         0.0919\sym{***}&         0.0114         &         0.0206\sym{*}  \\
               &       (0.0280)         &       (0.0163)         &       (0.0587)         &       (0.0103)         \\
2010     &         0.2277\sym{***}&         0.0862\sym{***}&         0.0127         &         0.0200\sym{*}  \\
               &       (0.0275)         &       (0.0157)         &       (0.0579)         &       (0.0099)         \\
2011     &         0.2336\sym{***}&         0.0921\sym{***}&        -0.0058         &         0.0291\sym{**} \\
               &       (0.0272)         &       (0.0156)         &       (0.0586)         &       (0.0098)         \\
2012     &         0.2649\sym{***}&         0.0858\sym{***}&        -0.0391         &         0.0201\sym{*}  \\
               &       (0.0270)         &       (0.0154)         &       (0.0569)         &       (0.0099)         \\
2013     &         0.2325\sym{***}&         0.0919\sym{***}&         0.0056         &         0.0265\sym{**} \\
               &       (0.0270)         &       (0.0158)         &       (0.0584)         &       (0.0099)         \\
2014     &         0.2449\sym{***}&         0.0967\sym{***}&         0.0077         &         0.0282\sym{**} \\
               &       (0.0272)         &       (0.0160)         &       (0.0582)         &       (0.0100)         \\
2015     &         0.2307\sym{***}&         0.0939\sym{***}&         0.0002         &         0.0307\sym{**} \\
               &       (0.0273)         &       (0.0158)         &       (0.0579)         &       (0.0099)         \\
2016     &         0.2523\sym{***}&         0.0921\sym{***}&        -0.0264         &         0.0294\sym{**} \\
               &       (0.0275)         &       (0.0158)         &       (0.0578)         &       (0.0099)         \\
2017     &         0.2342\sym{***}&         0.0937\sym{***}&        -0.0145         &         0.0343\sym{***}\\
               &       (0.0270)         &       (0.0154)         &       (0.0573)         &       (0.0098)         \\
2018     &         0.2358\sym{***}&         0.0935\sym{***}&        -0.0210         &         0.0355\sym{***}\\
               &       (0.0272)         &       (0.0154)         &       (0.0572)         &       (0.0098)         \\
2019     &         0.2519\sym{***}&         0.0977\sym{***}&        -0.0297         &         0.0383\sym{***}\\
               &       (0.0272)         &       (0.0154)         &       (0.0572)         &       (0.0098)         \\
\midrule
N              &          42,046         &          42,046         &           9,591         &          31,732         \\
\bottomrule
\end{tabular}
}

%% file: tables/A2_zerobias_125_1_july2024.tex
{
	\def\sym#1{\ifmmode^{#1}\else\(^{#1}\)\fi}
	\begin{tabular}{l*{4}{c}}
		\toprule
		&\multicolumn{1}{c}{(1)} &\multicolumn{1}{c}{(2)} &\multicolumn{1}{c}{(3)} &\multicolumn{1}{c}{(4)}  \\  \addlinespace
		&\multicolumn{1}{c}{Over Y/N}&\multicolumn{1}{c}{Diff of Logs}&\multicolumn{1}{c}{Diff of Logs ($+$)}&\multicolumn{1}{c}{Diff of Logs ($-$)}\\
		\midrule
		\textbf{Individual Characteristics} & & &  \\ \addlinespace
Age            &        -0.0004         &         0.0000         &         0.0002         &         0.0002         \\
               &       (0.0003)         &       (0.0002)         &       (0.0006)         &       (0.0001)         \\
Woman          &        -0.0210\sym{*}  &        -0.0160\sym{*}  &        -0.0193         &        -0.0068\sym{*}  \\
               &       (0.0087)         &       (0.0067)         &       (0.0262)         &       (0.0029)         \\
Years of education&        -0.0017         &         0.0007         &         0.0002         &         0.0016\sym{***}\\
               &       (0.0010)         &       (0.0009)         &       (0.0031)         &       (0.0004)         \\
\textit{Region (ref. cat: North)} & & & \\
East           &         0.0029         &        -0.0080         &        -0.0382\sym{*}  &         0.0002         \\
               &       (0.0105)         &       (0.0056)         &       (0.0149)         &       (0.0035)         \\
South          &        -0.0030         &         0.0092         &         0.0214         &         0.0040         \\
               &       (0.0104)         &       (0.0070)         &       (0.0219)         &       (0.0035)         \\
West           &         0.0106         &         0.0168\sym{*}  &         0.0236         &         0.0088\sym{**} \\
               &       (0.0097)         &       (0.0065)         &       (0.0206)         &       (0.0032)         \\
\addlinespace
\textit{Migration background (ref. cat: none)} & & & \\
Direct migration background&        -0.0163         &        -0.0098         &        -0.0153         &        -0.0034         \\
               &       (0.0125)         &       (0.0066)         &       (0.0153)         &       (0.0047)         \\
Indirect migration background&        -0.0075         &        -0.0019         &        -0.0040         &        -0.0007         \\
               &       (0.0162)         &       (0.0091)         &       (0.0225)         &       (0.0070)         \\
		\textit{Personality Traits} & & &  \\
Openness (std.)&         0.0017         &        -0.0016         &        -0.0037         &        -0.0013         \\
               &       (0.0045)         &       (0.0027)         &       (0.0079)         &       (0.0015)         \\
Conscientiousness (std.)&         0.0036         &        -0.0008         &        -0.0092         &         0.0008         \\
               &       (0.0046)         &       (0.0030)         &       (0.0089)         &       (0.0018)         \\
Extraversion (std.)&         0.0134\sym{**} &         0.0067\sym{**} &         0.0073         &         0.0006         \\
               &       (0.0041)         &       (0.0026)         &       (0.0082)         &       (0.0014)         \\
Agreeableness (std.)&        -0.0082         &        -0.0049         &        -0.0007         &        -0.0021         \\
               &       (0.0045)         &       (0.0031)         &       (0.0100)         &       (0.0015)         \\
Neuroticism (std.)&         0.0005         &        -0.0003         &        -0.0017         &        -0.0001         \\
               &       (0.0042)         &       (0.0027)         &       (0.0083)         &       (0.0013)         \\
	\textbf{Household Characteristics} & & & \\ \addlinespace
Married        &        -0.0000         &         0.0022         &         0.0234         &        -0.0057         \\
               &       (0.0091)         &       (0.0046)         &       (0.0122)         &       (0.0029)         \\
Household Size &        -0.0078\sym{*}  &        -0.0019         &         0.0146         &        -0.0037\sym{**} \\
               &       (0.0036)         &       (0.0027)         &       (0.0094)         &       (0.0013)         \\
Number of kids in HH&         0.0090\sym{*}  &         0.0023         &        -0.0172         &         0.0051\sym{**} \\
               &       (0.0045)         &       (0.0035)         &       (0.0118)         &       (0.0016)         \\
Partner        &         0.0047         &         0.0011         &        -0.0274         &         0.0089\sym{**} \\
               &       (0.0103)         &       (0.0056)         &       (0.0142)         &       (0.0034)         \\
Partner $\times$ Partner Inc in \texteuro1,000  &         0.0056\sym{***}&         0.0031\sym{**} &         0.0020         &         0.0014\sym{**} \\
               &       (0.0016)         &       (0.0012)         &       (0.0026)         &       (0.0005)         \\
		\bottomrule
		\multicolumn{4}{l}{\footnotesize Continued on next page}\\
	\end{tabular}
}						

%% file: tables/A2_zerobias_125_2_july2024.tex
{
	\def\sym#1{\ifmmode^{#1}\else\(^{#1}\)\fi}
	\begin{tabular}{l*{4}{c}}
		\toprule
		&\multicolumn{1}{c}{(1)} &\multicolumn{1}{c}{(2)} &\multicolumn{1}{c}{(3)} &\multicolumn{1}{c}{(4)} \\  \addlinespace
			&\multicolumn{1}{c}{Over Y/N}&\multicolumn{1}{c}{Diff of Logs}&\multicolumn{1}{c}{Diff of Logs ($+$)}&\multicolumn{1}{c}{Diff of Logs ($-$)}\\
		\midrule
\textbf{Job and Firm Characteristics} & & & \\ \addlinespace
White Collar Worker&        -0.0494\sym{***}&         0.0031         &         0.0310         &         0.0104\sym{***}\\
               &       (0.0087)         &       (0.0062)         &       (0.0184)         &       (0.0030)         \\
Union memberhsip&        -0.0322\sym{***}&        -0.0112\sym{*}  &         0.0210         &        -0.0075\sym{**} \\
               &       (0.0075)         &       (0.0056)         &       (0.0216)         &       (0.0028)         \\
\multicolumn{2}{l}{\textit{Working hours (ref. cat.: <35 hours)}} & &  \\
Full-time (35-44)&        -0.0688\sym{***}&        -0.0013         &         0.0088         &         0.0262\sym{***}\\
               &       (0.0080)         &       (0.0055)         &       (0.0183)         &       (0.0031)         \\
Over-time (45+)&        -0.0231\sym{*}  &         0.0129         &         0.0191         &         0.0242\sym{***}\\
               &       (0.0102)         &       (0.0067)         &       (0.0204)         &       (0.0035)         \\
		\textit{Firm size (ref. cat.: $<$10 employees)} & & & \\
10-49 employees&        -0.0346\sym{**} &        -0.0374\sym{***}&        -0.0497         &        -0.0155\sym{***}\\
               &       (0.0116)         &       (0.0101)         &       (0.0266)         &       (0.0036)         \\
50-249 employees&        -0.0490\sym{***}&        -0.0487\sym{***}&        -0.0603\sym{*}  &        -0.0193\sym{***}\\
               &       (0.0118)         &       (0.0104)         &       (0.0285)         &       (0.0038)         \\
250+ employees &        -0.0654\sym{***}&        -0.0641\sym{***}&        -0.0766\sym{**} &        -0.0299\sym{***}\\
               &       (0.0124)         &       (0.0101)         &       (0.0270)         &       (0.0040)         \\
\textit{Pay tercile firm (ref. cat.: 1st Tercile)} & & &  \\
2. Tertile paying firm&        -0.1140\sym{***}&        -0.0598\sym{***}&        -0.0264         &        -0.0261\sym{***}\\
               &       (0.0085)         &       (0.0056)         &       (0.0141)         &       (0.0027)         \\
3. Tertile paying firm&        -0.1949\sym{***}&        -0.1083\sym{***}&        -0.0254         &        -0.0560\sym{***}\\
               &       (0.0101)         &       (0.0074)         &       (0.0213)         &       (0.0035)         \\
\textit{Workforce composition} & & & \\
Share of highly qualified employees&         0.0031         &         0.0155         &        -0.0198         &         0.0169\sym{*}  \\
               &       (0.0192)         &       (0.0144)         &       (0.0525)         &       (0.0067)         \\
Share of female employees&        -0.0624\sym{**} &        -0.0629\sym{***}&        -0.1512\sym{***}&         0.0019         \\
               &       (0.0195)         &       (0.0151)         &       (0.0407)         &       (0.0057)         \\
Share of German employees&        -0.1309\sym{***}&        -0.0076         &         0.0539         &         0.0239         \\
               &       (0.0385)         &       (0.0212)         &       (0.0475)         &       (0.0133)         \\
Share of fulltime employees&        -0.1250\sym{***}&        -0.1113\sym{***}&        -0.2329\sym{***}&        -0.0027         \\
               &       (0.0175)         &       (0.0165)         &       (0.0444)         &       (0.0053)         \\
Mean age of employees in firm&        -0.0014\sym{*}  &        -0.0003         &         0.0001         &        -0.0000         \\
               &       (0.0007)         &       (0.0005)         &       (0.0013)         &       (0.0002)         \\
Constant         &         0.6478\sym{***}&         0.0812\sym{*}  &         0.3586\sym{***}&        -0.1795\sym{***}\\
               &       (0.0642)         &       (0.0369)         &       (0.1038)         &       (0.0211)         \\

		\midrule
		Firm sector controls & \checkmark  & \checkmark  & \checkmark  & \checkmark   \\
		Survey characteristics controls & \checkmark & \checkmark & \checkmark & \checkmark  \\
		Year FE & \checkmark & \checkmark & \checkmark & \checkmark  \\
		\midrule
Shapley decomposition: &   &   &   &    \\
\hspace{3mm} Individual characteristics &   &  \multicolumn{2}{c}{3.93\%}    &    \\
\hspace{3mm} Household characteristics &   &  \multicolumn{2}{c}{15.61\%}    &    \\
\hspace{3mm} Job and Firm characteristics &   &  \multicolumn{2}{c}{69.24\%}    &    \\
\hspace{3mm} Survey characteristics &   &  \multicolumn{2}{c}{0.61\%}    &    \\
\hspace{3mm} Year FE &   &  \multicolumn{2}{c}{10.59\%}    &    \\
		\midrule
N              &          42,046         &          42,046         &           9,704         &          32,110         \\
R2             &         0.0925         &         0.0734         &         0.0545         &         0.0613         \\
Mean dep. var. &         0.2308         &        -0.0678         &         0.1668         &        -0.1391         \\
		\bottomrule
	\end{tabular}
}

%% file: tables/A3_zerobias_5_1_july2024.tex
{
	\def\sym#1{\ifmmode^{#1}\else\(^{#1}\)\fi}
	\begin{tabular}{l*{4}{c}}
		\toprule
		&\multicolumn{1}{c}{(1)} &\multicolumn{1}{c}{(2)} &\multicolumn{1}{c}{(3)} &\multicolumn{1}{c}{(4)}  \\  \addlinespace
		&\multicolumn{1}{c}{Over Y/N}&\multicolumn{1}{c}{Diff of Logs}&\multicolumn{1}{c}{Diff of Logs ($+$)}&\multicolumn{1}{c}{Diff of Logs ($-$)}\\
		\midrule
		\textbf{Individual Characteristics} & & &  \\ \addlinespace
Age            &        -0.0004         &         0.0000         &         0.0002         &         0.0002         \\
               &       (0.0003)         &       (0.0002)         &       (0.0007)         &       (0.0001)         \\
Woman          &        -0.0207\sym{*}  &        -0.0160\sym{*}  &        -0.0196         &        -0.0069\sym{*}  \\
               &       (0.0086)         &       (0.0067)         &       (0.0270)         &       (0.0029)         \\
Years of education&        -0.0012         &         0.0007         &        -0.0001         &         0.0017\sym{***}\\
               &       (0.0010)         &       (0.0009)         &       (0.0032)         &       (0.0004)         \\
\textit{Region (ref. cat: North)} & & & \\
East           &        -0.0005         &        -0.0080         &        -0.0366\sym{*}  &        -0.0001         \\
               &       (0.0104)         &       (0.0056)         &       (0.0153)         &       (0.0036)         \\
South          &        -0.0052         &         0.0092         &         0.0241         &         0.0033         \\
               &       (0.0103)         &       (0.0070)         &       (0.0225)         &       (0.0036)         \\
West           &         0.0085         &         0.0168\sym{*}  &         0.0257         &         0.0077\sym{*}  \\
               &       (0.0095)         &       (0.0065)         &       (0.0212)         &       (0.0033)         \\
\addlinespace
\textit{Migration background (ref. cat: none)} & & & \\
Direct migration background&        -0.0156         &        -0.0098         &        -0.0177         &        -0.0041         \\
               &       (0.0124)         &       (0.0066)         &       (0.0158)         &       (0.0048)         \\
Indirect migration background&        -0.0084         &        -0.0019         &        -0.0033         &        -0.0013         \\
               &       (0.0162)         &       (0.0091)         &       (0.0230)         &       (0.0072)         \\
		\textit{Personality Traits} & & &  \\
Openness (std.)&         0.0020         &        -0.0016         &        -0.0042         &        -0.0015         \\
               &       (0.0045)         &       (0.0027)         &       (0.0082)         &       (0.0015)         \\
Conscientiousness (std.)&         0.0030         &        -0.0008         &        -0.0089         &         0.0012         \\
               &       (0.0046)         &       (0.0030)         &       (0.0092)         &       (0.0019)         \\
Extraversion (std.)&         0.0132\sym{**} &         0.0067\sym{**} &         0.0074         &         0.0003         \\
               &       (0.0040)         &       (0.0026)         &       (0.0084)         &       (0.0015)         \\
Agreeableness (std.)&        -0.0084         &        -0.0049         &        -0.0004         &        -0.0021         \\
               &       (0.0044)         &       (0.0031)         &       (0.0103)         &       (0.0015)         \\
Neuroticism (std.)&         0.0008         &        -0.0003         &        -0.0016         &        -0.0004         \\
               &       (0.0042)         &       (0.0027)         &       (0.0085)         &       (0.0013)         \\
	\textbf{Household characteristics} & & & \\ \addlinespace
Married        &         0.0000         &         0.0022         &         0.0240         &        -0.0059\sym{*}  \\
               &       (0.0090)         &       (0.0046)         &       (0.0125)         &       (0.0029)         \\
Household Size &        -0.0076\sym{*}  &        -0.0019         &         0.0155         &        -0.0040\sym{**} \\
               &       (0.0035)         &       (0.0027)         &       (0.0097)         &       (0.0013)         \\
Number of kids in HH&         0.0087         &         0.0023         &        -0.0179         &         0.0055\sym{***}\\
               &       (0.0044)         &       (0.0035)         &       (0.0121)         &       (0.0016)         \\
Partner        &         0.0043         &         0.0011         &        -0.0277         &         0.0104\sym{**} \\
               &       (0.0102)         &       (0.0056)         &       (0.0145)         &       (0.0035)         \\
Partner $\times$ Partner Inc in 1,000\texteuro &         0.0058\sym{***}&         0.0031\sym{**} &         0.0019         &         0.0013\sym{*}  \\
               &       (0.0016)         &       (0.0012)         &       (0.0027)         &       (0.0005)         \\
		\bottomrule
		\multicolumn{4}{l}{\footnotesize Continued on next page}\\
	\end{tabular}
}						

%% file: tables/A3_zerobias_5_2_july2024.tex
{
	\def\sym#1{\ifmmode^{#1}\else\(^{#1}\)\fi}
	\begin{tabular}{l*{4}{c}}
		\toprule
		&\multicolumn{1}{c}{(1)} &\multicolumn{1}{c}{(2)} &\multicolumn{1}{c}{(3)} &\multicolumn{1}{c}{(4)} \\  \addlinespace
			&\multicolumn{1}{c}{Over Y/N}&\multicolumn{1}{c}{Diff of Logs}&\multicolumn{1}{c}{Diff of Logs ($+$)}&\multicolumn{1}{c}{Diff of Logs ($-$)}\\
		\midrule
\textbf{Job and firm characteristics} & & & \\ \addlinespace
White collar worker&        -0.0504\sym{***}&         0.0031         &         0.0334         &         0.0108\sym{***}\\
               &       (0.0086)         &       (0.0062)         &       (0.0188)         &       (0.0030)         \\
Union memberhsip&        -0.0325\sym{***}&        -0.0112\sym{*}  &         0.0232         &        -0.0073\sym{**} \\
               &       (0.0075)         &       (0.0056)         &       (0.0221)         &       (0.0028)         \\
\multicolumn{2}{l}{\textit{Working Hours (ref. cat.: <35 hours)}} & &  \\
Full-time (35-44)&        -0.0583\sym{***}&        -0.0013         &         0.0019         &         0.0327\sym{***}\\
               &       (0.0078)         &       (0.0055)         &       (0.0186)         &       (0.0032)         \\
Over-time (45+)&        -0.0126         &         0.0129         &         0.0116         &         0.0306\sym{***}\\
               &       (0.0101)         &       (0.0067)         &       (0.0209)         &       (0.0036)         \\
		\textit{Firm Size (ref. cat.: $<$10 employees)} & & & \\
10-49 employees&        -0.0322\sym{**} &        -0.0374\sym{***}&        -0.0511         &        -0.0097\sym{**} \\
               &       (0.0115)         &       (0.0101)         &       (0.0279)         &       (0.0037)         \\
50-249 employees&        -0.0441\sym{***}&        -0.0487\sym{***}&        -0.0634\sym{*}  &        -0.0111\sym{**} \\
               &       (0.0117)         &       (0.0104)         &       (0.0298)         &       (0.0039)         \\
250+ employees &        -0.0582\sym{***}&        -0.0641\sym{***}&        -0.0828\sym{**} &        -0.0215\sym{***}\\
               &       (0.0123)         &       (0.0101)         &       (0.0283)         &       (0.0041)         \\
\textit{Pay tercile firm (ref. cat.: 1st Tercile)} & & &  \\
2. Tercile paying firm&        -0.1091\sym{***}&        -0.0598\sym{***}&        -0.0284         &        -0.0220\sym{***}\\
               &       (0.0084)         &       (0.0056)         &       (0.0146)         &       (0.0027)         \\
3. Tercile paying firm&        -0.1895\sym{***}&        -0.1083\sym{***}&        -0.0260         &        -0.0506\sym{***}\\
               &       (0.0100)         &       (0.0074)         &       (0.0221)         &       (0.0036)         \\
\textit{Workforce composition} & & & \\
Share of highly qualified employees&         0.0104         &         0.0155         &        -0.0261         &         0.0157\sym{*}  \\
               &       (0.0189)         &       (0.0144)         &       (0.0542)         &       (0.0068)         \\
Share of female employees&        -0.0629\sym{**} &        -0.0629\sym{***}&        -0.1547\sym{***}&         0.0056         \\
               &       (0.0194)         &       (0.0151)         &       (0.0421)         &       (0.0058)         \\
Share of German employees&        -0.1319\sym{***}&        -0.0076         &         0.0549         &         0.0249         \\
               &       (0.0383)         &       (0.0212)         &       (0.0484)         &       (0.0139)         \\
Share of fulltime employees&        -0.1217\sym{***}&        -0.1113\sym{***}&        -0.2397\sym{***}&         0.0000         \\
               &       (0.0174)         &       (0.0165)         &       (0.0459)         &       (0.0055)         \\
Mean age of employees in firm&        -0.0013         &        -0.0003         &         0.0001         &        -0.0001         \\
               &       (0.0007)         &       (0.0005)         &       (0.0014)         &       (0.0002)         \\
Constant         &         0.6160\sym{***}&         0.0812\sym{*}  &         0.3832\sym{***}&        -0.2027\sym{***}\\
               &       (0.0623)         &       (0.0369)         &       (0.1096)         &       (0.0220)         \\

		\midrule
		Firm sector controls & \checkmark  & \checkmark  & \checkmark  & \checkmark   \\
		Survey characteristics controls & \checkmark & \checkmark & \checkmark & \checkmark  \\
		Year FE & \checkmark & \checkmark & \checkmark & \checkmark  \\
		\midrule
Shapley decomposition: &   &   &   &    \\
\hspace{3mm} Individual characteristics &   &  \multicolumn{2}{c}{3.93\%}    &    \\
\hspace{3mm} Household characteristics &   &  \multicolumn{2}{c}{15.61\%}    &    \\
\hspace{3mm} Job and Firm characteristics &   &  \multicolumn{2}{c}{69.24\%}    &    \\
\hspace{3mm} Survey characteristics &   &  \multicolumn{2}{c}{0.61\%}    &    \\
\hspace{3mm} Year FE &   &  \multicolumn{2}{c}{10.59\%}    &    \\
		\midrule
N              &          42,046         &          42,046         &           9,375         &          30,979         \\
R2             &         0.0879         &         0.0734         &         0.0582         &         0.0571         \\
Mean dep. var. &         0.2230         &        -0.0678         &         0.1725         &        -0.1441         \\
		\bottomrule
	\end{tabular}
}

%% file: tables/04_shapeley_values_v5.tex
{
\def\sym#1{\ifmmode^{#1}\else\(^{#1}\)\fi}
\begin{tabular}{lc}
\toprule
Factor & All  \\
\midrule
\textbf{Individual characteristics} &  \\
Socio-demographic variables 	& 2.51\%  \\
Personality traits 				& 1.42\%  \\
\textbf{Household characteristics} & \\
Household characteristics 		& 15.61\%  \\
\textbf{Job and Firm characteristics} & \\
Job characteristics				& 1.31\%  \\				
Number of employees 			& 29.50\% \\
Wage tercile firm 				& 15.48\% \\
Sector							& 17.66\% \\
Workforce composition 			& 5.29\% \\
\textbf{Survey characteristics} & \\
Survey characteristics 			& 0.61\%   \\
\textbf{Year FE} &  \\
Year FE 			& 10.59\%    \\
\midrule
N &  42,046  \\
R2 & 0.07 \\
\bottomrule
\end{tabular}
}

%% file: Reporting_Bias_Caliendo_et_al.bbl
\begin{thebibliography}{72}
\providecommand{\natexlab}[1]{#1}

\bibitem[{Abowd and Stinson(2013)}]{AbowdStinson2013}
\textsc{Abowd, J.} and \textsc{Stinson, M.} (2013). {Estimating measurement
  error in annual job earnings: a comparison of survey and administrative
  data}. \textit{The Review of Economics and Statistics}, \textbf{95}~(5),
  1451--1467.

\bibitem[{Acemoglu(1997)}]{Acemoglu1997}
\textsc{Acemoglu, D.} (1997). {Training and innovation in an imperfect labour
  market}. \textit{The Review of Economic Studies}, \textbf{64}~(3), 445--464.

\bibitem[{Adda \textit{et~al.}(2017)Adda, Dustmann and Stevens}]{Addaetal17}
\textsc{Adda, J.}, \textsc{Dustmann, C.} and \textsc{Stevens, K.} (2017). {The
  career costs of children}. \textit{Journal of Political Economy},
  \textbf{125}~(2), 293--337.

\bibitem[{Angel \textit{et~al.}(2019)Angel, Disslbacher, Humer and
  Schnetzer}]{Angel2019}
\textsc{Angel, S.}, \textsc{Disslbacher, F.}, \textsc{Humer, S.} and
  \textsc{Schnetzer, M.} (2019). What did you really earn last year?:
  Explaining measurement error in survey income data. \textit{Journal of the
  Royal Statistical Society Series A}, \textbf{182}~(4), 1411--1437.

\bibitem[{Angel \textit{et~al.}(2018)Angel, Heuberger and Lamei}]{Angel2018}
\textsc{---}, \textsc{Heuberger, R.} and \textsc{Lamei, N.} (2018). Differences
  between household income from surveys and registers and how these affect the
  poverty headcount: Evidence from the {Austrian SILC}. \textit{Social
  Indicators Research}, \textbf{138}, 575--603.

\bibitem[{Antoni \textit{et~al.}(2023)Antoni, Beckmannshagen, Grabka, Keita and
  Trübswetter}]{Antoni2023}
\textsc{Antoni, M.}, \textsc{Beckmannshagen, M.}, \textsc{Grabka, M.~M.},
  \textsc{Keita, S.} and \textsc{Trübswetter, P.} (2023). {Befragungsdaten der
  SOEP-Core-, IAB-SOEP Migrationsstichprobe, IAB-BAMF-SOEP Befragung von
  Geflüchteten und SOEP-Innovationssample verknüpft mit administrativen Daten
  des IAB (SOEP-CMI-ADIAB) 1975-2020}. \textit{FDZ-Datenreport, 03/2023(de)}.

\bibitem[{Antoni \textit{et~al.}(2019)Antoni, Bela and Vicari}]{Antoni2019}
\textsc{---}, \textsc{Bela, D.} and \textsc{Vicari, B.} (2019). Validating
  earnings in the {G}erman {N}ational {E}ducational {P}anel {S}tudy.
  {D}eterminants of measurement accuracy of survey questions on earnings.
  \textit{methods, data, analyses}, \textbf{13}~(1), 32.

\bibitem[{Arni \textit{et~al.}(2014)Arni, Caliendo, Künn and
  Zimmermann}]{Arni2014}
\textsc{Arni, P.}, \textsc{Caliendo, M.}, \textsc{Künn, S.} and
  \textsc{Zimmermann, K.} (2014). {The IZA evaluation dataset survey: A
  scientific use file}. \textit{IZA Journal of European Labor Studies},
  \textbf{3}~(6), 1--20.

\bibitem[{Becker(1962)}]{Becker1962}
\textsc{Becker, G.~S.} (1962). {Investment in human capital: A theoretical
  analysis}. \textit{Journal of Political Economy}, \textbf{70}~(5), 9--49.

\bibitem[{Becker and Chiswick(1966)}]{Becker1966}
\textsc{---} and \textsc{Chiswick, B.~R.} (1966). Education and the
  distribution of earnings. \textit{American Economic Review},
  \textbf{56}~(1/2), 358--369.

\bibitem[{Bee \textit{et~al.}(2023)Bee, Mitchell, Mittag, Rothbaum, Sanders,
  Schmidt and Unrath}]{mittag2023}
\textsc{Bee, A.}, \textsc{Mitchell, J.}, \textsc{Mittag, N.}, \textsc{Rothbaum,
  J.}, \textsc{Sanders, C.}, \textsc{Schmidt, L.} and \textsc{Unrath, M.}
  (2023). National experimental wellbeing statistics. \textit{SEHSD Working
  Paper, 2023-02}.

\bibitem[{Bertrand \textit{et~al.}(2010)Bertrand, Goldin and
  Katz}]{Bertrandetal10}
\textsc{Bertrand, M.}, \textsc{Goldin, C.} and \textsc{Katz, L.~F.} (2010).
  {Dynamics of the gender gap for young professionals in the financial and
  corporate sectors}. \textit{American Economic Journal: Applied Economics},
  \textbf{2}~(3), 228–255.

\bibitem[{Bertrand \textit{et~al.}(2015)Bertrand, Kamenica and
  Pan}]{bertrand2015gender}
\textsc{---}, \textsc{Kamenica, E.} and \textsc{Pan, J.} (2015). Gender
  identity and relative income within households. \textit{The Quarterly Journal
  of Economics}, \textbf{130}~(2), 571--614.

\bibitem[{Bertrand and Mullainathan(2001)}]{bertrand2001people}
\textsc{---} and \textsc{Mullainathan, S.} (2001). Do people mean what they
  say? {I}mplications for subjective survey data. \textit{American Economic
  Review}, \textbf{91}~(2), 67--72.

\bibitem[{Blau and Kahn(2017)}]{blau2017gender}
\textsc{Blau, F.~D.} and \textsc{Kahn, L.~M.} (2017). The gender wage gap:
  Extent, trends, and explanations. \textit{Journal of Economic Literature},
  \textbf{55}~(3), 789--865.

\bibitem[{Bollinger(1998)}]{Bollinger1998}
\textsc{Bollinger, C.~R.} (1998). Measurement error in the current population
  survey: A nonparametric look. \textit{Journal of Labor Economics},
  \textbf{16}~(3), 576--594.

\bibitem[{Bollinger \textit{et~al.}(2019)Bollinger, Hirsch, Hokayem and
  Ziliak}]{Bollingeretal2019}
\textsc{---}, \textsc{Hirsch, B.~T.}, \textsc{Hokayem, C.} and \textsc{Ziliak,
  J.~P.} (2019). \textit{Journal of Political Economy}, \textbf{127}~(5),
  2143--2185.

\bibitem[{Bollinger \textit{et~al.}(2018)Bollinger, Hirsch, Hokayem and
  Ziliak}]{Bollinger2018}
\textsc{---}, \textsc{---}, \textsc{Hokayem, C.~M.} and \textsc{Ziliak, J.~P.}
  (2018). The good, the bad and the ugly: measurement error, non-response and
  administrative mismatch in the cps. \textit{Working Paper, Gatton College of
  Business, University of Kentucky}.

\bibitem[{Bollinger and Tasseva(2023)}]{bollinger2022income}
\textsc{---} and \textsc{Tasseva, I.~V.} (2023). {Income source confusion using
  the SILC}. \textit{Public Opinion Quaterly}, \textbf{87}~(S1), 542--574.

\bibitem[{Borghans \textit{et~al.}(2008)Borghans, Duckworth, Heckman and
  Ter~Weel}]{Borghans2008}
\textsc{Borghans, L.}, \textsc{Duckworth, A.}, \textsc{Heckman, J.} and
  \textsc{Ter~Weel, B.} (2008). {The economics and psychology of personality
  traits}. \textit{Journal of Human Resources}, \textbf{43}~(4), 972--1059.

\bibitem[{Bossler and Westermeier(2020)}]{Bossler2020}
\textsc{Bossler, M.} and \textsc{Westermeier, C.} (2020). Measurement error in
  minimum wage evaluations using survey data. \textit{IAB-Discussion Paper,
  11/2020}.

\bibitem[{Bound \textit{et~al.}(1994)Bound, Brown, Duncan and
  Rodgers}]{Bound1994}
\textsc{Bound, J.}, \textsc{Brown, C.}, \textsc{Duncan, G.} and
  \textsc{Rodgers, W.~L.} (1994). Evidence on the validity of cross-sectional
  and longitudinal labor market data. \textit{Journal of Labor Economics},
  \textbf{12}~(3), 345--368.

\bibitem[{Bound \textit{et~al.}(2001)Bound, Brown and Mathiowetz}]{Bound2001}
\textsc{---}, \textsc{---} and \textsc{Mathiowetz, N.} (2001). Chapter 59 -
  {M}easurement error in survey data. \textit{Handbook of Econometrics},
  vol.~5, Elsevier, pp. 3705--3843.

\bibitem[{Bound and Krueger(1991)}]{Bound1991}
\textsc{---} and \textsc{Krueger, A.} (1991). The extent of measurement error
  in longitudinal earnings data: Do two wrongs make a right? \textit{Journal of
  Labor Economics}, \textbf{9}~(1), 1--24.

\bibitem[{Boyce \textit{et~al.}(2010)Boyce, Brown and Moore}]{Boyceetal2010}
\textsc{Boyce, C.}, \textsc{Brown, G.} and \textsc{Moore, S.} (2010). {Money
  and happiness: Rank of income, not income, affects life satisfaction.}
  \textit{Psychological Science}, \textbf{21}~(4), 471--475.

\bibitem[{Bricker and Engelhardt(2008)}]{Bricker2008}
\textsc{Bricker, J.} and \textsc{Engelhardt, G.~V.} (2008). Measurement error
  in earnings data in the health and retirement study. \textit{Journal of
  Economic and Social Measurement}, \textbf{33}~(1), 39--61.

\bibitem[{Cheung and Lucas(2015)}]{CheungLucas2015}
\textsc{Cheung, F.} and \textsc{Lucas, R.} (2015). { When does money matter
  most? Examining the association between income and life satisfaction over the
  life course.} \textit{Psychological Aging}, \textbf{30}~(1), 120--135.

\bibitem[{Dauth and Eppelsheimer(2020)}]{dautheppelsheimer20}
\textsc{Dauth, W.} and \textsc{Eppelsheimer, J.} (2020). {Preparing the sample
  of integrated labour market biographies (SIAB) for scientific analysis: A
  guide}. \textit{Journal for Labour Market Research}, \textbf{54}~(10).

\bibitem[{Deaton(2003)}]{deaton2003health}
\textsc{Deaton, A.} (2003). Health, inequality, and economic development.
  \textit{Journal of Economic Literature}, \textbf{41}~(1), 113--158.

\bibitem[{Deaton(2008)}]{deaton2008income}
\textsc{---} (2008). Income, health, and well-being around the world: Evidence
  from the gallup world poll. \textit{Journal of Economic Perspectives},
  \textbf{22}~(2), 53--72.

\bibitem[{Destatis(2024)}]{destatis}
\textsc{Destatis} (2024). {Qualität der Arbeit - Gender Pay Gap}.

\bibitem[{Duncan and Hill(1985)}]{Duncan1985}
\textsc{Duncan, G.} and \textsc{Hill, D.} (1985). An investigation of the
  extent and consequences of measurement error in labor-economic survey data.
  \textit{Journal of Labor Economics}, \textbf{3}~(4), 508--532.

\bibitem[{Frijters \textit{et~al.}(2004)Frijters, Haisken-DeNew and
  Shields}]{Frijtersetal2004}
\textsc{Frijters, P.}, \textsc{Haisken-DeNew, J.~P.} and \textsc{Shields,
  M.~A.} (2004). {Money does matter! Evidence from increasing real income and
  life satisfaction in {East Germany} following reunification}.
  \textit{American Economic Review}, \textbf{94}~(3), 730--740.

\bibitem[{Frodermann \textit{et~al.}(2021)Frodermann, Schmucker, Seth and vom
  Berge}]{Frodermann2021}
\textsc{Frodermann, C.}, \textsc{Schmucker, A.}, \textsc{Seth, S.} and
  \textsc{vom Berge, P.} (2021). {Stichprobe der Integrierten
  Arbeitsmarktbiografien (SIAB) 1975-2019}. \textit{FDZ Datenreport, 01/2021
  (de)}.

\bibitem[{Fuchs \textit{et~al.}(1997)Fuchs, Krueger and
  Poterba}]{fuchs1997economists}
\textsc{Fuchs, V.~R.}, \textsc{Krueger, A.~B.} and \textsc{Poterba, J.} (1997).
  Why do economists disagree about policy? \textit{NBER Working Paper, 6151}.

\bibitem[{Goebel \textit{et~al.}(2019)Goebel, Grabka, Liebig, Kroh, Richter,
  Schr{\"o}der and Schupp}]{Goebel2019}
\textsc{Goebel, J.}, \textsc{Grabka, M.~M.}, \textsc{Liebig, S.}, \textsc{Kroh,
  M.}, \textsc{Richter, D.}, \textsc{Schr{\"o}der, C.} and \textsc{Schupp, J.}
  (2019). {The German Socio-Economic Panel (SOEP)}. \textit{Jahrb{\"u}cher
  f{\"u}r National{\"o}konomie und Statistik}, \textbf{239}~(2), 345--360.

\bibitem[{Goldin(2014)}]{Goldin14}
\textsc{Goldin, C.} (2014). {A grand gender convergence: Its last chapter}.
  \textit{American Economic Review}, \textbf{104}~(4), 1091--1119.

\bibitem[{Gottschalk(2005)}]{Gottschalk2005}
\textsc{Gottschalk, P.} (2005). Downward nominal-wage flexibility: Real or
  measurement error? \textit{The Review of Economics and Statistics},
  \textbf{87}~(3), 556--568.

\bibitem[{Gottschalk and Huynh(2005)}]{GottschalkHuynh2005}
\textsc{---} and \textsc{Huynh, M.} (2005). Validation study of earnings data
  in the {SIPP - Do} older workers have larger measurement error?
  \textit{Center for Retirement Research Working Papers, 2005-07}.

\bibitem[{Gottschalk and Huynh(2010)}]{Gottschalk2010}
\textsc{---} and \textsc{---} (2010). Are earnings inequality and mobility
  overstated? {T}he impact of nonclassical measurement error. \textit{The
  Review of Economics and Statistics}, \textbf{92}~(2), 302--315.

\bibitem[{Hurst \textit{et~al.}(2014)Hurst, Li and Pugsley}]{Hurstetal2014}
\textsc{Hurst, E.}, \textsc{Li, G.} and \textsc{Pugsley, B.} (2014). {Are
  household surveys like tax forms? Evidence from income underreporting of
  self-employed}. \textit{The Review of Economics and Statistics},
  \textbf{96}~(1), 19--33.

\bibitem[{Isaoglu(2010)}]{Isaoglu2010}
\textsc{Isaoglu, A.} (2010). Occupational affiliation data and measurement
  errors in the {G}erman {S}ocio-{E}conomic {P}anel. \textit{SOEP Papers on
  Multidisciplinary Panel Data Research, 318}.

\bibitem[{Juarez(2012)}]{Juarez2012}
\textsc{Juarez, F.} (2012). {SHAPLEY2: Stata module to compute additive
  decomposition of estimation statistics by regressors or groups of
  regressors}. \textit{Statistical Software Components S457543, Boston College
  Department of Economics}.

\bibitem[{Kapteyn and Ypma(2007)}]{Kapteyn2007}
\textsc{Kapteyn, A.} and \textsc{Ypma, J.} (2007). Measurement error and
  misclassification: A comparison of survey and administrative data.
  \textit{Journal of Labor Economics}, \textbf{25}~(3), 513--551.

\bibitem[{Kearl \textit{et~al.}(1979)Kearl, Pope, Whiting and
  Wimmer}]{kearl1979confusion}
\textsc{Kearl, J.~R.}, \textsc{Pope, C.~L.}, \textsc{Whiting, G.~C.} and
  \textsc{Wimmer, L.~T.} (1979). A confusion of economists? \textit{American
  Economic Review}, \textbf{69}~(2), 28--37.

\bibitem[{Kim and Solon(2005)}]{Kim2005}
\textsc{Kim} and \textsc{Solon, G.} (2005). Implications of mean-reverting
  measurement error for longitudinal studies of wages and employment.
  \textit{The Review of Economics and Statistics}, \textbf{87}~(1), 193--196.

\bibitem[{Kim and Tamborini(2014)}]{Kim2014}
\textsc{---} and \textsc{Tamborini, C.} (2014). Response error in earnings: An
  analysis of the survey of income and program participation matched with
  administrative data. \textit{Sociological Methods \& Research},
  \textbf{43}~(1), 39--72.

\bibitem[{Kleven \textit{et~al.}(2019{\natexlab{a}})Kleven, Landais, Posch,
  Steinhauer and Zweimüller}]{Klevenetal19b}
\textsc{Kleven, H.}, \textsc{Landais, C.}, \textsc{Posch, J.},
  \textsc{Steinhauer, A.} and \textsc{Zweimüller, J.} (2019{\natexlab{a}}).
  {Child penalties across countries: Evidence and explanations}. \textit{AEA
  Papers and Proceedings}, \textbf{109}, 122--126.

\bibitem[{Kleven \textit{et~al.}(2019{\natexlab{b}})Kleven, Landais and
  Sogaard}]{Klevenetal19a}
\textsc{---}, \textsc{---} and \textsc{Sogaard, J.~E.} (2019{\natexlab{b}}).
  {Children and gender inequality: Evidence from Denmark}. \textit{American
  Economic Journal: Applied Economics}, \textbf{11}~(4), 181--209.

\bibitem[{K{\"u}nn(2015)}]{kunn2015challenges}
\textsc{K{\"u}nn, S.} (2015). The challenges of linking survey and
  administrative data. \textit{IZA World of Labor, 214}.

\bibitem[{Madeira \textit{et~al.}(2022)Madeira, Margaretic, Martínez and
  Roje}]{Madeira2022}
\textsc{Madeira, C.}, \textsc{Margaretic, P.}, \textsc{Martínez, F.} and
  \textsc{Roje, P.} (2022). Assessing the quality of self-reported financial
  information. \textit{Journal of Survey Statistics and Methodology},
  \textbf{10}~(5), 1183--1210.

\bibitem[{Meijer \textit{et~al.}(2012)Meijer, Rohwedder and
  Wansbeek}]{Meijeretal2012}
\textsc{Meijer, E.}, \textsc{Rohwedder, S.} and \textsc{Wansbeek, T.} (2012).
  {Measurement Error in Earnings Data: Using a Mixture Model Approach to
  Combine Survey and Register Data}. \textit{Journal of Business \& Economic
  Statistics}, \textbf{30}~(2), 191--201.

\bibitem[{Meyer and Mittag(2019)}]{meyer2019using}
\textsc{Meyer, B.~D.} and \textsc{Mittag, N.} (2019). Using linked survey and
  administrative data to better measure income: Implications for poverty,
  program effectiveness, and holes in the safety net. \textit{American Economic
  Journal: Applied Economics}, \textbf{11}~(2), 176--204.

\bibitem[{Meyer \textit{et~al.}(2024)Meyer, Mittag and Wu}]{Meyer2024race}
\textsc{---}, \textsc{---} and \textsc{Wu, D.} (2024). \textit{Race, Ethnicity,
  and Measurement Error}, University of Chicago Press.

\bibitem[{Mincer(1974)}]{Mincer1974}
\textsc{Mincer, J.} (1974). {Age and Experience Profiles of Earnings}. In
  J.~Mincer (ed.), \textit{Schooling, Experience, and Earnings}, National
  Bureau of Economic Research, pp. 64--82.

\bibitem[{Moore \textit{et~al.}(2000)Moore, Stinson and
  Welniak}]{Mooreetal2000}
\textsc{Moore, J.}, \textsc{Stinson, L.} and \textsc{Welniak, E.} (2000).
  {Income measurement error in surveys: a review}. \textit{Journal of Official
  Statistics}, \textbf{16}~(4), 331--361.

\bibitem[{Mulligan and Rubinstein(2008)}]{Mulligan2008}
\textsc{Mulligan, C.} and \textsc{Rubinstein, Y.} (2008). {Selection,
  investment, and women's relative wages over time}. \textit{The Quarterly
  Journal of Economics}, \textbf{123}~(3), 1061--1110.

\bibitem[{Pencavel(1986)}]{Pencavel1986}
\textsc{Pencavel, J.} (1986). {Chapter 1: Labor supply of men: A survey}.
  \textit{Handbook of Labor Economics}, vol.~1, Elsevier, pp. 3--102.

\bibitem[{Pischke(1995)}]{Pischke1995}
\textsc{Pischke, J.-S.} (1995). Measurement error and earnings dynamics: Some
  estimates from the {PSID} validation study. \textit{Journal of Business \&
  Economic Statistics}, \textbf{13}~(3), 305--314.

\bibitem[{Prati(2017)}]{prati2017hedonic}
\textsc{Prati, A.} (2017). Hedonic recall bias. {Why} you should not ask people
  how much they earn. \textit{Journal of Economic Behavior \& Organization},
  \textbf{143}, 78--97.

\bibitem[{Richter and Schupp(2015)}]{RichterSchupp2015}
\textsc{Richter, D.} and \textsc{Schupp, J.} (2015). {The SOEP Innovation
  Sample (SOEP IS)}. \textit{Schmollers Jahrbuch}, \textbf{135}~(3), 389--399.

\bibitem[{Roth and Slotwinski(2021)}]{roth2021gender}
\textsc{Roth, A.} and \textsc{Slotwinski, M.} (2021). Gender norms and income
  misreporting within households. \textit{CESifo Working Paper, 7298}.

\bibitem[{Sakshaug and Kreuter(2012)}]{SakshaugKreuter2012}
\textsc{Sakshaug, J.} and \textsc{Kreuter, F.} (2012). {Assessing the Magnitude
  of Non-Consent Biases in Linked Survey and Administrative Data}.
  \textit{Survey Research Methods}, \textbf{6}~(2), 113--122.

\bibitem[{Sandmo(1970)}]{Sandmo1970}
\textsc{Sandmo, A.} (1970). {The effect of uncertainty on saving decisions}.
  \textit{The Review of Economic Studies}, \textbf{37}~(3), 353--360.

\bibitem[{Schmillen \textit{et~al.}(2024)Schmillen, Umkehrer and von
  Wachter}]{schmillen2024measurement}
\textsc{Schmillen, A.}, \textsc{Umkehrer, M.} and \textsc{von Wachter, T.}
  (2024). Measurement error in longitudinal earnings data: Evidence from
  germany. \textit{Journal for Labour Market Research}, forthcoming.

\bibitem[{Shorrocks(2013)}]{Shorrocks2013}
\textsc{Shorrocks, A.} (2013). {Decomposition procedures for distributional
  analysis: A unified framework based on the Shapley value}. \textit{Journal of
  Economic Inequality}, \textbf{11}, 99--126.

\bibitem[{SOEP(2022)}]{SOEP2022}
\textsc{SOEP} (2022). {Sozio-oekonomisches Panel (SOEP), Version 37, data from
  1984-2020 (SOEP-Core v37, EU-Edition)}.

\bibitem[{SOEP-IS(2020)}]{SOEPIS2020}
\textsc{SOEP-IS} (2020). {SOEP Innovation Sample (SOEP-IS), data from
  1998-2019}.

\bibitem[{Solon(1992)}]{solon1992intergenerational}
\textsc{Solon, G.} (1992). {Intergenerational income mobility in the United
  States}. \textit{American Economic Review}, \textbf{32}~(3), 393--408.

\bibitem[{Stantcheva(2022)}]{Stantcheva2022}
\textsc{Stantcheva, S.} (2022). How to run surveys: A guide to creating your
  own identifying variation and revealing the invisible. \textit{NBER Working
  Paper, No. 30527}.

\bibitem[{Stüber \textit{et~al.}(2023)Stüber, Grabka and
  Schnitzlein}]{Stueber2023}
\textsc{Stüber, H.}, \textsc{Grabka, M.} and \textsc{Schnitzlein, D.} (2023).
  {A tale of two data sets: Comparing German administrative and survey data
  using wage inequality as an example}. \textit{Journal for Labour Market
  Research}, \textbf{57}~(8).

\bibitem[{Valet \textit{et~al.}(2019)Valet, Adriaans and
  Liebig}]{valet2018comparing}
\textsc{Valet, P.}, \textsc{Adriaans, J.} and \textsc{Liebig, S.} (2019).
  Comparing survey data and administrative records on gross earnings:
  nonreporting, misreporting, interviewer presence and earnings inequality.
  \textit{Quality \& Quantity}, \textbf{53}, 471--491.

\end{thebibliography}
